%% file: thesis.tex
\documentclass[12pt,twoside,openright,a4paper]{book}
\usepackage[english]{babel}
\usepackage{phdthesis}

\usepackage[table,xcdraw]{xcolor}

\usepackage{cite}
\usepackage[pdftex]{graphicx}
   \graphicspath{{figs/}}
   \DeclareGraphicsExtensions{.pdf,.jpg,.png}
\usepackage[colorinlistoftodos]{todonotes}
\usepackage{comment}
\usepackage{enumitem}
\usepackage{soul}
\usepackage{hyperref}
\hypersetup{
    colorlinks,
    citecolor=black,
    filecolor=black,
    linkcolor=black,
    urlcolor=black
}

\usepackage{notoccite}

\usepackage{bytefield}

\usepackage{psboxit}
\PScommands

\usepackage[splitrule]{footmisc} 
\usepackage{fancyhdr} 
\usepackage{bibentry}
\usepackage{xspace}
\usepackage{latexsym}    
\usepackage{url}
\usepackage{array}
\usepackage{times}
\usepackage{amsmath,epsfig,amsfonts}
\usepackage{subfig}
\usepackage[toc,page]{appendix}
\usepackage{latexsym}

\usepackage{minitoc}

\usepackage[left=3.25cm, right=2.25cm, vmargin=3.5cm, twoside]{geometry}
\usepackage{algorithm}
\usepackage{algpseudocode}
\usepackage{amsthm}
\usepackage{rotating}
\usepackage[avantgarde]{quotchap}

\usepackage{nomencl}

\makenomenclature 

\usepackage{amssymb} 

\newcommand{\commentPHD}[2]{}

\usepackage[]{subfig}
\usepackage{float}   

\usepackage{nomencl}
\makenomenclature
\newfloat{map}{tbph}{lom}
\restylefloat*{map} 
\floatstyle{plain}	
\floatname{map}{Table}
\newcommand{\be}{\begin{equation}}
\newcommand{\ee}{\end{equation}}
\newcommand{\beqn}{\begin{eqnarray}}
\newcommand{\eeqn}{\end{eqnarray}}
\newcommand{\bea}{\begin{eqnarray}}
\newcommand{\eea}{\end{eqnarray}}
\newcommand{\beas}{\begin{eqnarray*}}
\newcommand{\eeas}{\end{eqnarray*}}

\newsubfloat[position=bottom,listofformat=subsimple]{map}

\usepackage{setspace}

\newcommand{\dnp}{DNP\xspace}
\newcommand{\apenet}{APEnet\xspace}
\newcommand{\apenetp}{APEnet+\xspace}
\newcommand{\apenetv}{APEnet+ V5\xspace}
\newcommand{\apelink}{APElink\xspace}
\newcommand{\aperouter}{APErouter\xspace}
\newcommand{\apepacket}{APEpacket\xspace}
\newcommand{\apephy}{APEphy\xspace}
\newcommand{\apeni}{APE Network Interface\xspace}

\newcommand{\pcie}{PCIe\xspace}
\newcommand{\nvidia}{NVIDIA\xspace}

\newcommand{\quong}{QUonG\xspace}
\newcommand{\ie}{\textit{i.e.}\xspace}

\newcommand{\exanest}{ExaNeSt\xspace}
\newcommand{\comm}[1]{}

\newcommand{\nanet}{NaNet\xspace}

\newcommand{\realtime}{real-time\xspace}

\newcommand{\gpudirect}{GPUDirect\xspace}

\newcommand{\rdma}{RDMA\xspace}

\newcommand{\exanet}{ExaNet\xspace}

\usepackage{acronym}

\begin{document}

\pagestyle{empty}
\input{sections/frontespizio}\setlength{\baselineskip}{1.60\baselineskip}
\pagenumbering{roman}
\onehalfspacing
\setlength{\headheight}{28pt}


\dominitoc
\tableofcontents   
\listoftables
\listoffigures	

\chapter*{Acronyms}
\begin{acronym}
\acro{HPC}{High Performance Computing}
\acro{FFT}{Fast Fourier Transform}
\acro{DFT}{Discrete Fourier Transform}
\acro{FPGA}{Field Programmable Gate Arrays}
\acro{SoC}{System on a Chip}
\acro{HBM}{High Memory Bandwidth}
\acro{GPGPU}{General Purpose Graphical Processing Units}
\acro{VLSI}{Very Large Scale Integration}
\acro{HDL}{Hardware Description Language}
\acro{UDP}{User Datagram Protocol}
\acro{IP}{Internet Protocol}
\acro{APE}{Array Processor Experiment}
\acro{DNP}{Distributed Network Processor}
\acro{DOR}{Dimension Ordered Routing}
\acro{VCT}{Virtual Cut Through}
\acro{DMA}{Direct Memory Access}
\acro{RDMA}{Remote Direct Memory Access}
\acro{HEP}{High Energy Physics}
\acro{NIC}{Network Interface Card}
\acro{GbE}{Gigabit Ethernet}
\acro{PCIe}{Peripheral Component Interconnect Express}
\acro{MAC}{Medium Access Control}
\acro{PHY}{Physical Layer}
\acro{SFP}{Small Form-Factor Pluggable}
\acro{QSFP}{Quad Small Form-Factor Pluggable}
\acro{DIT}{Decimation in Time}
\acro{DIF}{Decimation in Frequency}
\acro{CLB}{Configurable Logic Block}
\acro{LUT}{Look Up Table}
\acro{DSP}{Digital Signal Processing}
\acro{FLOPS}{Floating Point Operations per Second}
\acro{BRAM}{Block RAM}

\end{acronym}

\mainmatter
\pagenumbering{arabic}
\pagestyle{fancy}


\input{sections/introduction}

\input{sections/network}

\input{sections/3d_fft}

\input{sections/arch_model}

\input{sections/results}

\input{sections/conclusions}


\cleardoublepage
\phantomsection
\addcontentsline{toc}{chapter}{Bibliography}  
\bibliographystyle{IEEEtran}
\bibliography{bibliography}

\end{document}

%% file: sections/frontespizio.tex


\begin{center}

{\includegraphics[width=3.5cm]{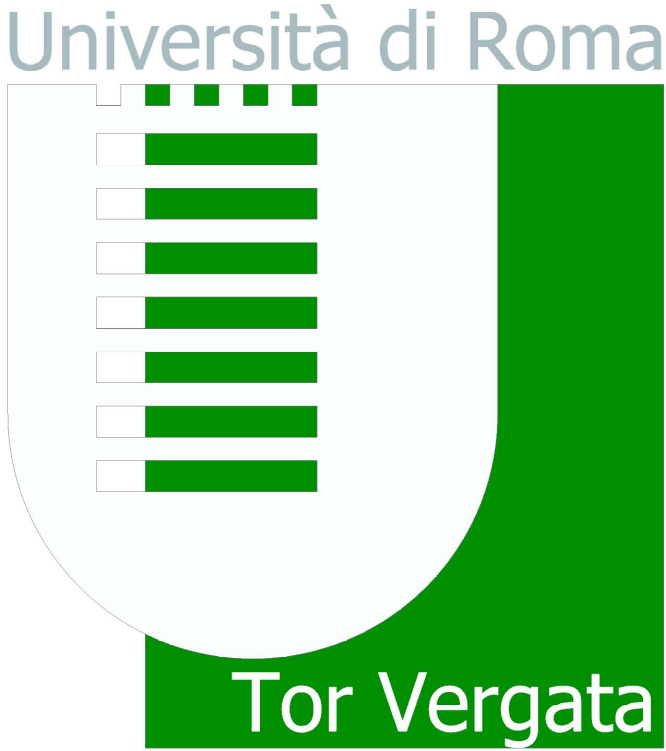} }

\vspace*{1.5cm}
\LARGE
{\sc A multi-FPGA High Performance Computing System for 3D FFT-based numerical simulations}\\

\vspace*{1.5 cm}
\LARGE
{Roberto Ammendola}

\vspace*{1.5 cm}
\large 
{\sc PhD Course in Electronic Engineering}
\vspace*{0.7 cm}

\vspace*{0.7 cm}
\Large
{\sc Universit\`a di Roma Tor Vergata} \\

\vspace*{0.4 cm}
\normalsize
{\sc Department of Electronic Engineering}

\end{center}

\vskip 2.5cm

{\sc Tutor: \hspace*{8.0 cm} Pierpaolo Loreti }

{\sc Course Director: \hspace*{5.72 cm} Aldo Di Carlo}




\vspace*{1 cm}

\begin{center}
\large
{a.a. 2017/2018}\\

\vspace*{0.5 cm}
{XXXI Ciclo}\\
\end{center}
\hrule


\newpage

\large
{This thesis was evaluated by the two following external referees:}\\

\begin{center}

  \vspace*{1.0cm}
  
  \large
  {\sc Piero Vicini}\\
  \normalsize
  {Senior Researcher at Istituto Nazionale di Fisica Nucleare, Sezione di Roma}\\
  
  \vspace*{2.0 cm}
  
  \large
  {\sc Salvatore Pontarelli}\\
  \normalsize
  {Researcher at Consorzio Nazionale Interuniversitario per le Telecomunicazioni}\\

\end{center}

\vspace*{2.0 cm}
{The time and effort of the external referees in evaluating this thesis, as well as their valuable and constructive suggestions, are very much appreciated and greatly acknowledged. }\\

\vskip 0.2cm
\hrule
\vskip 20 pt

\newpage


  	
  
	




%
	

%% file: sections/introduction.tex
\chapter{Introduction}
\minitoc

In the field of High Performance Computing, communications among processes represent a typical bottleneck for massively parallel scientific applications. Object of this research is the development of a network interface card with specific offloading capabilities that could help large scale simulations in terms of communication latency and scalability with the number of computing elements.

Until the early 2000s, general purpose single-core CPU-based systems were the processing systems of choice for HPC applications. They replaced exotic supercomputing architectures because they were inexpensive, and performance scaled with frequency in line with Moore's Law. After the mid-2000s the multi-core architecture era started as the only viable solution to keep up with predicted performance scaling. It is around year 2010 that CPU-based systems augmented with hardware accelerators as co-processors started to emerge as an alternative to CPU-only systems. This has opened up opportunities for accelerators, mainly General Purpose Graphics Processing Units (GPGPUs) to advance HPC to previously unattainable performance levels \cite{sundararajan2010high}.

Since then, programmable device technology (namely Field Programmable Gate Arrays, or FPGA), while sharing the same silicon complexity of a GPGPU, has struggled to emerge as a real accelerator competitor, mainly due (i) to the lack of well-established high level synthesis tools, (ii) higher costs and slower lead times, (iii) an actually poor result in terms of time-to-solution \cite{weber2011comparing}. By mitigating these negative aspects, the FPGA technology started becoming known and widespread lately by leveraging its own peculiarities, which are the re-configurable computing approach and the high power efficiency \cite{georgopoulos2019energy} \cite{muslim2017efficient}.

Moreover nowadays FPGAs, thanks to the variety of embedded on-chip resources, allow offloading of increasingly complex tasks: not only for the pure computational part on an algorithm, but also for the communication part in the case of distributed parallel systems \cite{SANCHEZCORREA201841}.

In particular in this thesis a specific computational task has been addressed, the three-dimensional Fast Fourier Transform (3D FFT), which is peculiarly weighty for the interconnection network when parallel systems are involved. The main goal of this study is finding a clever way to move part of the computational weight closer to the network, in order to exploit the communication patterns peculiarities and eventually take advantage of data reuse within the process of transmission.

\section{FPGA as accelerator engines}

Modern HPC systems are evaluating inclusion of FPGAs as components of their system architectures because they can combine effective hardware acceleration capabilities and dedicated communication facilities in a single device. The resulting design is suitable to effectively execute distributed tasks in computer clusters. An actual example is Microsoft Catapult \cite{caulfield2016cloud}, a data center able to act as an HPC system thanks to the introduction of FPGA accelerators. In this context, FPGAs also allow optimizing the data exchange among the hardware acceleration modules thanks to the direct connections supported by the network controllers which are integrated into the programmable hardware.


FPGA devices today make available a great amount of configurable logic and in addition to this they recently started to implement an always increasing amount of fixed specialized hardware blocks, which include floating point arithmetic engines, several Mbits of embedded memory blocks, memory controllers for external RAM, dozens of high speed transceivers for integrated low latency board-to-board communication with standard protocols. They also started to integrate embedded processors, such as ARM, with hardwired peripherals for storage (SATA) and networking (Ethernet), so that they are often referred to as System on a Chip (SoC) devices. Moreover, devices with in-package memory chips, such as High Bandwidth Memory (HBM), are becoming available as on latest generation GPGPUs; the tremendous memory bandwidth that HBM will bring to FPGA devices will impact the scenario for accelerators in the near future. 

Furthermore, the range of problem size of typical interest requires local memory size in the order of GB, which makes impossible to use only on-chip memory to buffer data during the 3D FFT calculation. Thus we can not avoid to use an off-chip memory in order to satisfy the buffering requirements (being external memory banks or in-package HBM).



\section{A case study: Computational Fluid Dynamics}


A widely used algorithm in the simulations of physical phenomena is the Multidimensional FFT and in particular the three-dimensional FFT \cite{cochran1967fast} that are employed in solving the partial differential equations of physical models, such as the Navier-Stokes equations that describe the motion of turbulent fluids \cite{chorin1968numerical} \cite{davidson2004turbulence} or Newtonian mechanics equations of Molecular Dynamics \cite{car1985unified}. Simulation of such problems requires the use of HPC systems since the size of the problem grows rapidly and this is mainly due to three reasons:
\begin{itemize}
    \item the dimension of the problem increases with the cube of the FFT size;
    \item the representation of the single data point value typically requires a double precision floating point complex number (\textit{i.e.} 32, 64 or 128 bits);
    \item the single point usually comprises different components (\textit{e.g.} if the point represents the velocity it has three components per point).
\end{itemize}

\begin{figure}[!t]
\centering
\includegraphics[width=\columnwidth]{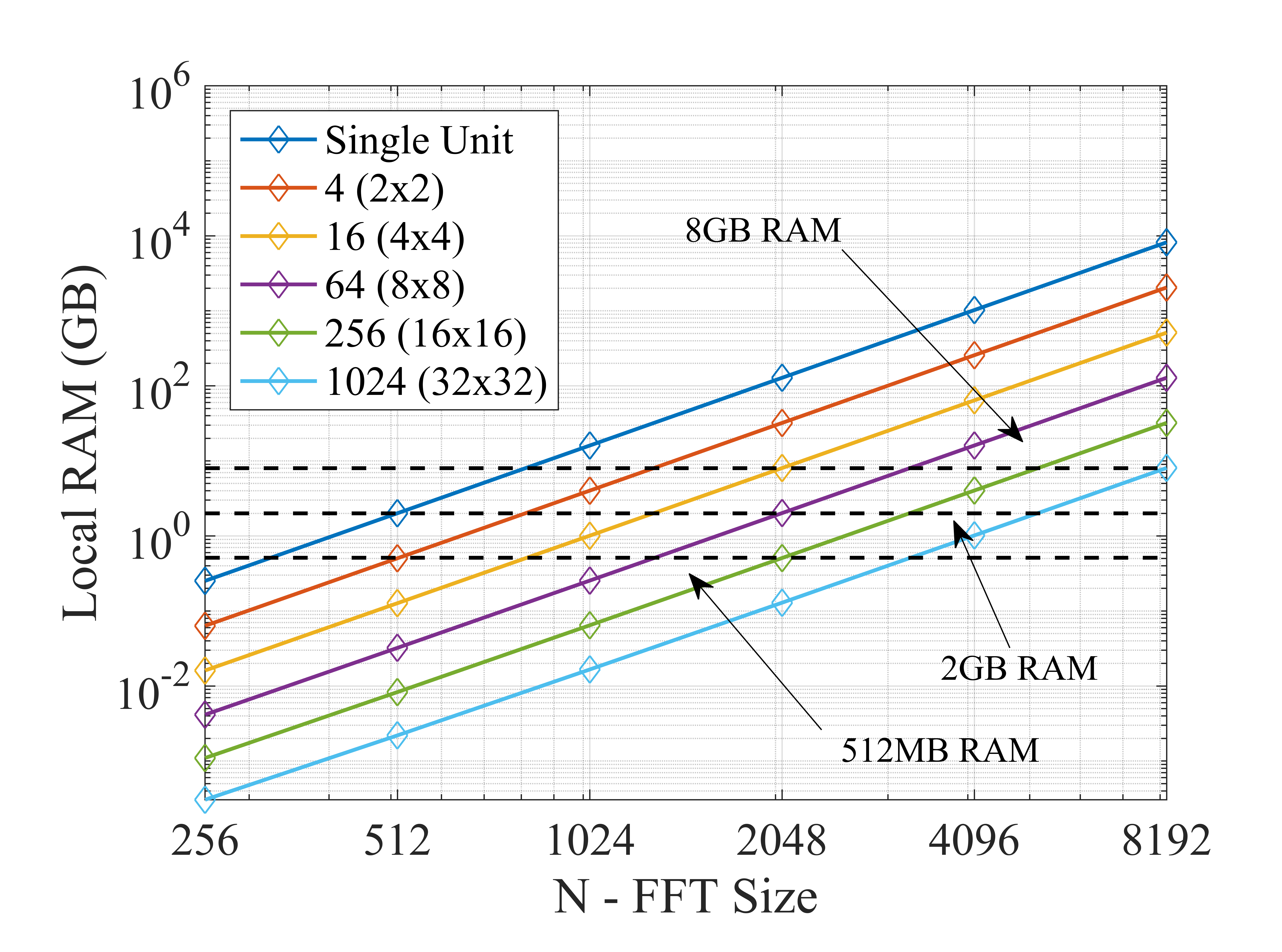}
\caption{Required local RAM per node for various cluster sizes increasing the FFT size.}
\label{fig:local_ram}
\end{figure}

Fig.~\ref{fig:local_ram} reports the required RAM trend on each computing node when solving a 3D FFT of size N for different computer cluster configurations, starting from the single unit, up to a 1024 nodes. The required memory per node is compared with the off-chip memory sizes in the FPGAs development kits currently available on the market (512 MB, 2GB and 8GB). The figure shows that the single unit needs handling about 0.25 GB for $N = 256$ or up to 1024 GB for $N = 4096$, indicating that the trend is not sustainable.

This implies that typical physical problems have to be tackled on HPC computer clusters where multiple compute units work in parallel in order to distribute the data domain. For example, considering an FPGA with 8GB of local RAM such as the Xilinx UltraScale+ VU37P, we can execute an $N = 8192$ points 3D FFT only using a computing cluster composed of 1024 FPGAs .

\begin{figure}[!t]
\centering
\includegraphics[width=\columnwidth]{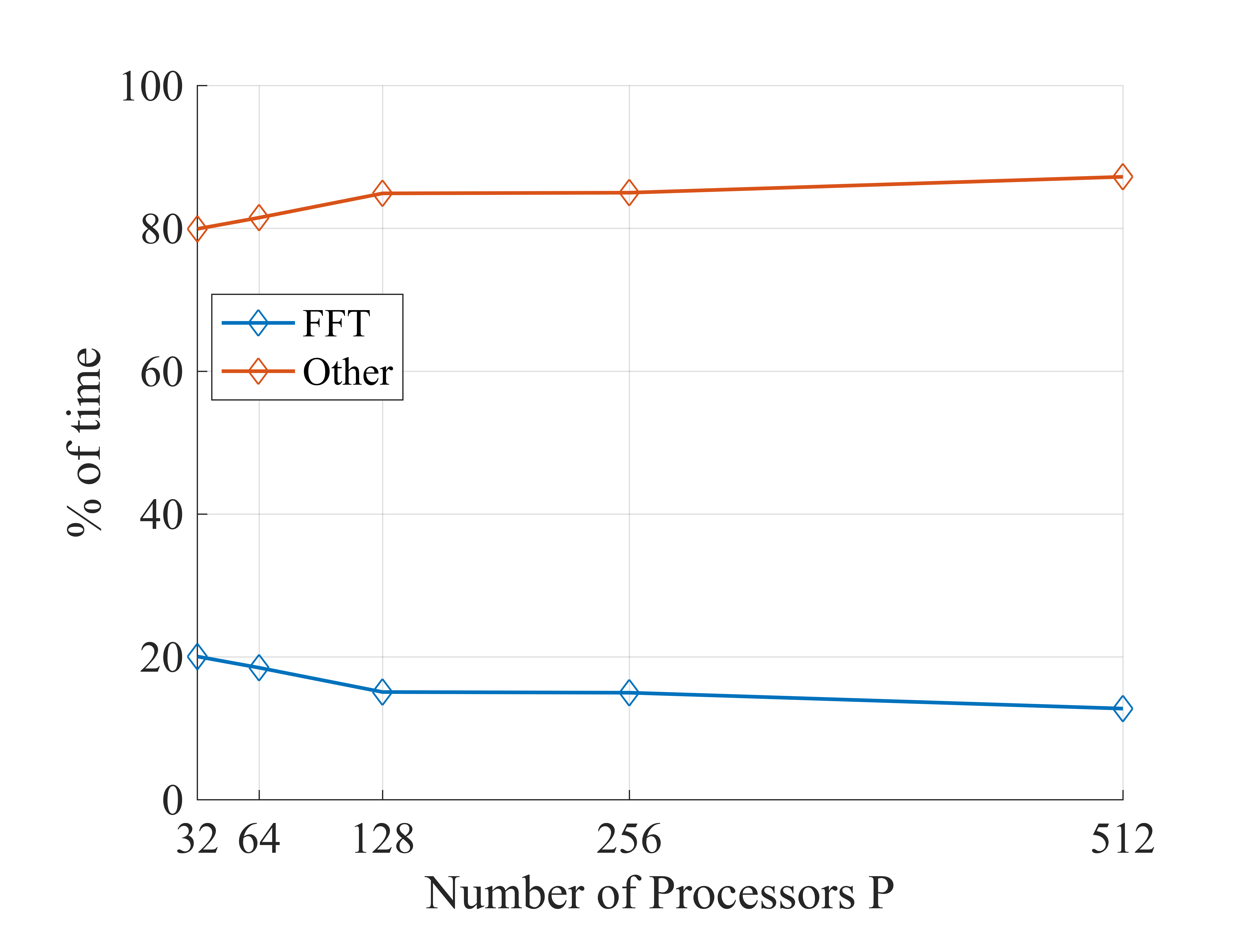}
\caption{Percentage of time for the FFT and the other computations for Navier-Stokes step increasing the number of processors.}
\label{fig:real_sim}
\end{figure}

To better quantify the impact of the FFT in physics simulations, in particular regarding  we have performed a computation of the Navier-Stokes equations for a $1024 \times 1024 \times 1024$ volume on an increasing number of processors. The computation was performed by NewTurb cluster \cite{lanotte2015turbulence}, a 26-node Intel Xeon based cluster, for a total of 624 computing cores equipped with a 40 Gbps Infiniband interconnect network, placed at Physics Department in the University of Roma Tor Vergata. In Fig.~\ref{fig:real_sim} we report the percentage of execution time dedicated to the FFT and inverse FFT computation and the percentage of the other computing activities. As can be noted, the percentage of time dedicated to the calculation decreases with the number of processors used while the percentage of FFT computing time increases.

A typical HPC system architecture to solve 3D FFT-based applications is a massively parallel computing platform built upon commodity multi-core processors eventually coupled with floating-point efficient accelerators, like GPGPUs. While the 3D FFT problem is both computational and communication intensive, on such platforms it is clear that the communication task becomes the bottleneck in strong scaling. The architecture we propose is aiming to overcome this bottleneck, by adding a dedicated network directly attached to each node accelerators, which in turn is an FPGA device.

\section{Related Work}
Historically, finding optimal solutions for implementing the multidimensional
Fast Fourier Transform algorithm on parallel High Performance Computing platforms
has always been an active field of research. We  can find three main paths
used to approach the problem, one being the development of fully custom machines, an other
is developing optimal algorithm for clusters of x86-based platforms and the latter
based on programmable logic devices used as custom accelerators.

\subsection{Custom ASIC platforms}
Back in the '90, as in \cite{morante1999parallel} where a comparative analysis between the APE
family supercomputers and Cray T3E is done, scarcity of hardware resources
used to force developers to find a general strategy in order to solve efficiently the parallelized
problem. In particular the lack of hardware-specific integer addressing capabilities made
any redistribution of data among nodes a great loss of efficiency, thus imposing a d-dimensional
data domain decomposition for a d-dimensional FFT problem and forcing data communication
within each 1-d FFT stages. It is shown how the optimal asymptotic scaling behaviour given by the formula:
$aN^3/P \log_2(N^3/P) + bN^3/P^{(d-1)/d} $
can be achieved on the above mentioned platforms, with a,b being machine-specific measurable constants.

More recently, in \cite{young200932x32x32} a fixed size special-purpose machine is presented
(P=512 and N=32,64) with the algorithmic techniques that leverage the architectural features. Designed 
specifically for molecular dynamics simulations, with 32-bit fixed point arithmetic, the complex 3D data domain
is initially divided in subcubes. Then three communication and computation phases are performed alternatively along the three spatial axes, showing a fixed scheme of communication patterns. Each computing node is equipped with a 1-d FFT engine, composed of a four-core single-stage radix-2 butterflies. Successive generation of this
custom machine \cite{shaw2014anton} is no longer FFT-based as computational kernel.

Porting 3D FFT code onto IBM BlueGene/L \cite{eleftheriou2005performance} has clearly shown how
on these machines the main concern is on the communication cost of the transpose phase, rather than the 
calculation phase. Results are shown for grid sizes up to $128^3$ with acceptable speed-up up to 16384 nodes.

\subsection{Algorithm optimizations on standard platforms}

In \cite{calvin1996implementation} are presented general methods to efficiently design mono- and bi-dimensional
FFTs on parallel computers with different network topologies. Complexity cost is analytically expressed, discovering that best results are on meshes of computing nodes from the point of view of the network topology,
and reaching the maximum overlap of communication and computation by using pipelining technique from the algorithm point of view.

Several works examine specifically the problem for 3D real transforms. 

Authors in \cite{dmitruk2001scalable} clearly model the per processor execution time in commodity x86 clusters
with the formula: $t_c \frac{5}{2} \frac{N^3\log_2(N^3)}{P} + t_a 3 \frac{N^3}{P} + t_w 2 \frac{N^3}{P} + t_s 2P $
where the parameters $t_c, t_a, t_w$ and $t_s$ represent respectively the single processor FFT computation time per word, the time for a memory to memory copy of a word, the time for transmission of a single word and the startup time for a message. Here the algorithm is a 1D domain decomposition involving P processors, and the model is validated for N up to 256 and P up to 16.

Starting with the work in \cite{pekurovsky2012p3dfft} it is nowadays well known that the best scalability and efficiency performance are achieved with a 2D domain decomposition. Here the model for the total execution time is
$N^3 [2.5 \log(N)/(PF) + bm/(P \sigma_{mem}) + cm/(2 \sigma_{bi}(P))] $, where $m$ is the size of each array element in bytes, $\sigma_{bi}(P)$ is the bisection bandwidth of the network portion containing P tasks, $c$ is a parameter expressing the network contention, $b$ represents the number of memory accesses per data element, $\sigma_{mem}$ is the memory bandwidth per task per node, $F$ represents the floating point operations per second available on the hardware. The main factor affecting performance is clearly identified in the network bisection bandwidth for this approach on examined architectures.

Further optimizations (e.g. overlapping computation with transposition) are shown in \cite{ayala2013parallel}, especially for larger size of N and on specific parallel computer architectures.

\subsection{FPGA-based platforms}

Building 1D FFT engines on FPGA devices (or VLSI custom chip) has been a research focus mainly due to signal processing applications, 
hence a wide literature can be found on designing pipeline FFT processors (\cite{he1996new}, \cite{garrido2009pipelined}, \cite{arioua2011vhdl}, \cite{garrido2013pipelined}, \cite{mao2017continuous}, \cite{nguyen2017fpga}), with several costs and benefits analysis on various possible engine architectures.
Several options are valid, such us the size of the basic butterfly (radix$-2^2 -2^3$ or $-2^4$), frequency
or time decimation, reordering strategies, etc., thus impacting on the quantity of resource used and sustainable operating frequency.

In particular, regarding double precision floating point arithmetic, an extensive analysis on architecture
has been made in \cite{hemmert2005analysis}. Architectures are classified in three main typologies: parallel,
pipelined and parallel-pipelined (or hybrid) and each one is modeled in terms of time to complete an FFT calculation and memory size requirements.

On recent platforms, but in single precision floating point arithmetic, it is viable to think of a full 3D transform on a single FPGA, as in \cite{humphries20143d}, for N up to 64 by using only on-chip RAM and multiple instances of the FFT IP core (up to 64). Data is organized in RAM per planes, in order to optimize memory access for x-axis and y-axis FFT computations and account for a higher latency on the z-axis. Different FFT IPs can work in parallel on different memory addresses, in order to exploit maximum floating point resources.

Assembling networks of FPGAs for 3D FFTs is done in \cite{sheng2014design}, with a 3D-torus low-latency interconnect, in order to increase the global size of the problem. Results of simulated latency for N up to 128 and P up to 512 are shown.

In \cite{sanaullah2018fpga} can be found latest comparison of performance of 3D FFT implementation on CPU, GPU and FPGA, the latest being better both using a standard IP Core design and an OpenCL-HDL design on the FPGA. The OpenCL-HDL approach allows to use even less resource than the standard approach, as well as shorter execution time. Results are shown for N=16,32 and 64, using single precision arithmetic.

\subsection{Differences on our approach}

With respect to the literature analysis, our approach is based on four main pillars: (i) arithmetic is double precision floating point, as requested by scientific simulations; (ii) global problem size needs to scale up to N=4096 and beyond; (iii) the basic butterfly engine is a radix-2 decimation in frequency algorithm, placed in a fully pipelined chain for the 3D FFT computation, including the communications among separated computing elements; (iv) a performance model is proposed in order to better balance computational resources, memory bandwidth and network bandwidth.

\section{Structure of the thesis}

The work of this thesis is organized as follow.

In Chapter~\ref{chap:network} a brief introduction of the previous work is done, in order to outline the research context where this work is placed. The network-related intellectual property cores are presented as possible building blocks to the system we envision.

In Chapter~\ref{chap:fft} the addressed computational problem is described as well as the proposed solution strategy. The Fast Fourier Transform algorithm is presented with its mathematical foundations and a timing model is introduced.

In Chapter~\ref{chap:model} the 3D FFT architectural models are presented and a quantitative comparison is made among the proposed alternatives.

In Chapter~\ref{chap:results} is discussed the implementation of the proposed FFT engine with the performance characterization and results. Also the design of the other architectural blocks is described, in order to foresee the performance of the complete system.

%% file: sections/network.tex
\chapter{Communication Protocol Offload}
\label{chap:network}
\minitoc
\section{Introduction}

The work in this thesis rise from several years of consolidate research and deployment activity in the field of developing custom computing machines for numerical simulations of great interest in modern physics. This activity has lead lately to the development of dedicated network interconnect facilities in the field of HPC and collaborating in the development of custom network interface cards for real-time data acquisition, tested in the fields of a High Energy Physics experiments at CERN, namely NA62 \cite{gil2017beam}, and space experiments, namely CSES-Limadou \cite{ALFONSI2017187}.

The aim of this chapter is to present the building blocks composing the network part  which can be used in the design of the HPC system subject of this thesis.

In Section \ref{sec:apenet} we describe the \apenet interconnect family, specialized in building 3D-Torus networks for off-the-shelf computing nodes. In Section \ref{sec:nanet} we present the real-time data acquisition board used in the NA62 experiment, in order to enable on-line GPU computing between detector data stream and low-level trigger processor. This project has lead to the development of a multi-rate multi-platform 1/10/40/100 UDP/IP Gigabit Ethernet core, described in \ref{sec:updip}.

\section{APENet} \label{sec:apenet}
\subsection{History of the project}

The Array Processor Experiment (APE) is a project aimed at custom designs for HPC, started by the Istituto Nazionale di Fisica Nucleare (INFN) and partnered by a number of physics institutions all over the world; since its start in 1984, it has developed four generations of custom machines (APE~\cite{albanese:1987}, ape100~\cite{battista:1993}, APEmille~\cite{aglietti:1998} and apeNEXT~\cite{belletti:2006}).
Leveraging the acquired \mbox{know-how} in networking and \mbox{re-employing} the gained insights, a \mbox{spin-off} project called \apenet developed an interconnect board based on FPGA that allows to assemble a PC cluster with multi-dimensional torus topologies using \mbox{off-the-shelf} computing components.

The design of \apenet interconnect is easily portable and can be configured for different environments:
\begin{itemize}
    \item the \apenet~\cite{ammendola2005apenet} was the first \mbox{point-to-point}, \mbox{low-latency}, \mbox{high-throughput} network interface card for Lattice Quantum ChromoDynamics (LQCD) dedicated clusters;
    \item the Distributed Network Processor~\cite{DNP2012} (\dnp) was one of the key elements of RDT (\mbox{Risc+DSP+DNP}) chip for the implementation of a tiled architecture in the framework of the EU FP6 SHAPES project~\cite{shapes4278510};
    \item the \apenetp Network Interface Card, based on an Altera Stratix IV FPGA, was used in a hybrid, GPU-accelerated x86\_64 cluster \quong~\cite{ammendola2011quong:long} with a 3D toroidal mesh topology, able to scale up to $10^{4}\div10^{5}$ nodes in the framework of the EU FP7 EURETILE project. \apenetp was the first  device to directly access the memory of the \nvidia GPU providing \gpudirect RDMA capabilities and experiencing a boost in GPU to GPU latency test;
    \item the updated \apenetp network IP named \exanet --- \ie routing logic and link controller --- is responsible for data transmission at Tier~0/1/2 in the framework of H2020 \exanest project \cite{katevenis2018next}. 
\end{itemize}

Table~\ref{tab:apegen} summarizes the \apenet families comparing the main features.

\begin{table}[!t]
\footnotesize
\centering
\begin{tabular}{|c|c|c|c|c|c|}
\hline
                            & \textbf{\apenet}    & \textbf{DNP}    & \textbf{\apenetp} & \textbf{\apenetv} & \textbf{\exanet}\\
\hline
\textbf{Year}               & 2003                & 2007            & 2012              & 2014              & 2017       \\
\hline                                                                                                      
\textbf{FPGA}               & Altera Stratix  & ASIC            & Altera Stratix IV & Altera Stratix V  & Xilinx Ultrascale+     \\
\hline                                                                                                      
\textbf{BUS}                & \mbox{PCI-X}        & \mbox{AMBA-AHB} & \pcie gen2        & \pcie gen3        & AXI        \\
\hline                                                                                                      
\textbf{Computing}          & Intel CPU           & RISC+DSP        & \nvidia GPU       & \nvidia GPU       & ARM+FPGA   \\ 
\hline                                                                                                      
\textbf{Bandwidth}          & 6.4~Gbps            &                 & 34~Gbps           & 45~Gbps           & 32~Gbps    \\
\hline                                                                                                      
\textbf{Latency}            & $6.5\mu$s           &                 & $4\mu$s           & $5\mu$s         & $1.1\mu$s  \\
\hline                                                                                                      
\end{tabular}
\caption{The \apenet interconnect family road-map in the years}
\label{tab:apegen}
\end{table}

\subsection{\apenet Interconnect architecture}

\begin{figure}[!t]
\centering
    \includegraphics[width=.6\textwidth]{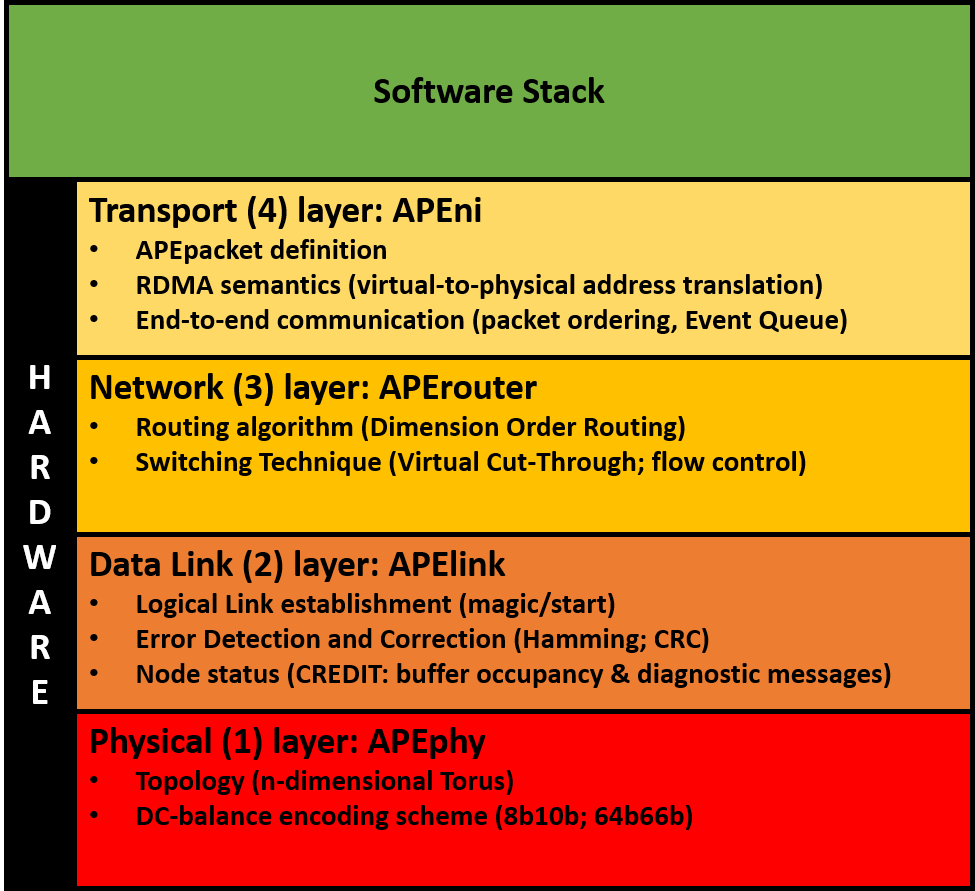}
    \caption{The layered architecture of \apenet}
    \label{fig:apelayer}
\end{figure}

The \apenet interconnect has at its core the \dnp acting as an
offloading network engine for the computing node, performing internode
data transfers;
the \dnp has been developed as a parametric Intellectual Property
library; there is a degree of freedom in choosing some fundamental
architectural features, while others can be customized at
\mbox{design-time} and new ones can be easily added.
The \apenet architecture is based on a layer models, as shown in
Figure~\ref{fig:apelayer}, including physical, data-link, network, and
transport layers of the OSI model.


The physical layer --- \textbf{\apephy} --- defines the data encoding
scheme for the serialization of the messages over the cable and shapes
the network topology.
\apephy provides \mbox{point-to-point} bidirectional,
\mbox{full-duplex} communication channels of each node with its
neighbours along the available directions (\ie the connectors
composing the \mbox{I\/O} interface).
\apephy is strictly dependent on the embedded transceiver system
provided by the available FPGA.
It is normally based on a customization of tools provided by the FPGA
vendor --- \ie \mbox{DC-balance} encoding scheme, deskewing, alignment
mechanism, byte ordering, equalization, channel bonding.
In \apenetp and \apenetv, four bidirectional lanes, bonded into a single
channel with usual \mbox{8b\/10b} encoding, \mbox{DC-balancing} at
transmitter side and byte ordering mechanisms at receiver side, allow
to achieve the target bandwidth (34~Gbps~\cite{ammendola:2012:NSS} and
45~Gbps~\cite{APEnetTwepp:2014} respectively).


The data-link layer --- \textbf{\apelink} --- establishes the logical
link between nodes and guarantees reliable communication, performing
error detection and correction.
\apelink~\cite{APEnetTwepp:2013} is the INFN proprietary
\mbox{high-throughput}, \mbox{low-latency} data transmission protocol
for direct network interconnect based on \mbox{word-stuffing}
technique, meaning that the data transmission needs submission of a
magic word every time a control frame is dispatched to distinguish it
from data frames.
The \apelink manages the frame flow by encapsulating the packets into
a light, \mbox{low-level} protocol.
Further, it manages the flow of control messages for the upper layers
describing the status of the node (\ie health status and buffer
occupancy), and transmitted through the \apelink protocol.


The network layer --- \textbf{\aperouter} --- defines the switching
technique and routing algorithm.
The Routing and Arbitration Logic manages dynamic links between blocks
connected to the switch.
The \aperouter applies a dimension order
routing~\cite{Dally:1987:Deadlock} (DOR) policy: it consists in
reducing to zero the offset between current and destination node
coordinates along one dimension before considering the offset in the
next dimension.
The employed switching technique --- \ie when and how messages are
transferred along the paths established by the routing algorithm,
\textit{de facto} managing the data flow --- is Virtual
\mbox{Cut-Through}~\cite{Kermani79virtualcut-through:} (VCT): the
router starts forwarding the packet as soon as the algorithm has
picked a direction and the buffer used to store the packet has enough
space.
The \mbox{deadlock-avoidance} of DOR routing is guaranteed by the
implementation of two virtual channels~\cite{dally:1990} for each
physical channel.

\begin{figure}[!t]
\centering
    \includegraphics[width=.6\textwidth]{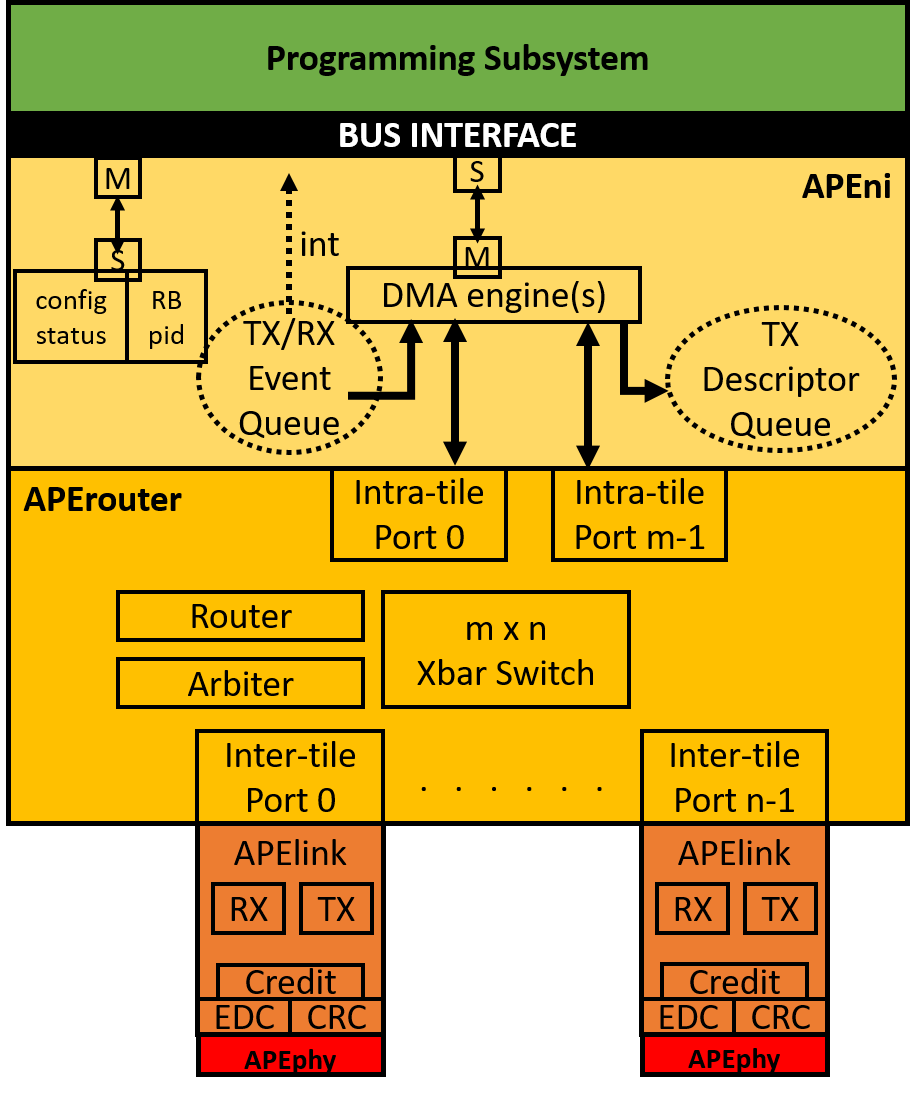}
    \caption {A block diagram of \apenet architecture}
    \label{fig:internals}
\end{figure}


The transport layer --- \textbf{\apeni} --- defines \mbox{end-to-end}
protocols and the \apepacket.
The \apeni block has basically two main tasks: on the transmit data
path, it gathers data coming in from the bus interfacing the
programming subsystem, fragmenting the data stream into packets ---
\apepacket --- which are forwarded to the relevant destination ports,
depending on the requested operation;
on the receive side, it implements PUT and GET semantics providing
hardware support for the RDMA (Remote Direct Memory Access) protocol
that allows to transfer data over the network without explicit support
from the remote node's CPU.

The full \apeni offers a \mbox{register-based} space for configuration
and status signalling towards the host.
Further, it offers a variable size region for defining a number of
ring buffers, each one linked to an OS process accessing the device.
These regions are typically accessed in slave mode by the host, which
is master (\mbox{read/write} of single 32-bit based registers).
A certain number of DMA engines are used to move data to and from the
device, plus other additional services: a TX descriptor queue (to issue buffer
transfers from host to device) and an event queue (to notify different
kind of completions to host). Single or Multiple DMA engines could manage
the same \mbox{intra-tile} port.

The block diagram of the \apenet interconnect architecture is shown
in Figure~\ref{fig:internals}.

\section{NaNet}
\label{sec:nanet}
The \nanet project goal is the design and implementation of a family of \mbox{FPGA-based} \pcie Network Interface Cards for High Energy Physics to bridge the \mbox{front-end} electronics and the software trigger computing nodes.

The design includes a network stack protocol offload engine yielding a very stable communication latency, a feature making \nanet suitable for use in \realtime contexts; \nanet \gpudirect \rdma capability, inherited from the \apenetp 3D torus NIC dedicated to HPC systems~\cite{APEnetChep:2012}, extends its \mbox{realtime-ness} into the world of GPGPU heterogeneous computing.
\nanet features multiple link technologies to increase the scalability of the entire system allowing for lowering the numerosity of PC farm clusters.
The key characteristic is the management of custom and standard network protocols in hardware, in order to avoid OS jitter effects and guarantee a deterministic behaviour of communication latency while achieving maximum capability of the adopted channel.
Furthermore, it integrates a processing stage which is able to reorganize data coming from detectors on the fly, in order to improve the efficiency of applications running on computing nodes.
Ad hoc solutions can be implemented according to the needs of the experiment (data decompression, reformatting, merge of event fragments).

Finally, data transfers to or from application memory are directly managed avoiding bounce buffers.
\nanet accomplishes this \mbox{zero-copy} networking by means of an hardware implemented memory copy engine that follows the RDMA paradigm for both CPU and GPU --- this latter supporting the \gpudirect V2/RDMA by \nvidia to minimize the I/O latency in communicating with GPU accelerators.
The quirks in the interactions of this engine with the bulky virtual memory management of the GNU/Linux host are smoothed out by adopting a proprietary Translation \mbox{Look-aside} Buffer based on Content Addressable Memory~\cite{ammendola:2013:FPT}.

\begin{figure}[!t]
\centering
\includegraphics[width=\textwidth]{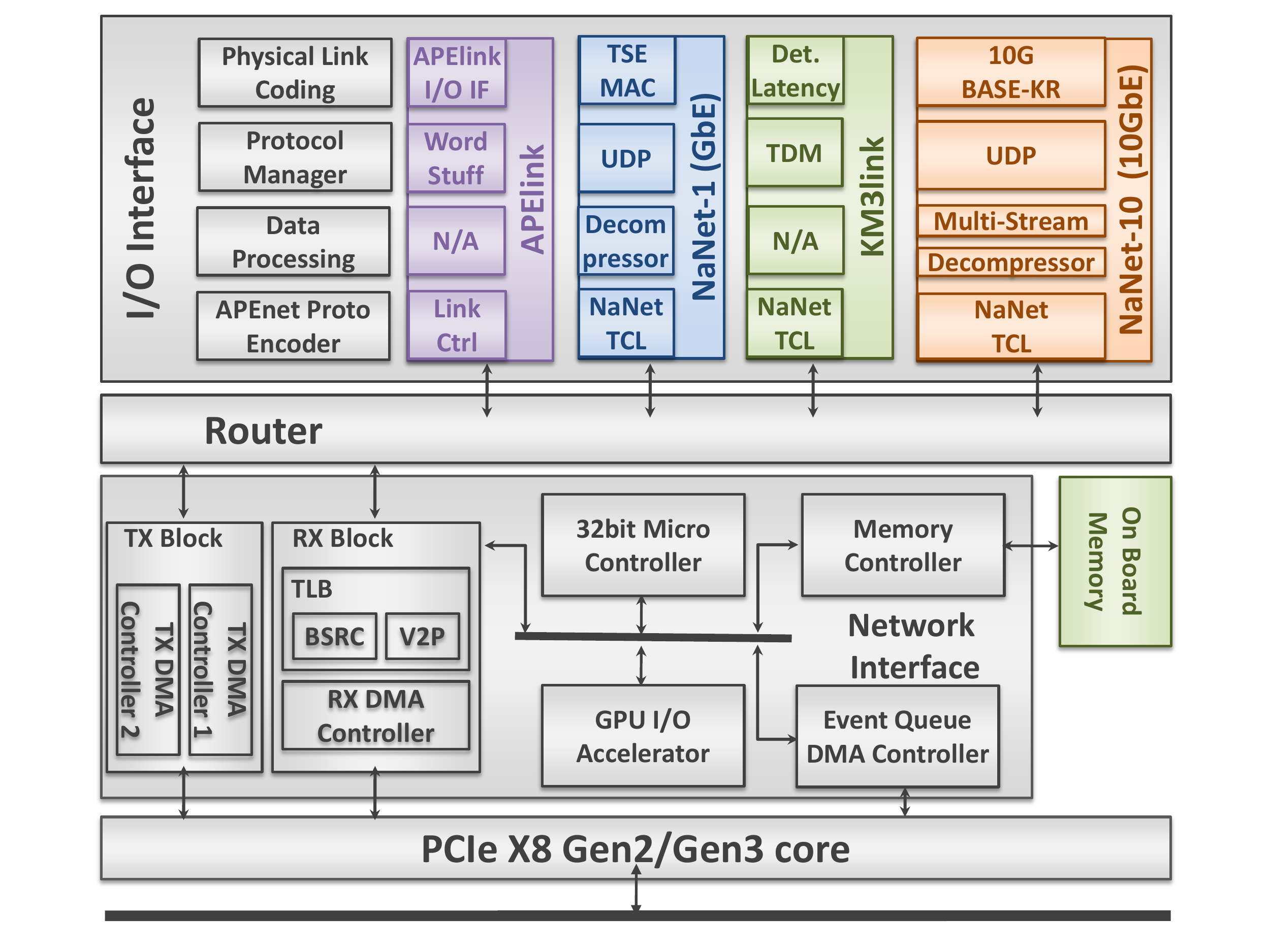}
\caption{The NaNet PCIe Network Interface Card family architecture.}
\label{fig:nanet}
\end{figure}

The hardware organization of the \nanet device (shown in Fig~\ref{fig:nanet}) clearly derives from the \apenet developments described in Section~\ref{sec:apenet}, with the main differentiation affecting the network channel communication protocols.

One of the key features of the NaNet design is the I/O interface; it consists of 4 elements: Physical Link Coding, Protocol Manager, Data Processing, APEnet Protocol Encoder. The Physical Link Coding covers the Physical and Data Link Layers of the OSI model, managing transmission of data frames over a common local media and performing error detection. The Protocol Manager covers the Network and Transport Layers, managing the reliability of communication data flow control. A processing stage, Data Processing, that applies some function to the data stream in order to ease the work for the applications running on the computing node, can be enabled on the fly. Finally, the APEnet Protocol Encoder performs a protocol translation to a format more suited for PCIe DMA memory transaction.

Several network protocols have been implemented in order to adapt on occasion at the experiment needs:
\begin{itemize}
    \item NaNet-1  offers a GbE channel with a Triple Speed Ethernet MAC core together with an external PHY layer; it has been the first card to implements a direct communication mechanism with NVIDIA GPUs (GPUDirect V2/RDMA).
    \item NaNet3 implements 4 I/O optical channels with deterministic latency and bandwidth up to 2.5 Gbps for the data transport and slow control management of the experiment.
    \item NaNet-10  is  a  four-ports  10GbE  PCIe  Network  Interface  Card  designed  for low-latency  real-time  operations  with  GPU  systems; it is the  third  generation  of  the  NaNet  NIC  family, representing an  upgrade  to  the  NaNet-1  board.
\end{itemize}

\section{Hardware UDP/IP protocol engine}
\label{sec:updip}

The \nanet family boards leverage on standard Ethernet Medium Access Control (MAC) sublayer and Ethernet Physical Layer (PHY) core, that are common in the IP core libraries for main FPGA vendors. In the latest FPGA generations the transceiver offerings cover the gamut of state of the art high speed protocols, including Ethernet in all its variant even to 100 Gbps data rate and announced compliance with 200 and 400 Gpbs data rate on the coming soon next generation.

\begin{figure}[!t]
\centering
\includegraphics[width=\textwidth]{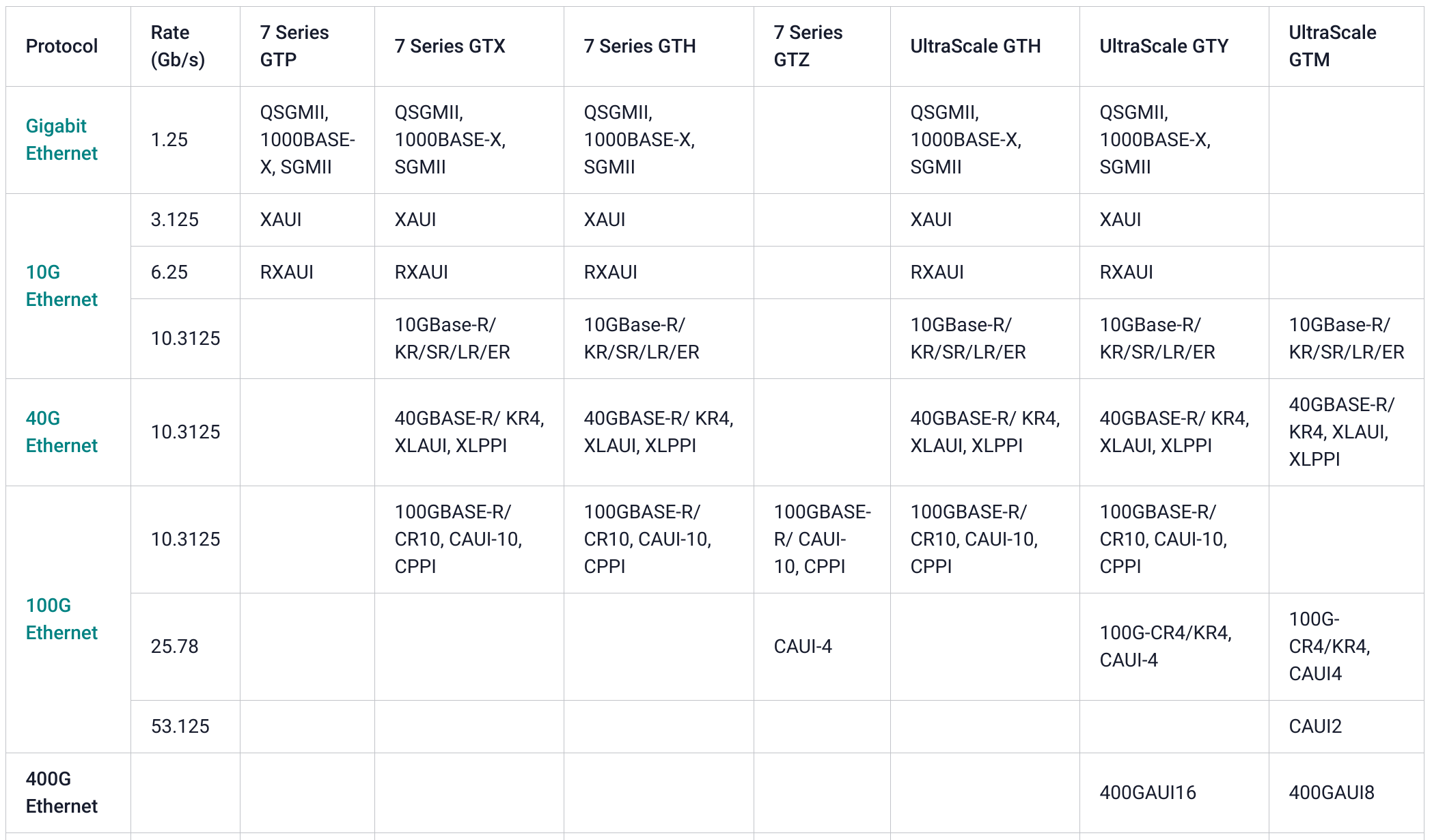}
\caption{An overview of supported Ethernet PHY technologies for latest generation of Xilinx FPGAs \cite{xilinx_phy}.}
\label{fig:phy}
\end{figure}

These supported communication protocols can be easily used with consolidated and widespread channel link technologies such as SFP, SFP+, QSFP+, QSFP28, which are electrically compatible with FPGA transceivers.

On top of the PHY and MAC we added the Internet Protocol (IPv4) layer and the User Datagram Protocol (UDP) layer. The advantage of building an UDP-based NIC on an FPGA is the possibility to use a standard and widespread communication protocol, together with the capability of having a clock cycle accurate latency control, enabling real time payload data extraction and processing.

The work of this thesis has given the opportunity to unify these developments into a single multi-protocol 1/10/40 Gb cross-platform Ethernet core, including future developments towards 100 Gb Ethernet and beyond.

Given the very special purpose of our NIC, the choice of UDP over the more reliable TCP protocol has been driven by several reasons, in particular:
\begin{itemize}
    \item a light-weight protocol is needed;
    \item communication is primarily point-to-point among well defined sender peers and receiving peers;
    \item source and destination UDP ports are enough to guarantee a payload tagging needed from the hardware-equivalent application layer;
    \item congestion control is performed by design, \textit{i.e.} by controlling maximum needed data rate.
\end{itemize}

\begin{figure}[!t]
\centering
\includegraphics[width=\textwidth]{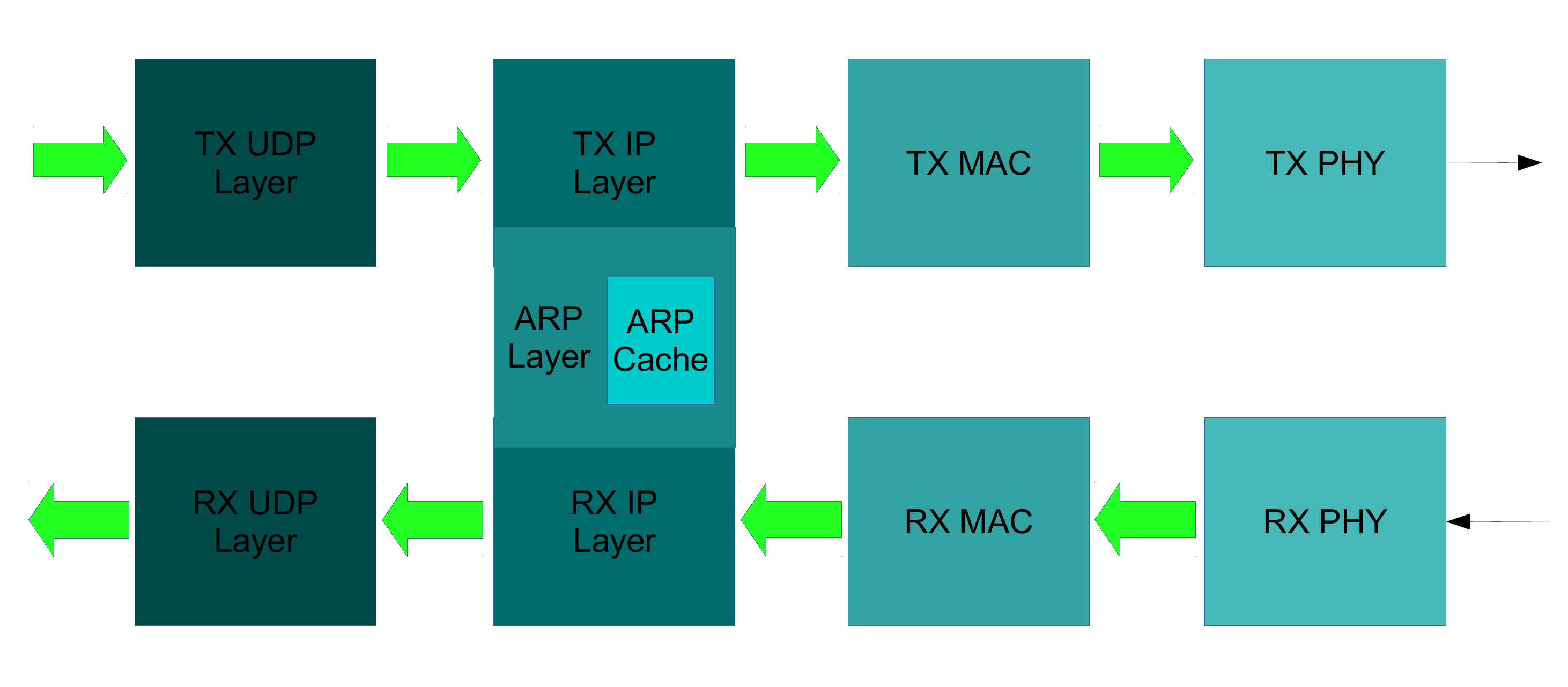}
\caption{UDP/IP logic block architecture.}
\label{fig:upd}
\end{figure}

The core provides full encoding and decoding of UDP, IPv4 and ARP protocols, with a logic organization depicted in Fig.~\ref{fig:upd}. It was initially derived and adapted from the FPGA-proven 1 Gbps UDP/IP open core \cite{udpip}. At the front-end side it exposes an AXI-like streaming interface with a datapath width that depends on the requested protocol data rate. In Tab.~\ref{tab:udpip_cores} we list the interface difference for the various supported data rates. Several registers are exposed for UDP header settings (\textit{e.g.} source or destination port and destination IP address) both in the transmit and receive side. The NIC IP and MAC address are also fully customizable. 

We recall in the following the Ethernet Packet structure, together with the IP and UDP headers:

\begin{bytefield}[bitwidth=3.5mm, bitheight=10mm]{32}
    \bitheader{0-31}\\
    \begin{rightwordgroup}{Ethernet Header}
        \wordbox[tlr]{1}{Preamble} \\
        \bitbox[blr]{24}{} & \bitbox{8}{\tiny{Start of Frame Delimiter}}\\
        \wordbox[tlr]{1}{MAC Destination} \\
        \bitbox[blr]{16}{} & \bitbox[tlr]{16}{MAC Source}\\
        \wordbox[blr]{1}{} \\
        \bitbox{32}{802.1Q Tag} \\
        \bitbox{16}{Ether Type}
    \end{rightwordgroup} \\
    \begin{rightwordgroup}{IP Header}
        \bitbox{4}{\small{Version}} &
        \bitbox{4}{\small{Header length}} &
        \bitbox{8}{\tiny{Differentiated Services}} &
        \bitbox{16}{Total Length}\\
        \bitbox{16}{Identification} &
        \bitbox{3}{\small{Flags}} &
        \bitbox{13}{Fragment Offset}\\
        \bitbox{8}{\small{Time to Live}} &
        \bitbox{8}{Protocol} &
        \bitbox{16}{Header Checksum}\\
        \bitbox{32}{Source Address} \\
        \bitbox{32}{Destination Address} \\
        \bitbox{32}{Options (if Header Length $> 5$)}
    \end{rightwordgroup} \\
    \begin{rightwordgroup}{UDP Header}
        \bitbox{16}{Source Port} & \bitbox{16}{Destination Port}\\
        \bitbox{16}{Payload Length} & \bitbox{16}{Checksum}
    \end{rightwordgroup} \\
    \begin{rightwordgroup}{Payload}
    \wordbox{3}{Data}
    \end{rightwordgroup} \\
    \begin{rightwordgroup}{Ethernet Trailer}
        \bitbox{32}{Frame Check Sequence} \\
        \wordbox{3}{Interpacket Gap} 
    \end{rightwordgroup} \\
\end{bytefield}

The Preamble and Start of Frame Delimiter fields in the Ethernet Header are in charge of the underlying MAC core, as well as the Ethernet Trailer. The rest of the fields are in charge of our core, that assembly them according to the datapath width.

The  core  offers  ARP  level  functionalities,with a 256-entries cache for IP-to-MAC address translation.  Underlying ARP communication is  automatic  when  first  packet  transfer  occurs  and  sender  and  receiver  mutually  exchange information  about  their  own  IP  and  MAC  addresses.   There  is  no  data  buffering  internally, allowing minimal latency between the Data Processing block and the Physical layer. For this reason packet segmentation and reassembly are not supported.

As of now, this core has been used for the trigger primitive transmission of the Liquid Krypton Calorimeter in the NA62 experiment at CERN\cite{salamon2018na62}, in the 1 Gb version. In the same experiment, the 10 Gb version has been used in the \nanet framework \cite{ammendola2018real} in order to send the Ring Imaging Cherenkov detector data towards a GPU unit for real-time generation of trigger primitives. During the two years of technical stop at CERN (2019-2021), we are working to upgrade it with the 40 Gb version and eventually place it in the upgraded version of the Level-0 Trigger Processor (L0TP) \cite{AMMENDOLA20191}.

Lately, the 1 Gb version has been used also in the development of the data acquisition system for the Electric Field Detector located in the Italian-Chinese collaboration CSES-Limadou satellite \cite{badoni2018high}. Here, the hardware generated multiple data stream is tagged making use of the UDP Source and Destination Port, in order to have the most lightweight possible standard protocol and no additional custom application-level overhead.

\begin{table}[]
    \centering
    \begin{tabular}{|c|c|c|c|c|}
        \hline
        Data Rate & Datapath Width & Clock Frequency & Link Technology & Platform Used \\
        (Gbps) & (bit) & (MHz) & & \\
        \hline
        1 & 8 & 125 & SFP/SGMII & Altera \& Xilinx \\
        10 & 64 & 156.25 & SFP+ & Altera \\
        40 & 128/256 & 322.22 & QSFP+ & Altera \\
        100 & 512 & 322.22 & QSFP28 & Xilinx \\
        \hline
    \end{tabular}
    \caption{UDP/IP core interface specifications.}
    \label{tab:udpip_cores}
\end{table}

%% file: sections/3d_fft.tex
\chapter{Three-dimensional Fast Fourier Transform}
\label{chap:fft}
\minitoc

In this Chapter we introduce the mathematical definition of the three-dimensional Fast Fourier Transform. We propose a general computing architecture in order to distribute the problem on such a parallel machine and some possible ways to organize the computing tasks. Then, we describe the algorithm used in the FFT engine and we finally derive from it an architecture. Some metrics are defined in order to describe this architecture.

\section{Introduction}

The multidimensional FFT is an optimized implementation of the multidimensional Discrete Fourier Transform (DFT). In the most general case, supposing D dimensions and being $\textbf{x}[n_1, n_2, \dots n_D]$ a N-tuple of D-dimensional vectors of complex numbers, the DFT can be defined as:

\begin{equation}
    \textbf{X}[k_1,k_2, \dots, k_D] = \sum_{n_1=0}^{N_1-1}\left(W_N^{n_1 k_1} \sum_{n_2=0}^{N_2-1}\left(W_N^{n_2 k_2} \dots \sum_{n_D=0}^{N_D-1}W_N^{n_D k_D} \textbf{x}[n_1,n_2,\dots,n_D] \right) \right)
    \label{eq:3dfft}
\end{equation}
with the so-called twiddle factor $W_N$ defined as:
\begin{equation}
    \begin{array}{l}
    W_N^{n_r k_r} = e^{-i \frac{2\pi n_r k_r}{N}} \\
    r \in \{1,2,\dots,D \}
    \end{array}
\end{equation}
and $\textbf{x}$ being a vector field in the spatial domain and $\textbf{X}$ its corresponding Fourier space. Output indices run from $k_i=0,1,\dots, N_{i-1}$ and each dimension have a size of $N_i$.
In our case of interest $D=3$ and we rename the three dimensions in x, y and z.
Analyzing Eq. \ref{eq:3dfft}, it can be noted that a three-dimensional FFT can be computed as a nested sequence of uni-dimensional FFTs for the $N_x\times N_y \times N_z$-points data domain and the order in which the three dimensions are elaborated is not relevant. We also assume that $N_x=N_y=N_z=N$ in the following for the sake of simplicity.

The Inverse FFT is the operation that allows the data-set back from frequency domain to time domain. From a computational point of view the inverse and the forward transforms are substantially very similar, the only difference is an overall $1/N$ scaling factor and twiddle factor $W_N$ that are complex conjugate. There are several ways to re-use forward algorithms to compute the inverse transform, as in \cite{duhamel1988computing}. We will not cover the inverse transform in this work, but it is clear that the arguments are equivalent.

\section{Distributing the 3DFFT}

\subsection{Node Architecture}

\begin{figure}[!t]
\centering
\includegraphics[width=.8\textwidth]{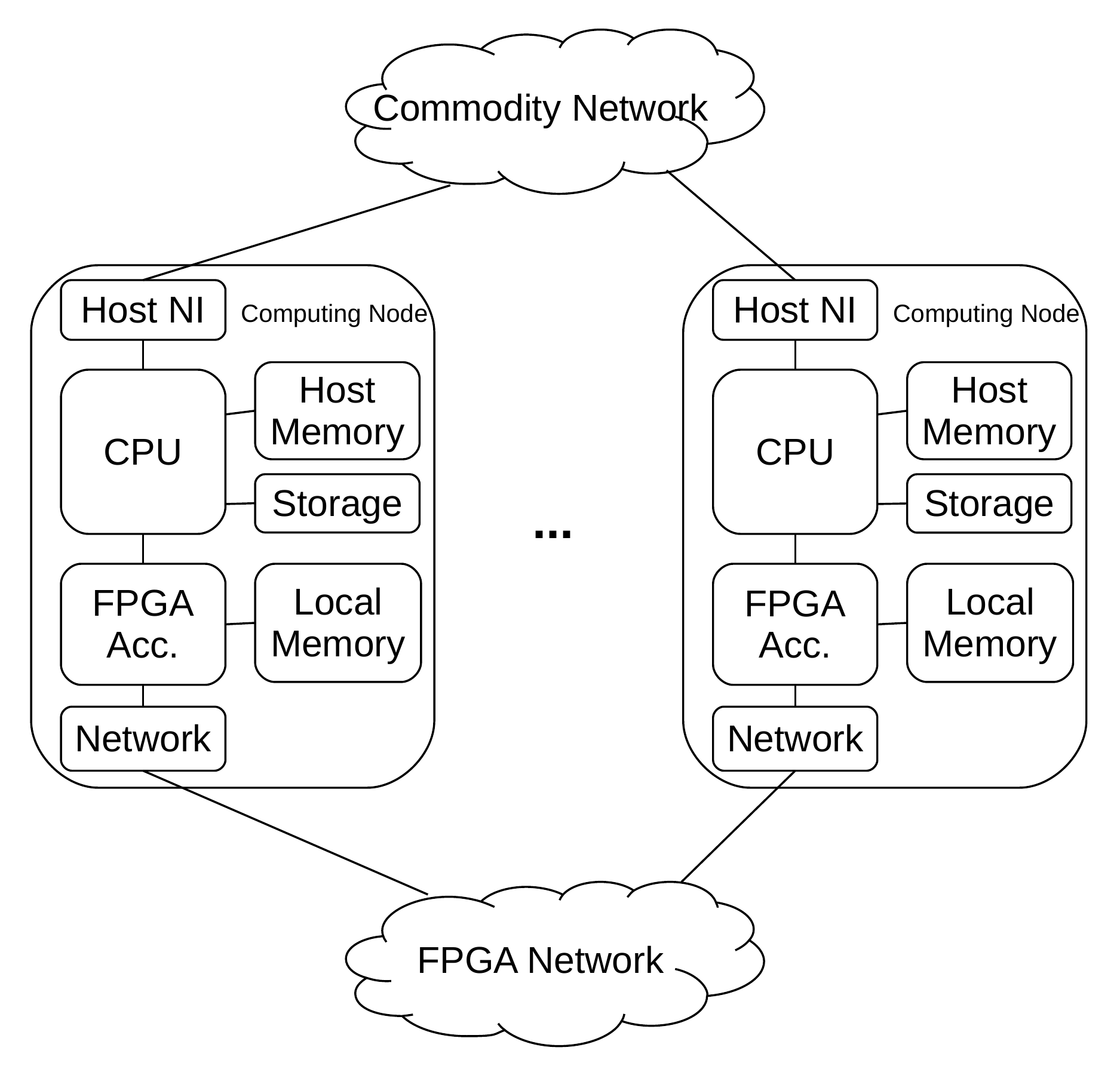}
\caption{A general architecture of an HPC system composed of P processing elements equipped with FPGAs used as computing and networking acceleration devices.}
\label{gen_computing_arch}
\end{figure}
In Fig. \ref{gen_computing_arch} we report the reference architecture considered in this work that includes $P$ distributed computing units.  Each node comprises its own CPU, memory, local storage and a standard network for management, storage sharing and application-related network traffic. Each CPU is connected with an FPGA accelerator that has its own local memory and network interface. In the proposed architecture, the dedicated network interfaces are directly attached to the FPGA and managed by internal controllers in order to avoid the communication latency typically introduced by the Operating System and all software layers; furthermore using an hard-wired controller, due to the completely deterministic nature of the 3D FFT problem (\textit{i.e.} given $N$ and $P$ the communication patterns are repeatedly constant) it is possible to optimize the inter-nodes communication at a clock-cycle accurate level.

The CPU and FPGA units can be even though as single SoC devices, for denser and higher power efficient implementations.

\subsection{Distributed Computing Logic}
At application startup time, data is typically generated or fetched from local/remote storage and loaded into host memory of each $P$ computing node. When the 3D FFT operation is globally called within the application, data is transferred to FPGA, which can use directly the local memory to buffer data and the network interface to exchange data in the FPGA dedicated network. At the end of the FFT phase data is written back on host memory, to be used by host processors for local computations. This scheme allows great flexibility in the non-FFT part of the application code, which runs on ARM or x86 processors with standard tools, leaving to the FPGA accelerators the hardware optimized FFT part.

\subsection{Data domain decomposition}
There are several ways to decompose the data grid among the compute units for parallel implementations of FFT, and the optimal solution is strictly related to the network topology of the cluster. Due to the three dimensional nature of the data domain, 1D (slices), 2D (pencils) and 3D (subcubes) decompositions are possible.

The 1D domain decomposition has been widely selected on commodity architectures due to the use of FFT optimized libraries such as FFTW (\cite{frigo1998fftw}). However \cite{pekurovsky2012p3dfft} and \cite{ayala2013parallel}  demonstrated in that a 2D decomposition allows to overcome scalability limitations that are intrinsically present with a 1D domain decomposition as long as $P$ is large enough.

On large HPC installations, which often have custom multi-dimensional torus interconnects, such as in \cite{eleftheriou2005performance} or \cite{young200932x32x32} a 3D decomposition is typically attained, in order to fully exploit the available network bandwidth. However if there are no specific data locality properties to exploit in the domain decomposition (i.e. in the local computation phase), it is clear that the two dimensional solution provides the higher scalability. In fact, the 3D decomposition is always adding an initial communication phase in order to re-shape the data distribution to a 2D decomposition indeed.

\begin{figure}[!t]
\centering
\includegraphics[width=\columnwidth]{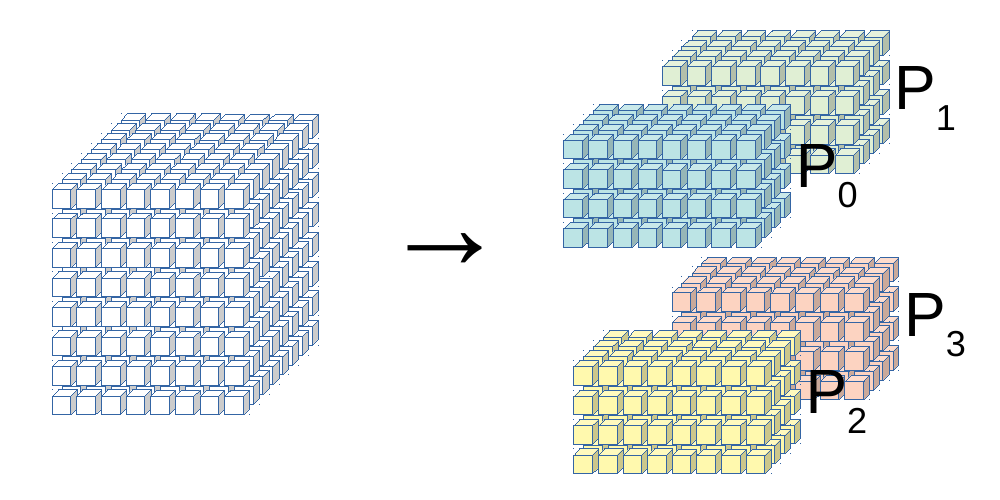}
\caption{2D domain decomposition. Each cube representing a data point of the size of 2 double precision floating point numbers. The initial data volume is shared equally in P processes along two Cartesian coordinates.}
\label{domain_dec}
\end{figure}

For this reason in the present work we only considered the 2D domain decomposition, shown with an example in Fig. \ref{domain_dec}. In this drawing we represent how an initial cube of size $N_x N_y N_z$ is distributed by equally partitioning $N_y$ and $N_z$. $N_x$ data is kept all within the same process.
More generally, if we define a grid that comprises $P_u$ and $P_v$ processor along the two Cartesian coordinates, and $P=P_u P_v$ being the total number of processing elements of the cluster, we will have that each process owns a pencil of size $N_x \frac{N_y}{P_u} \frac{N_z}{P_v}$.

\subsection{Task organization}

\begin{figure}[!t]
\centering
\includegraphics[width=\columnwidth]{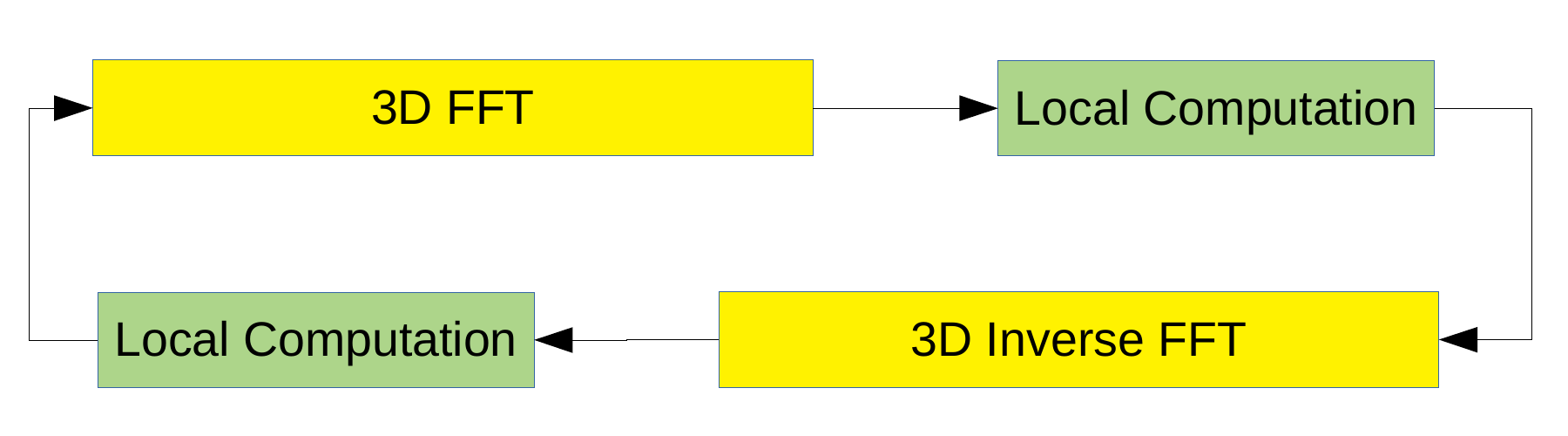}
\caption{A complete calculation step.}
\label{comp_step}
\end{figure}

We can think of different ways to implement the sequence of the three 1D FFTs in the three dimensions (X, Y and Z) in turn. After the 3D FFT phase, the complete calculation step foresee a local computation phase, then a complete inverse 3D FFT phase, and than again a local computation phase. The local computation phases are performed in parallel among computing units nodes and don't need communications. The whole step is repeated thousands of times in order to simulate the timing evolution of the problem. Fig. \ref{comp_step} shows the computational cycle of each time step. The inverse Fourier transform can be easily computed using the forward FFT engine adding a $1/N$ scaling factor and conjugating the imaginary part; we won't debate the inverse transform implementation here.

At a global level, with a 2D data domain decomposition, the X transform can proceed independently on each processing node because data on the X dimension of the grid resides entirely in the local host memory and each one has its own assigned portion of the array. When data is non-local, that means that it is divided across processor boundaries, the most efficient approach (\cite{foster1997parallel}, \cite{pekurovsky2012p3dfft}) is to reorganize the data array by a global transposition. This is called the \textit{transpose} method, in opposition to the \textit{distributed} method, where the 1D transform is performed in parallel with data exchange occurring at each inner calculation step; it is clear that the distributed approach has an higher cost in terms of data volume communication, as well as an higher hardware complexity due to a necessary fine grained global synchronization. 

\begin{figure}[!t]
\centering
\includegraphics[width=\columnwidth]{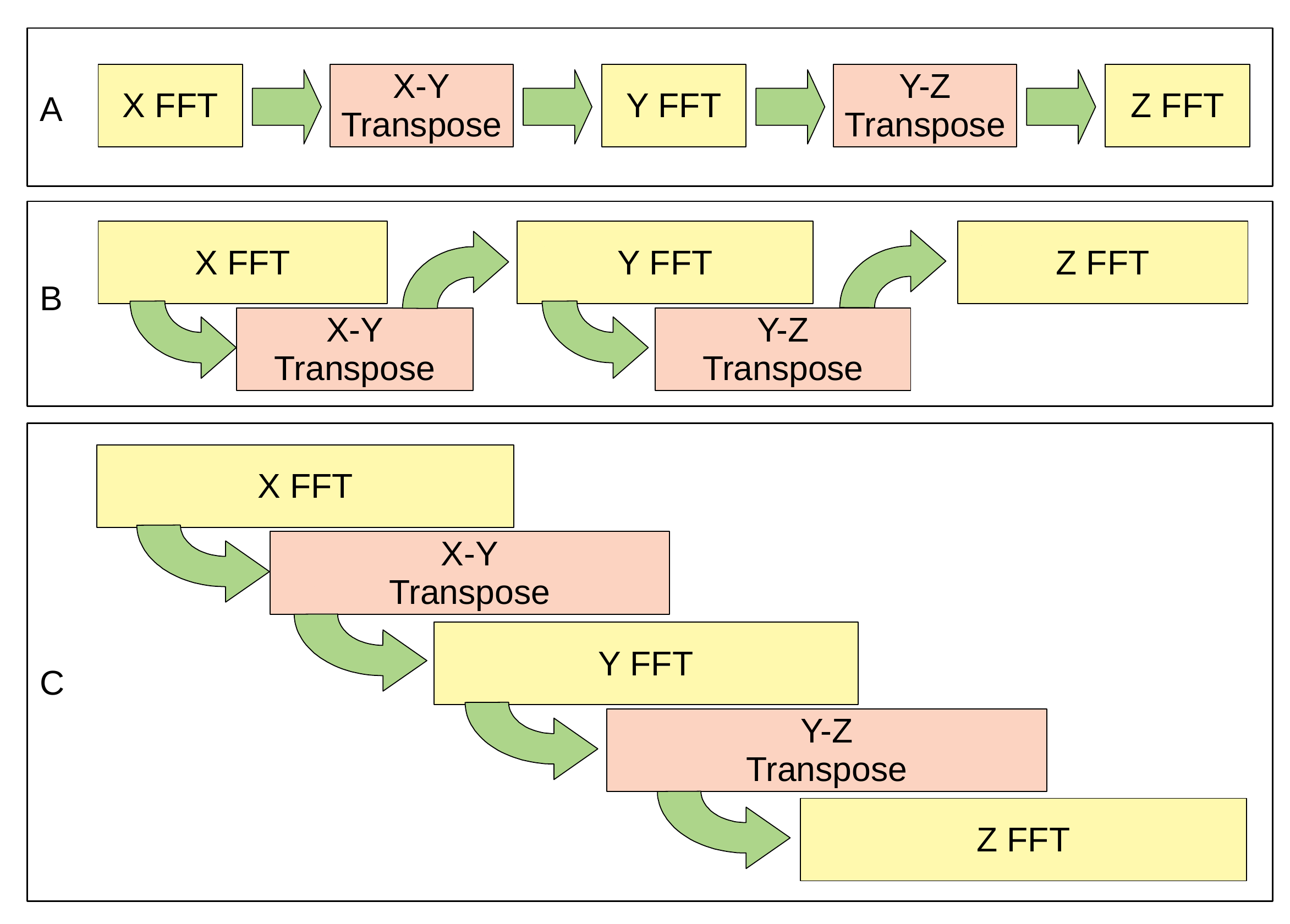}
\caption{Different approaches on organizing tasks on a global level.}
\label{comp_step_org}
\end{figure}

In our approach we follow the transpose method, meaning that the three transform phases are interspersed by two global transpose phase in order to first rearrange data from X to Y pencils, and then from Y to Z pencils. In such a way, the 1D transform phases are always performed within a local subset of the data volume. In Fig. \ref{comp_step_org} three possible strategies are depicted in order to organize the transform and transpose tasks: 

\begin{enumerate}[label=\Alph*)] 
\item all tasks are performed sequentially and no overlap is possible; this is typical on general purpose traditional computing architectures with no particular effort in overlapping communication and computation is made;
\item FFT calculation and transpose operations are overlapped, but FFT are evaluated sequentially; This solution exploit the fact that data after being processed can be exchanged between the nodes; technically this can be achieved using specific communication offload engine and/or using an asynchronous communication paradigm; in this model part of the communication time is masked by computational time; Note that if the communication is sufficient fast, FFT calculation tasks can be executed with out any interruption.
\item in this solution we consider all tasks are overlapped, as long as data dependency allows it; this solution extends the previous case and assume that nodes have enough computational power to executes part of the dimension in parallel. This model clearly needs specialized engines both for communication and for FFT computations, and allows a lower computational time at the cost of extra hardware resources. 
\end{enumerate}

Due to the intrinsic nature of FPGAs of parallelize task in a pipelined way, we can sketch a general architecture and a model of the computing engine which is able to implement case B and case C of Fig. \ref{comp_step_org}, hereafter respectively named \textit{sequential} and \textit{pipelined} architecture.

Moreover, further architectural optimizations are possible when considering all the components of the vector fields to be transformed.

\subsection{Data type} 

Typically, physical observables that compose the data set of the numerical simulations are real-valued vector fields, with one to three components each. The Fourier transform applies on single vectors of $N$ points, each point having a size $s=8$ bytes for double precision arithmetic in our case. 
For the property of complex conjugate symmetry ($X_k = X^*_{N-k}$) valid for real-valued Fourier transforms, after the first X FFT step the output data is complex but only the first $N/2 + 1$ complex points are significant, as the remaining points are redundant. This yields that during the X transform phase inputs are real vectors of size $sN$ and outputs can be reduced to complex vectors of size $2s(N/2+1)$, which is approximately identical.
Subsequent Y and Z FFT steps work in a complex to complex transform domain, thus the size of the vector is maintained.

The inverse transform works in a reverse way: data is complex for the inverse Z and Y transforms, and for the X the full $2sN$ vector size is recovered in order the ensure the Hermitian property of the FFT on real data, so that in the end of the last step data can be reduced ta a vector of real valued points of size $sN$.

In terms of data volume for each component and for each process we can write that the local volume V is:
\begin{equation}
    V = s N^3/P
\end{equation}
After the first transform phase the data volume becomes:
\begin{equation}
    V' = 2s N^2(N+2)/2P = s (N^3 + 2N^2)/P
\end{equation}
This leads to a small asymmetry in the data distribution after the X transform, and thus to a slight load imbalance, but it can be considered reasonably small for large data volumes, as in our case.

\begin{figure*}[!t]
\centering
\includegraphics[width=\textwidth]{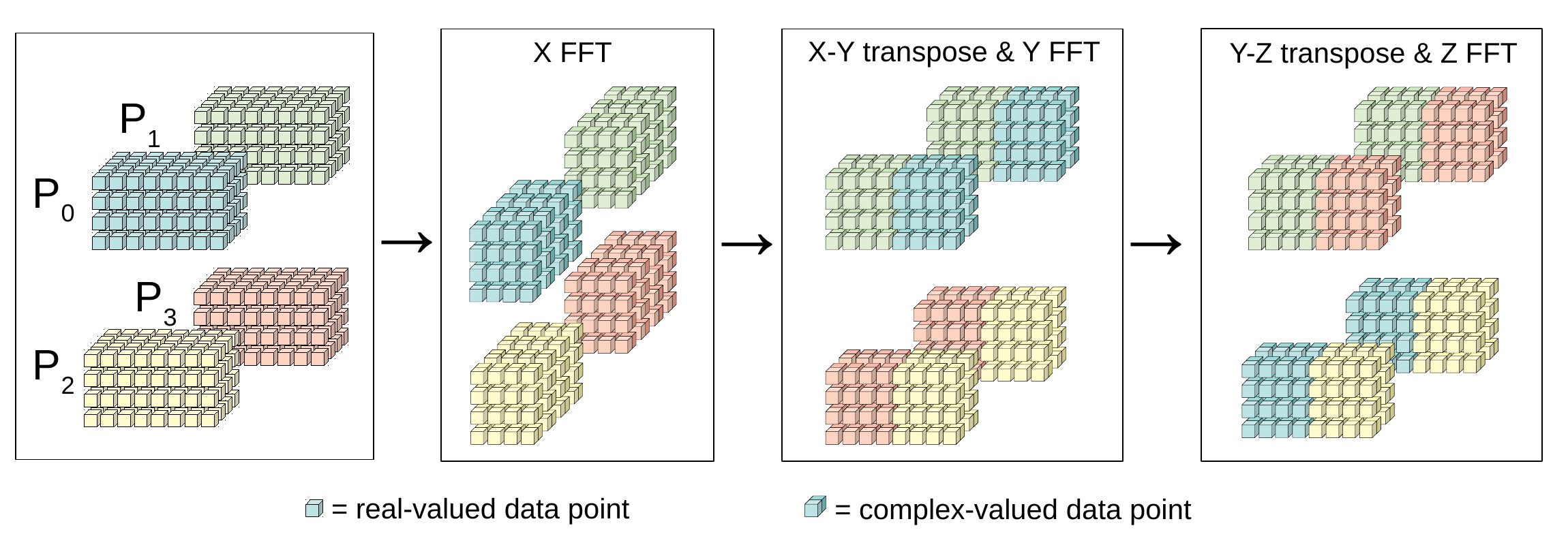}
\caption{Data type and data volume changes within the 3D FFT computational steps. $N/2+1$ is visually approximated with $N/2$ for the sake of simplicity.}
\label{data_exchage}
\end{figure*}

In order to visualize the data type cycle, an example is done in Fig.\ref{data_exchage}, where a $8^3$ data volume is divided in a $P=2 \times 2$ processing nodes grid.

\subsection{Network topology}

Network topology considerations follow directly from the 2D data domain decomposition. It is worth noting that during the X-Y transpose phase, data is exchanged among processes along the same row (in the sense of the first Cartesian axis of $P_u$ processing elements); moreover during the Y-Z transpose data is exchanged among processes along the same column (meaning the second Cartesian axis of $P_v$ processing elements). In this way rows and columns never exchange data traffic and can live on separated networks, one for each row and column.
In order to exploit this property, we foresee a couple of possible solutions for the network topology: a 2D torus network, or a 2D mesh of external switch components. We will compare how these two types of solutions have an impact on global performances and characterization.

\section{The uni-dimensional algorithm}

The Cooley-Tukey FFT \cite{cooley1969fast} algorithm is the most commonly used to calculate the DFT and the inverse DFT, and it is possible with the assumption $N=r^S$ ($r$ states the radix of the algorithm and typically $r=2$ in most of implementations). It is a divide and conquer algorithm that recursively breaks down a DFT into many smaller DFTs. The algorithm can be applied in two versions: the decimation in time (DIT) or the decimation in frequency (DIF). In the following, we are going to use the DIF version.

\begin{figure}[!t]
\centering
\includegraphics[width=.7\textwidth]{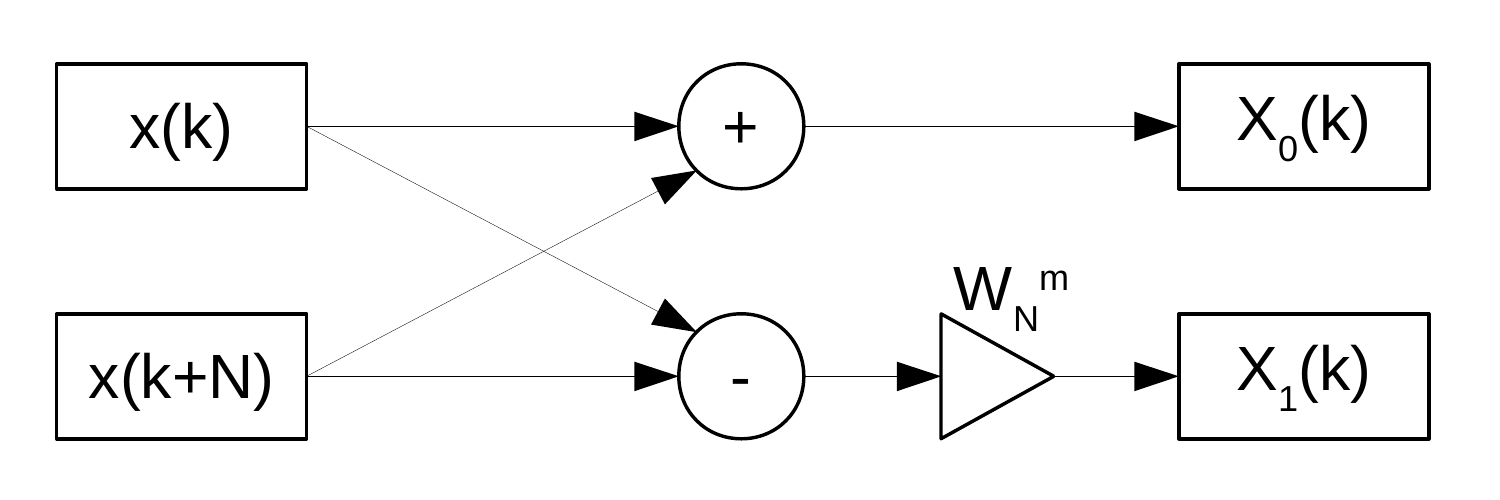}
\caption{Representation of basic butterfly operation for radix-2 FFT algorithm.}
\label{fft_2}
\end{figure}

The mono-dimensional FFT computation can be written as follows:

\begin{equation}
    \begin{split}
        X[k] & = \sum_{n=0}^{N-1} W_N^{kn}x[n] = \sum_{n=0}^{N/2-1} W_N^{kn}x[n] + \sum_{n=N/2}^{N-1} W_N^{kn}x[n] \\
        X[k] & = \sum_{n=0}^{N/2-1} W_N^{kn}x[n] + \sum_{n=0}^{N/2-1} W_N^{k(n+\frac{N}{2})}x[n+\frac{N}{2}] \\
        X[k] & = \sum_{n=0}^{N/2-1} (x[n] + (-1)^k x[n+\frac{N}{2}]) W_N^{kn}
    \end{split}
\end{equation}
where in the last step $W_N^{N/2)k}=(-1)^k$ was used. If we consider the even and odd part of $X[k]$ separately, we obtain:

\begin{equation} \label{even}
    X[2r] = \sum_{n=0}^{N/2-1} (x[n] + x[n+\frac{N}{2}]) W_{N/2}^{rn}
\end{equation}

\begin{equation} \label{odd}
    X[2r+1] = \sum_{n=0}^{N/2-1} (x[n] - x[n+\frac{N}{2}]) W_{N}^{n} W_{N/2}^{rn}
\end{equation}
where $r=0,1,\dots,N/2-1$.
As can be noted, the proposed formulation allows dividing the even and the odd part of the FFT thus reducing its size of the FFT by a factor 2 iteratively at every step. The basic computational kernel that can be defined from Eq.\ref{even} and Eq.\ref{odd} is the following:

\begin{equation}
    \begin{split}
         X_0(k) & = x(k)+x(k+\frac{N}{2})  \\
         X_1(k) & = [(x(k)-x(k+\frac{N}{2})]W_N^n
    \end{split}
    \label{eq_butterfly}
\end{equation}
and it can be represented with the structure shown in Fig.~\ref{fft_2}, often referred to as a "butterfly", where a couple of data words $x(k)$ and $x(k+N/2)$ are presented in input and a couple of data words $X_0(k)$ and $X_1(k)$ are produced at output. The complete FFT can be built from this basic butterfly kernel  building a structure of $N/2$ butterflies at each stage for a total of $log_2(N)$ stages.

\begin{figure}[!t]
\centering
\includegraphics[width=\textwidth]{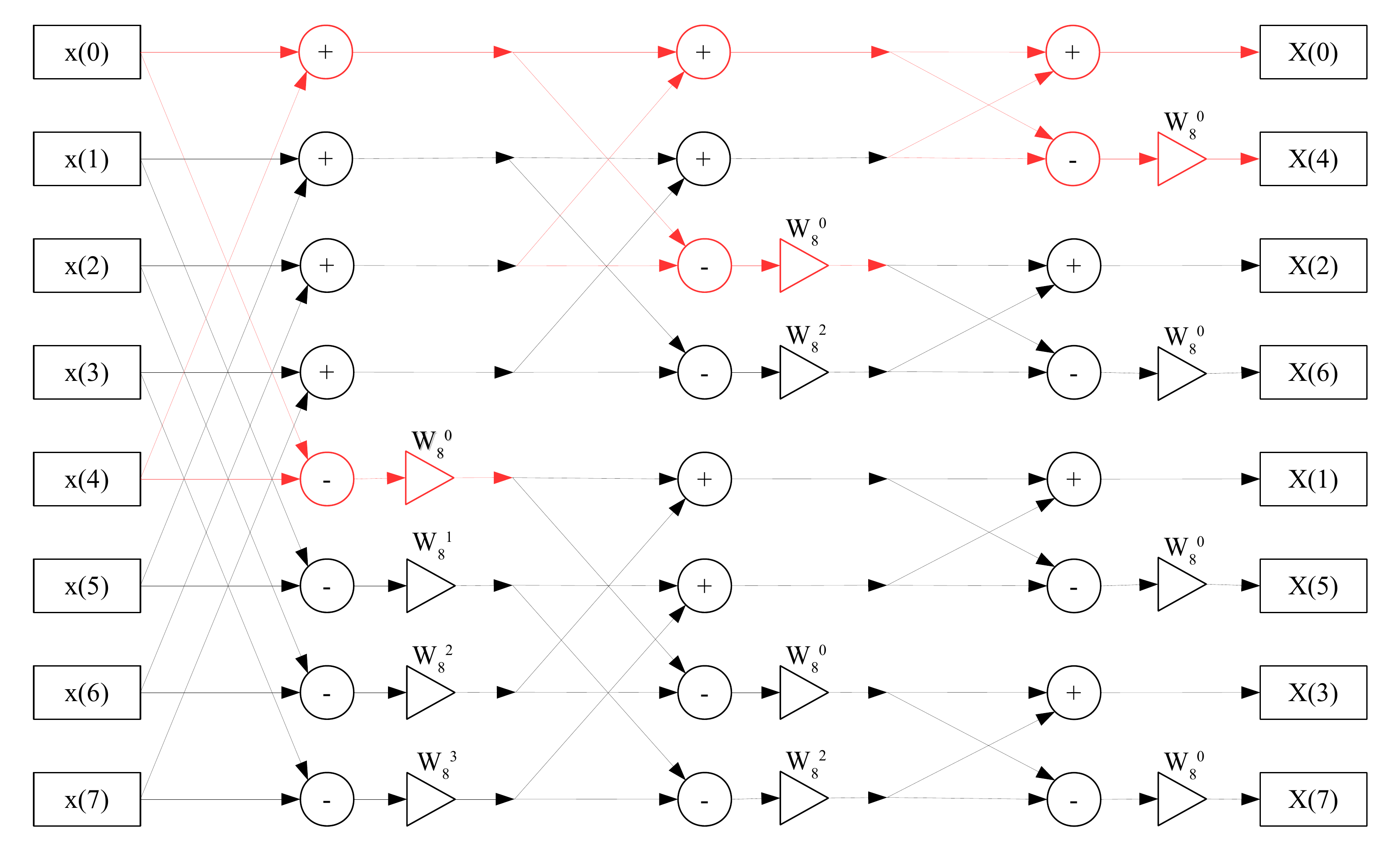}
\caption{N=8 example of a radix-2 FFT algorithm with decimation in frequency, using Fig.~\ref{fft_2} butterfly unit as building block.}
\label{8point_dif}
\end{figure}

In Fig.~\ref{8point_dif} we show an example of an 8-point FFT. Each stage includes $N/2 = 4$ butterfly units and the total number of stages is $log_2 N = 3$. Each butterfly unit operates on a pair of results from the previous stage. The distance between the two samples selected decreases by a factor two every stage. Finally, the output values have to be reordered.

From a computational cost point of view, the overall result is the reduction of number of operations from $O(N^2)$ to $O(N\log_r(N)$. The measure of number operations that need to be performed in order to execute a certain calculation is often referred to as computational complexity, and the case of FFT is a typical example of valuable reduction.

\section{1D FFT architecture modelling}

We can build a timing model starting from the butterfly unit, which has a characteristic calculation latency $l_{but}$ defined as the time a couple of data takes from the input to the output and the data throughput $B_{FFT}$ measuring the amount of data flowing through the butterfly units in the clock unit $t_{clk}$. When assembling a row of $log_2(N)$ butterfly units, in order to follow a pipelined architectural approach (\cite{he1996new}, \cite{garrido2009pipelined}, \cite{hemmert2005analysis}), the N points input data is time-multiplexed, two at each time unit, along with the proper twiddle factor $W_N^m$.  Between each step, an additional data reordering latency must be taken into account, namely $l_{reord}(S)$, that depends on the particular stage $S$. $l_{but}$ and $l_{reord}$ are strongly depending on the implementation and their values are detailed in Chapter \ref{chap:results}.

Summarizing, the time to compute a single FFT of size N with this architecture is then the sum of the time for the first data couple to traverse the entire chain of $S=log_2(N)$ butterflies $l_{FFT} = (l_{but}+l_{reord}(S)) log_2(N)$ with the time for all the $N/2$ couples of data to be processed in the pipeline, thus resulting:
\begin{equation}
    T_{FFT} = l_{FFT}  + t_{clk} N/2
\end{equation}

In the most general case, data are double precision complex numbers, thus if $2s$ is the size of the data, and we can write the data throughput as:
\begin{equation}
    B_{FFT}=4s/t_{clk}
\end{equation}
meaning that 2 complex data words are consumed and produced at each clock cycle when the pipeline engine is in full swing.

\begin{figure}[!t]
\centering
\includegraphics[width=\columnwidth]{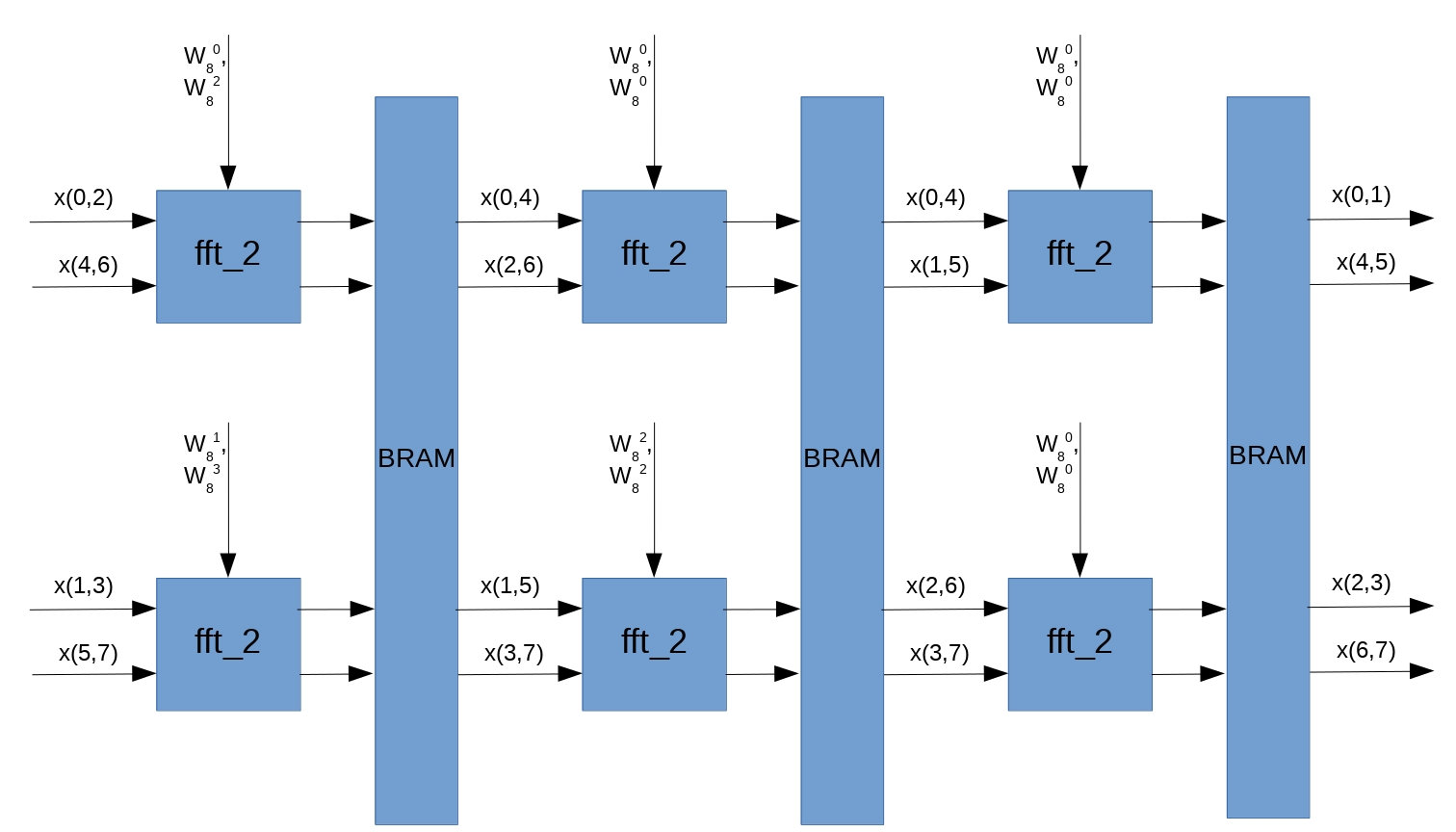}
\caption{Architecture of the 1D FFT parallel-pipelined engine with R rows and $\log_2(N)$ stages. The on-chip reordering table block represents the switching and delaying logic needed among each stage in order to map the radix-2 butterfly DIF tree as in Fig.\ref{8point_dif}.}
\label{parallel-pipeline}
\end{figure}

It is possible to use multiple instance of a pipelined engine, working in a parallel-pipelined fashion (as in \cite{hemmert2005analysis}, or \cite{garrido2013pipelined}), and represented as rows of butterfly units in Fig. \ref{parallel-pipeline}.

Being $R \leq N/2$ the number of possible row engines, in this case the model for the FFT calculation time and the data throughput becomes:
\begin{equation}
    T_{FFT} = l_{FFT} + t_{clk}\frac{N}{2R}
    \label{eq:TFFT}
\end{equation}
\begin{equation}
    B_{FFT}=4sR/t_{clk}
     \label{eq:BFFT}
\end{equation}

This FFT engine can be used also in the case of real-valued inputs (as in the X transform phase), even though it results to be sub-optimal in terms of FPGA resource usage. In this work we will not use real or complex valued optimized engines for X, Y and Z transforms because we prefer a more general and flexible architecture in this moment.
Nevertheless there are architectural considerations that a real-valued data transform phase will lead to, in terms of input and output data throughput and execution time.

%% file: sections/arch_model.tex
\chapter{3D FFT computing models}
\label{chap:model}
\minitoc

In this Chapter we describe in detail the components of the 3D FFT computing architecture, and how it is possible to break up the computational thread into specific hardware tasks. These tasks can be organized in several ways and they are described in terms of timing performance, data throughput and memory usage.

\section{3D FFT architectural blocks}

\begin{figure}[!t]
\centering
\includegraphics[width=\columnwidth]{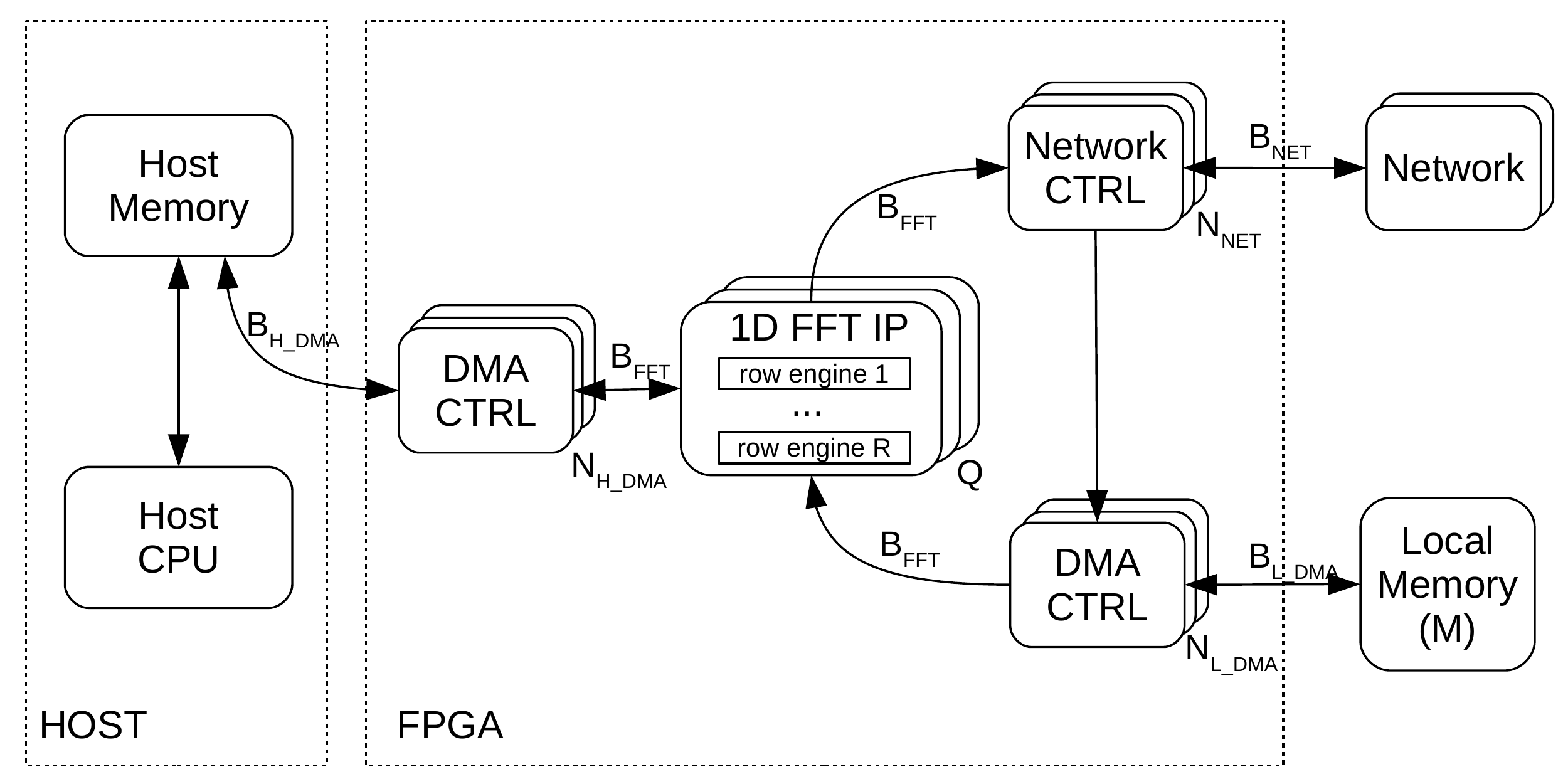}
\caption{Computing node architecture.}
\label{node_arch}
\end{figure}

Assuming a 2D data domain decomposition on P computing nodes, each one owns pencils of $N^3/P$ data fields of size $s=8$ bytes. The architecture of the logic blocks that reside in the FPGA device of each node is depicted in Fig. \ref{node_arch}. It is composed of four main blocks types:
\begin{itemize}
    \item DMA Controllers towards the host memory: responsible for read and write to the memory which is also accessible from the host processor, whether it is an embedded CPU for a SoC system or a standard x86 in other cases; in the first case the DMA controllers work on dedicated AXI buses that interface the Programming Logic (PL) side of the FPGA with the Processing System (PS) side; in the other case they are strictly tight to a PCIe endpoint in order to access host memory.
    \item 1D FFT IP cores: there can be a certain number Q of FFT engines, each one assembled with R row engines. The choice of Q and R depends in principle by the amount of available hardware resources on the FPGA device, and it will impact on the other logic blocks consequently.
    \item Network Controllers: dispatch and collect data from the FFT engines to the network of other computing nodes; part of the data is kept local. 
    \item DMA Controllers towards the local memory: gather the data from the Network Controller block, store it in local memory, and send it back to the FFT engines; data writing is responsible also for transposition, which means writing data units with stride $N$ or $N^2$. 
\end{itemize}

Due to the pipelined architecture of the FFT engines, data must continuously flow from host memory through the hardware blocks, until it end up in host memory again. As already said for the FFT core engine, each block introduces an intrinsic latency, namely $l_{DMA}$ (assuming host DMA and local DMA have roughly the same value) and $l_{COMM}$ for the network communication blocks. Furthermore we assume that each block is able to consume and produce the data volume $V$ with at least $B_{FFT}$ data throughput. Assuming with approximation that the data volume is $V=sN^3/P$ at each transform step, we can than write the total traversal time:
\begin{equation}
    T_{DMA} = l_{DMA} + V/B_{FFT} 
\end{equation}
\begin{equation}
    T_{FFT} = l_{FFT} + V/B_{FFT} 
\end{equation}
\begin{equation}
    T_{COMM} = l_{COMM} + V/B_{FFT} 
\end{equation}

In the following section we will see how it is possible to overlap and orchestrate these phases.


\section{Hardware tasks}

In order to build a performance and resource usage model for the proposed architectures, it is possible to break up the main computational thread into specific hardware operations, such as DMA transactions, data communications and so on, in order to explicit the effective parallelism that a custom programmable hardware can exploit. Thus we can identify this set of hardware operations:
\begin{enumerate}[label=\Alph*)]
    \item DMA Read from host memory of X pencils: data is read in burst from main memory of the system in (consecutive) chunks of N points, organized in X pencils. 
    \item X FFT: the volume $V$ of real-valued N-point data set is transformed along the X axis.
    \item X-Y fold communication: data is exchanged among $P_u$ nodes along the same row of the grid; a factor $1/P_u$ is kept local, the rest $(P_u-1)/P_u$ is sent and received with the other $P_u-1$ nodes.
    \item DMA Write on local memory with stride N: in order to perform X to Y transposition data is written on local memory with stride N. 
    \item DMA Read from local memory of Y pencils: data is read from memory ordered in Y pencils with stride 1.
    \item Y FFT: the volume $V'$ of complex N-point data set is transformed along the Y axis.
    \item Y-Z fold communication: like previous communication phase but data is exchanged along the $P_v$ nodes along the same column of the grid.
    \item DMA Write on local memory with stride $N^2$: like previous DMA write phase for the Y to Z transposition with a $N^2$ stride writing.
    \item DMA Read from local memory of Z pencils: data is read from memory ordered in Z pencils with stride 1.
    \item Z FFT: the volume $V'$ of complex N-point data set is transformed along the Z axis.
    \item DMA Write on host memory of Z pencils: the last phase is writing back the data into host memory.
\end{enumerate}

Considering the intrinsic streaming property of DMA transactions and data communications, we can outline two kinds of approach depending if the three FFT phases are performed sequentially or in a deep pipeline way. Moreover, due to the vector nature of the data domain, there can be a per-component parallel or streaming approach. We discuss these approach in terms of benefits and costs.

\section{Sequential vs pipelined architecture}

\begin{figure}[!t]
\centering
\includegraphics[width=\columnwidth]{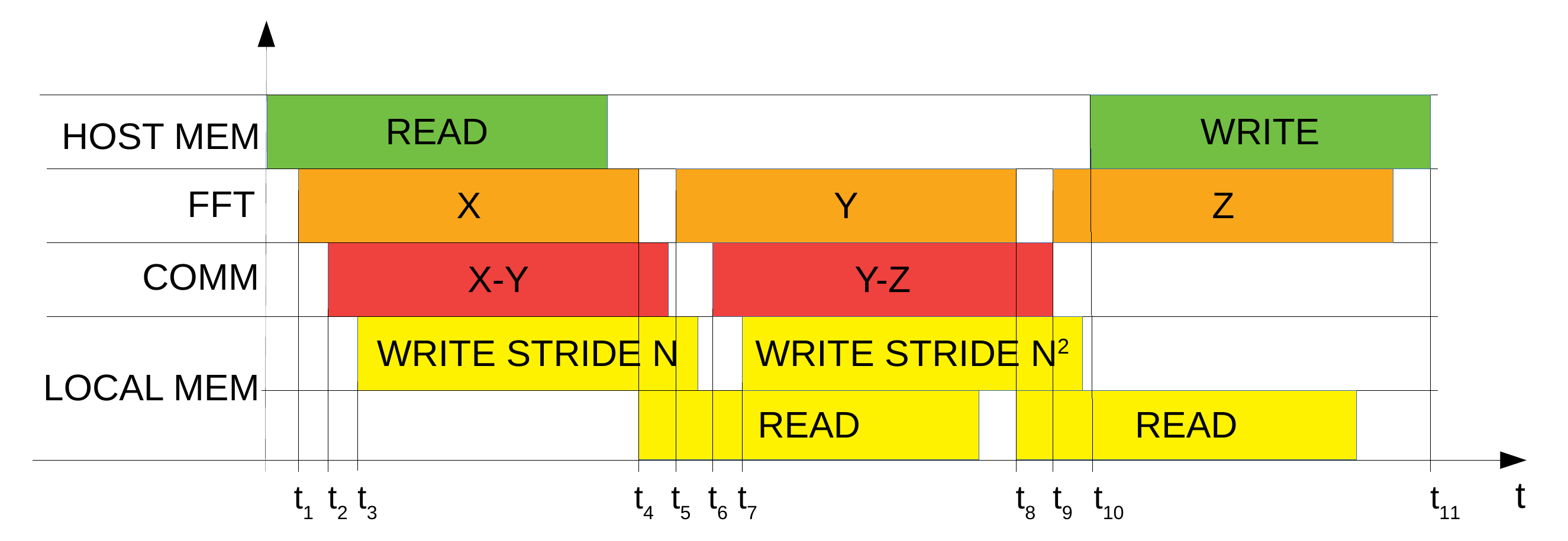}
\caption{Sequential architecture timeline details.}
\label{timeline_seq}
\end{figure}

\begin{figure}[!t]
\centering
\includegraphics[width=\columnwidth]{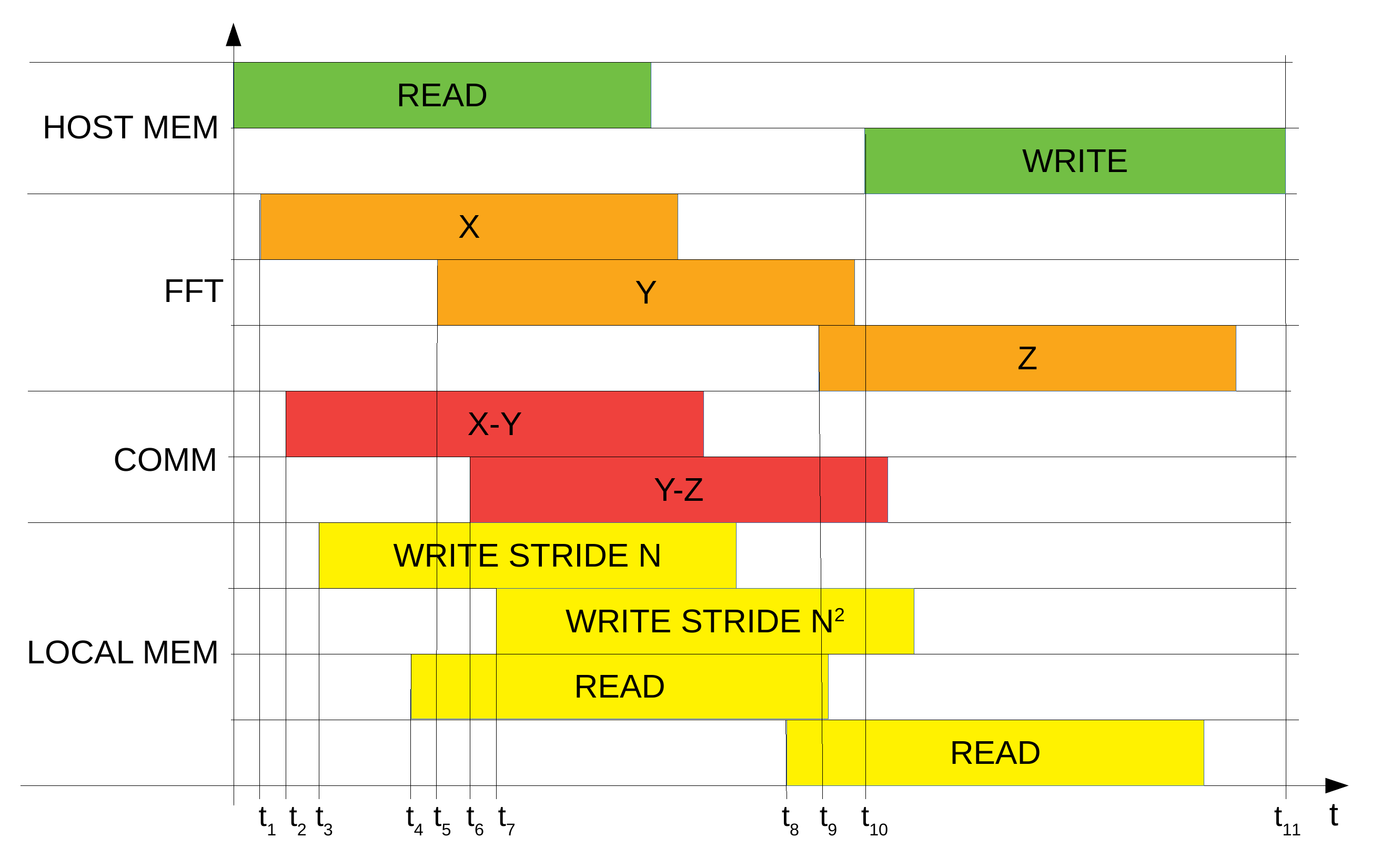}
\caption{Pipelined architecture timeline details.}
\label{timeline_pipe}
\end{figure}

\subsection{Sequential architecture}
In this case the architecture is completely sequential in the FFT computation, with $Q=k$ FFT engines working in turns on X, Y and Z transform of the incoming data, being $k$ a positive integer. The sequence of the single operations are broken up in Fig.\ref{timeline_seq}; in this architecture the Q engines work in parallel on different chunks of the X pencil, thus one DMA engine towards the host memory is needed for each FFT engine, as well as 2 concurrent DMA engines towards local memory and a single network controller. In numbers we have: $N_{H\_DMA} = k$ $N_{L\_DMA} = 2k$ $N_{NET} = k$.

The sequence of operations is as follow:
\begin{enumerate}
    \item at $t_0=0$ DMA Read from Host ram of X pencils;
    \item at $t_1=l_{DMA}$ data is available in input at the FFT engine, performing X transform;
    \item at $t_2=t_1 + l_{FFT}$ data start to exit from the FFT engine and is handled by the network controller;
    \item at $t_3=t_2 + l_{COMM}$ data exchanged through the network start to be written in local memory in Y pencils;
    \item at $t_4=t_1 + T_{FFT}^X = t_1 + l_{FFT} + t_{clk}N^3/2PRQ$ entire data set is X-transformed and Y pencils can be read from local memory;
    \item at $t_5=t_4 + l_{DMA}$ Y transform can start;
    \item at $t_6=t_5 + l_{FFT}$ data flow to the network controller;
    \item at $t_7=t_6 + l_{COMM}$ data start to be written in local memory in Z pencils;
    \item at $t_8=t_5 + T_{FFT}^Y = t_5 + l_{FFT} + t_{clk}(N^3+2N^2)/4PRQ$ entire data set is Y-transformed and Z pencils can be read from local memory;
    \item at $t_9=t_8 + l_{DMA}$ Z transform can start;
    \item at $t_{10}=t_9 + l_{FFT}$ data flow to the host memory controller;
    \item at $t_{11}=t_{10} + l_{DMA} + t_{clk}(N^3+2N^2)/4PRQ$ all data is written back into host memory.
\end{enumerate}

The total execution time of the complete 3D FFT with this architecture can be then summarized as:
\begin{equation}
    T_{tot}^{seq} = 4 l_{DMA} + 3 l_{FFT} + t_{clk}N^3/2PRQ + 2 t_{clk}(N^3+2N^2)/4PRQ
\end{equation}
For N large enough, the latencies of each block and the $N^2$ term (introduced by the real-to-complex transform with the $N/2+1$ term) become negligible, thus we can write:
\begin{equation}
     T_{tot}^{seq}  \simeq 2 t_{clk}N^3/2PRQ
\end{equation}
The data throughput needed to keep filled the engines in pipeline is two complex data words per cycle, which is:
\begin{equation}
    B^{seq} = 4sRQ/t_{clk} 
    \label{eq:bseq}
\end{equation}
While the data throughput $B^{seq}$ applies for each hardware block, DMA engines, FFT engines and network controller, data throughput in the DMA Read from Host operation is halved, due to the real-valued type of data. Instead, a further consideration is that the DMA Write to Host operation, that in this architecture is performed by the same DMA engine, still requires $B^{seq}$ bandwidth, as data is complex-valued at the end of the transform phases.
The requested bandwidth $B^{net}$ in the network of the processing nodes grid among rows and columns in the $u$ and $v$ axis is given by the amount of data that need to be exchanged at the maximum data throughput:
\begin{equation}
    \begin{array}{l}
         B^{net}_u=B^{seq}(P_u-1)/P_u\\
         B^{net}_v=B^{seq}(P_v-1)/P_v
    \end{array}
    \label{eq:bnet}
\end{equation}

The overlap between the DMA Write operation and the subsequent DMA Read operation is minimal, meaning that both the Y pencils and the Z pencils need to be stored at the same time in the local memory; thus, the local memory size $M_{tot}$ result to be:
\begin{equation}
    M_{tot}^{seq} = 2V'= 2s(N^3+2N^2)/P 
    \label{eq:mseq}
\end{equation}

\subsection{Pipelined architecture}
In the pipelined architecture, the aim is to build an even deeper pipeline among the X,Y and Z FFT engines, leveraging data reuse between them as soon as possible. In fact data dependency allows that the Y FFT (the $t_5$ time-point in the sequential case) can start as soon as $N/2P$ N-point X FFT (a plane in the 3D domain) are performed and stored on local memory. Similarly, Z FFT can start as soon as the whole volume (minus one plane) $N(N-1)/2P$ N-point Y FFT are performed and stored.

Being $k$ a positive integer that measure the multiplicity of independent FFT engines that can be instanced, there are $Q = 3k$ independent FFT engines,  $N_{H\_DMA} = k$ host DMA controllers, $N_{L\_DMA} = 4k$ host DMA controllers and $N_{NET} = 2k$ network controllers.

The sequence of operations is depicted in Fig. \ref{timeline_pipe} and is as follow:
\begin{enumerate}
    \item at $t_0=0$, DMA Read from Host ram;
    \item at $t_1=l_{DMA}$, data is available in input at the FFT engine, performing X transform;
    \item at $t_2=t_1 + l_{FFT}$, data start to exit from the FFT engine and is handled by the network controller;
    \item at $t_3=t_2 + l_{COMM}$, data exchanged through the network start to be written in local memory in Y pencils;
    \item at $t_4=t_3 + t_{clk}N^2/2PRk$, first plane is X-transformed and Y pencils can be read form local memory;
    \item at $t_5=t_4 + l_{DMA}$, Y transform can start;
    \item at $t_6=t_5 + l_{FFT}$, data flow to the network controller;
    \item at $t_7=t_6 + l_{COMM}$, data start to be written in local memory in Z pencils;
    \item at $t_8=t_7 + t_{clk}(N-1)N^2/4PRk$, the first N-1 planes are Y-transformed and Z pencils can be read from local memory;
    \item at $t_9=t_8 + l_{DMA}$, Z transform can start;
    \item at $t_{10}=t_9 + l_{FFT}$, data flow to the host memory controller;
    \item at $t_{11}=t_{10} + l_{DMA} + t_{clk}N^3/4PRk$, all data is written back into host memory;
\end{enumerate}
It is important to note that in this scheme the Y FFT engine (which has half of the N-point data set to process) will stall, meaning that Z transform can start only when X transform is finished, thus $t_8=t_2 + t_{clk}N^3/2PRk$.
With this constraint the result for the total execution time for the proposed pipelined architecture is:
\begin{equation}
     T_{tot}^{pipe} = 3 l_{DMA} + 2 l_{FFT} + t_{clk}N^3/4PRk + t_{clk}N^3/2PRk
\end{equation} 
that for N large enough it becomes:
\begin{equation}
    T_{tot}^{pipe}  \simeq 3 t_{clk}N^3/4PRk
\end{equation}

In order to prevent Y FFT engine to stall, it is possible to modify the architecture by increasing the number of FFT engines for the X transform by $2k$, while the Y and Z engine are still $k$.
With this solution the execution time is further reduced to:
\begin{equation}
     T_{tot}^{pipe}  \simeq t_{clk}N^3/2PRk
     \label{eq:tpipe}
\end{equation}
at the cost of increased resource needed for the FFT engines ($Q=4k)$. Due to the data type asymmetry introduced by the real-valued data domain, the $B^{pipe}$ as defined earlier is still valid.

The required data throughput among the hardware blocks is the same as in the sequential architecture, as well as the network bandwidth along the $u$ and $v$ axis:
\begin{equation}
    \begin{array}{l}
        B^{pipe} = 4s Rk/t_{clk}\\
        B^{net}_u=B^{pipe}(P_u-1)/P_u\\
        B^{net}_v=B^{pipe}(P_v-1)/P_v
    \end{array}
\end{equation}

With this architecture, there is a reduction in the requirement for temporary storage for Y pencils, because Y FFT can start right after the first plane of data is received into memory. We can set a minimum memory size requirement to 2 $N$ by $N/2P_u$ planes of complex data (used in a round-robin way to keep separate incoming and outgoing data), leading to a size of $2s N^2/P_u$ bytes for the Y pencils; for Z pencils we still need the complete volume $V'= s(N^3+2N^2)/P$ bytes. In total the local memory occupancy is thus:
\begin{equation}
    M_{tot}^{pipe} = s(N^3+2N^2)/P + 2s N^2/P_u
\end{equation}

\section{Parallel vs streaming vector processing}

As data to be transformed is typically a set of vector fields with $\mu$ total number of components, and since local computational phases must wait for all components to be transformed, we can approach the problem in two different ways: retrieving and computing $\mu$ components in parallel, or building a deeper pipeline and compute the $\mu$ components sequentially.

\subsection{Parallel vector processing}
When processing multiple components concurrently, the total time to compute the transform of $\mu$ components is the same as $T_{tot}^{seq}$ and $T_{tot}^{pipe}$ for a single component whether a sequential or a pipelined architecture is used, but there must be a subsequent increase of hardware resources by a factor of $\mu$, as well as the data throughput and local memory occupancy.

\subsection{Streaming vector processing}
\begin{figure}[!t]
\centering
\includegraphics[width=\columnwidth]{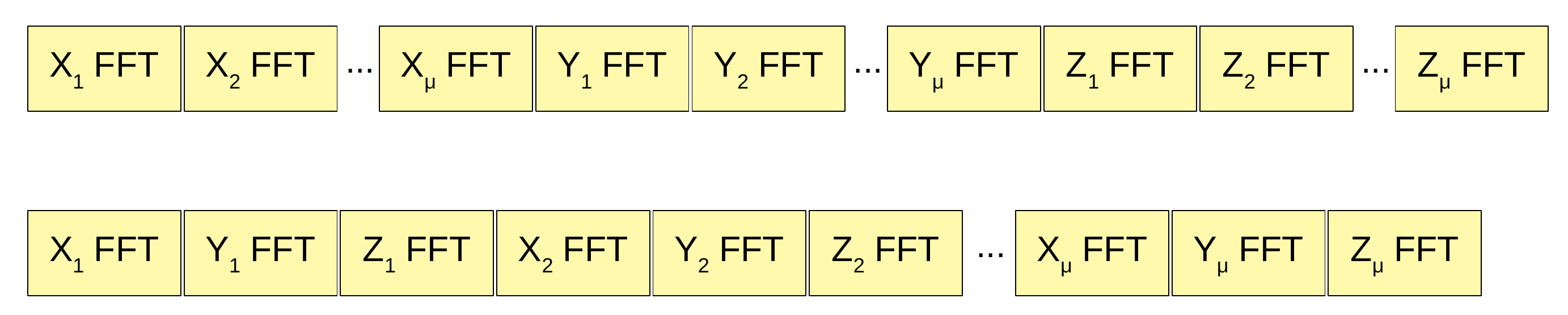}
\caption{Streaming flow of data.}
\label{streaming_arch}
\end{figure}

In order to build up a data stream, there are two possible choices: process the data per-component (first all X transforms $X_1$, $X_2$, ..., $X_\mu$, then $Y_1$, $Y_2$, ..., $Y_\mu$ and then $Z_1$, $Z_2$, ..., $Z_\mu$) or per-dimension (first the transforms of the first component $X_1$, $Y_1$, $Z_1$, then $X_2$, $Y_2$, $Z_2$, ending with $X_\mu$, $Y_\mu$, $Z_\mu$, as shown on Fig. \ref{streaming_arch}.

It is clear that the main difference between those two approaches is the local memory consumption. In the first case (per-component) all $\mu$ components need to be stored, while in the second only one component at a time is needed. The other metrics are identical.

\section{Deeper pipelined architectures}
When combining the previous described sequential and pipelined architectures for a single vector component, with the per-dimension streaming approach (which gives the most advantages in terms of resource utilization), we can finally outline the last two architectures: sequential and pipelined with streaming vector processing.

\subsection{Sequential streaming}
\begin{figure*}[!t]
\centering
\includegraphics[width=\textwidth]{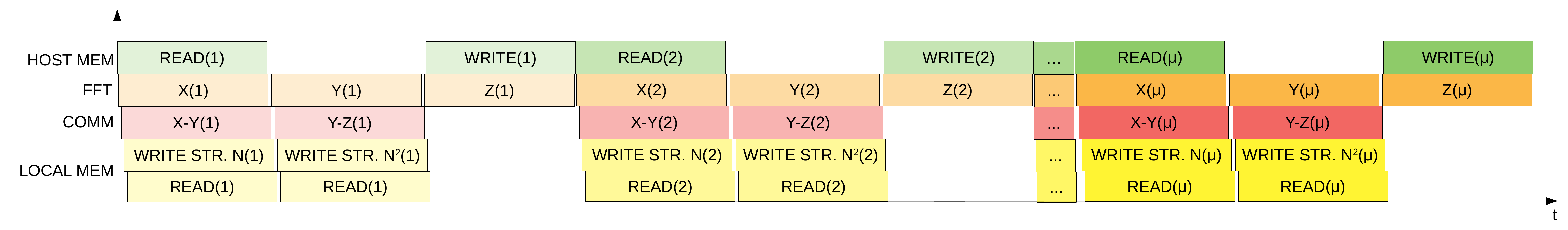}
\caption{Sequential architecture with streaming vector processing timeline details.}
\label{timeline_stream_pipe}
\end{figure*}

As shown in Fig. \ref{timeline_stream_pipe}, this architecture is a generalization of the sequential architecture for $\mu$ components. 

We can then characterize the total execution time, in the approximation of N large, as:
\begin{equation}
    T_{tot}^{seq}(\mu)  \simeq 2 \mu t_{clk}N^3/2PRQ
\end{equation}
while the other metrics are the same as in the equations \ref{eq:bseq}, \ref{eq:bnet} and \ref{eq:mseq}.

\subsection{Pipelined streaming}
In order to minimize execution time, while keeping the requested data throughput to a minimal value, the other solution is a pipelined architecture with streaming vector processing.

\begin{figure}[!t]
\centering
\includegraphics[width=\columnwidth]{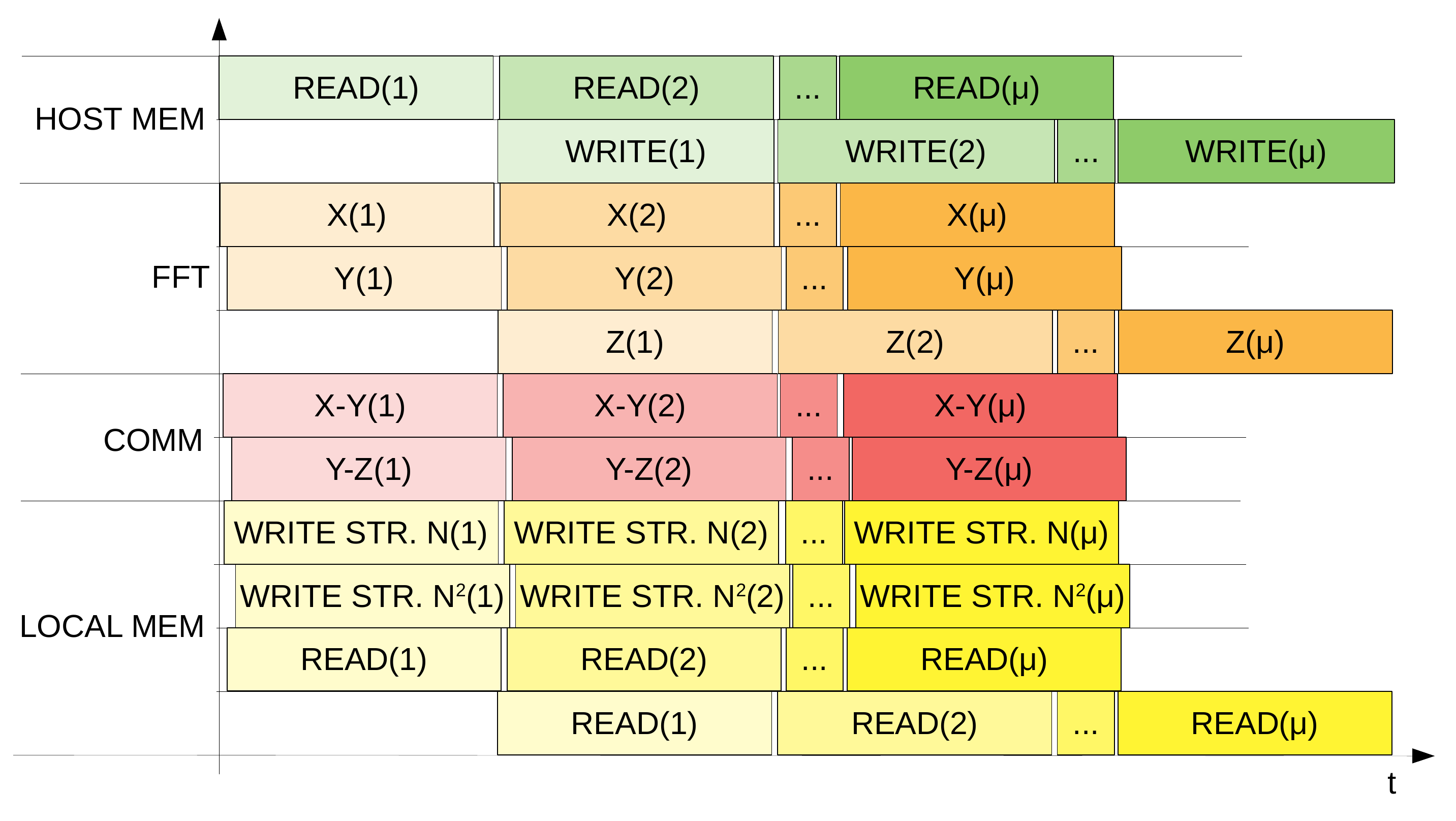}
\caption{Pipelined architecture with streaming vector processing timeline details.}
\label{timeline_deep_pipe}
\end{figure}

With this approach, the $\mu$ components are computed in a pipelined way, reusing the existing engines with an higher efficiency than in previous approaches. In fact, the timing diagram, depicted in Fig. \ref{timeline_deep_pipe}, shows how the components processing is unrolled along the time axis with all hardware blocks being active for most of the time. In order to prevent pipeline stall for the FFT engines, we use the trick of doubling the number of X FFT engines, with respect to Y and Z engines. In details, given $k$ the multiplicity of independent engines being instanced, we have $Q = 4k$ FFT engines,  $N_{H\_DMA} = 2k$ host DMA controllers, $N_{L\_DMA} = 4k$ host DMA controllers and $N_{NET} = 2k$ host DMA controllers.

Following from Eq.\ref{eq:tpipe} we can then write, for N large enough, the total execution time approximately as:
\begin{equation}
    T_{tot}^{pipe}  \simeq (\mu+1) t_{clk}N^3/4PRk
    \label{eq:pipe_model}
\end{equation}
while the required data throughput is still
\begin{equation}
    B^{pipe} = 4s Rk/t_{clk}
\end{equation}
which is the same for the sequential approach.

Local memory occupancy is independent from $\mu$, but affected by the streaming approach, in the sense that now a size of $2V'$ bytes is needed to buffer Z pencils, which is twice the size described previously, in order to avoid contentions in the continuous read/write operations on memory. In summary we have then:
\begin{equation}
    M_{tot}^{pipe} = 2s(N^3+2N^2)/P + 2s N^2/P_u
\end{equation}

\section{Discussion}


We resume the architectural solutions we discussed above in Table \ref{archi_table}. The two solutions (sequential Fig.\ref{timeline_stream_pipe}) and pipelined Fig. \ref{timeline_deep_pipe}) are compared to the parallel approach for $\mu$ components choosing, as an example, $k=1$.
The sequential approach minimize resource usage, while the pipelined approach minimize the execution time at the cost of a larger amount of resources, but still keeping the required data throughput to the lowest level. 
It is clear that, keeping the number of instanced FFT engines $Q$ constant (as long as the hardware resource allows it) the parallel approach gives the same results in terms of lower calculation time of the sequential architecture with $Q=\mu$; on the other side it requires a larger local memory size. This leads to the consideration that a parallel solution along the $\mu$ components is not worth the cost.
The most balanced solution, in terms of resources needed and execution time, seems to be the pipelined architecture with a minimal impact on memory occupancy and in terms of required data throughput.
In this case moreover, the two separate instances of the network controller fully adapt to the 2D topology of the network: the X-Y fold communication is performed along the row of the processors grid; concurrently and with maximum overlap the Y-Z fold communication is performed along the columns.

\begin{table}[]
    \centering
    \begin{tabular}{l|c|ccc}
         & & Sequential & Pipelined & Parallel \\
         \hline
        Total time & $T_{tot}$ & $2\mu$ & $(\mu+1)/2$ & 2\\
        Req. bandwidth & B & 1 & 1 & $\mu$\\
        Local RAM Size & $M_{tot}$ & $\sim 2$ & $\sim 2$ & $\sim 2\mu$\\
        \# Local DMA Controller & $N_{L\_DMA}$ & 2 & 4 & $2\mu$\\
        \# Host DMA Controller & $N_{H\_DMA}$ & 1 & 2 & $\mu$\\
        \# FFT engine & $Q$ & 1 & 4 &$\mu$\\
        \# Network Controller & $N_{NET}$ & 1 & 2 &$\mu$\\\hline
    \end{tabular}
    \caption{Comparison for $k=1$ of architectural features of sequential, pipelined and parallel approach. $T_{tot}$ is the total time spent in a complete 3D FFT for $\mu$ components, in unit of $t_{clk}N^3/2P$ seconds; B is the data throughput needed without stalling pipelines, in unit of $4s/t_{clk}$ bytes/s; local RAM size is in unit of $sN^3/P$ bytes (with a $O(N^2)$ approximation).}
    \label{archi_table}
\end{table}

A further comparison of sequential and pipelined architecture is made on Table \ref{seq_vs_pipe}, this time fixing the number of instanced FFT engines, Q, to be for example 4 for both architecture. It is clear that, even though that now the sequential approach is slightly better in terms of compute time, this costs a much higher throughput, which is something that in the pipelined architecture is kept minimal. In the case of large availability of hardware resource in order to implement a large number of FFT engines, the pipelined approach is the one that makes the best use of the available data throughput, which is often a scarce resource.

\begin{table}[]
    \centering
    \begin{tabular}{l|c|cc}
         & & Sequential & Pipelined \\
         \hline
        Total time & $T_{tot}$ & $\mu/2$ & $(\mu+1)/2$ \\
        Req. bandwidth & B & 4 & 1\\ 
        Local RAM Size & $M_{tot}$ & $\sim 2$ & $\sim 2$\\\hline
    \end{tabular}
    \caption{Comparison of sequential and pipelined architecture with fixed Q=4.}
    \label{seq_vs_pipe}
\end{table}

%% file: sections/results.tex
\chapter{Hardware implementation and results}
\label{chap:results}
\minitoc

Hardware implementation of the architectural blocks proposed has been performed on Xilinx Ultrascale+ family, which is the first to integrate in-package HBM memory for this vendor.

Single precision Fourier transforms are available as vendor IP libraries, with a great level of optimization, though the pipelined version is implemented only with a single data word input/output per cycle. As we are interested in double precision implementation, effort has been done in order to have an FFT core in double precision, yet with a limited optimization, but based on a parallel-pipelined architecture in order to guarantee a scalable implementation.

Preliminary results of this work have been published in \cite{IJASEIT8308}, but here we present more extensive results and details about the implementation and the hardware characterization.

\section{FPGA Radix-2 implementation}
\begin{figure}[!t]
\centering
\includegraphics[width=\columnwidth]{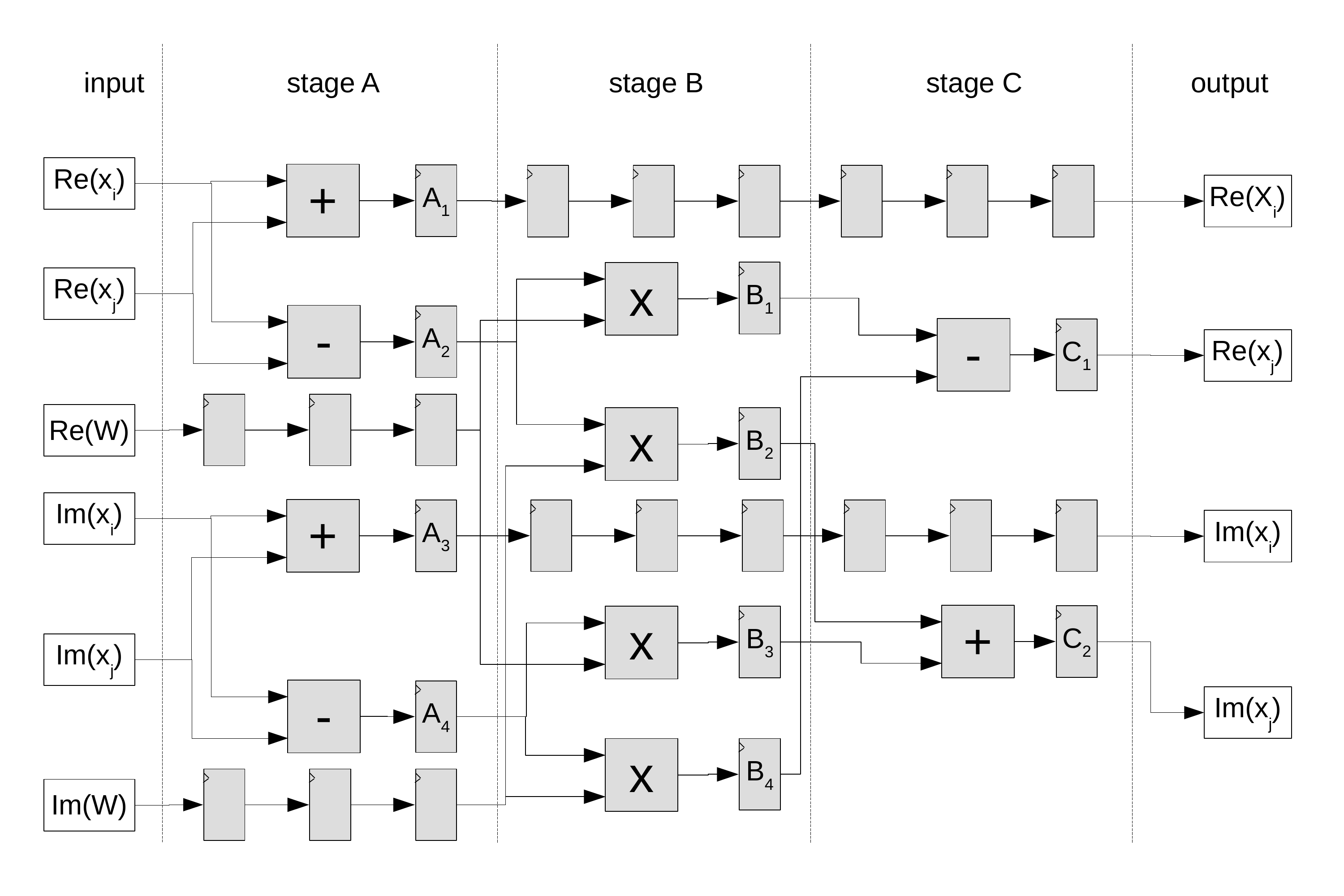}
\caption{Hardware architecture of the FFT radix-2 butterfly as logically expressed in Fig.~\ref{fft_2}.}
\label{butterfly}
\end{figure}

In this section we detail the design of the Radix-2 module depicted in Fig. \ref{fft_2} that will be used as main building block for the full mono-dimensional FFT implementation. Writing explicitly the real and imaginary parts of Eq.\ref{eq_butterfly} representing the inner butterfly FFT operation, we have:
\begin{equation}
    \begin{array}{l}
         \Re(X_i) = \Re(x_i)+\Re(x_j)\\
         \Im(X_i) = \Im(x_i)+\Im(x_j)\\
         \Re(X_j) = \Re(W_N)(\Re(x_i)-\Re(x_j))-\Im(W_N)(\Im(x_i)-\Im(x_j))\\
         \Im(X_j) = \Im(W_N)(\Re(x_i)-\Re(x_j))+\Re(W_N)(\Im(x_i)-\Im(x_j))
    \end{array}
\end{equation}
where $x_i$ and $x_j$ represent the butterfly inputs, $X_i$ and $X_j$ the outputs and $W$ is the twiddle factor. A direct hardware implementation of these equations for FPGA is reported in Fig. \ref{butterfly}  where the complex operations are broken up in real part and imaginary part double precision adders and multipliers. In order to maximize data throughput the hardware architecture makes use of all concurrent operations, for a total of 6 adders and 4 multipliers. Other approaches are possible, as reviewed in \cite{joshi2015fft}, targeting a specific optimization, such as low power consumption \cite{ayinala2012pipelined}, low area usage \cite{garrido2013pipelined}, or real-valued only signals \cite{garrido2009pipelined}.
In our implementation each operator has its characteristic (and parametric) internal latency and a streaming interface (in input and output ports), and they are typically available from IP libraries of the FPGA vendors, even in double precision format with fully IEEE-754 standard compliance. This means that a balanced chain of operators perfectly fits in a pipelined architecture, where after an initial delay (\textit{i.e.} latency) data is available at each clock cycle.

Calculation phases are organized in three stages (A, B and C in the following). In Stage A the following operations are performed in parallel:
\begin{equation*}
    \begin{array}{cc}
        A_1 & = \Re(x_i)+\Re(x_j)\\
        A_2 & = \Re(x_i)-\Re(x_j)\\
        A_3 & = \Im(x_i)+\Im(x_j)\\
        A_4 & = \Im(x_i)-\Im(x_j)
    \end{array}
\end{equation*}
while $\Re(W_N)$ and $\Im(W_N)$ are delayed to stage B on registers chains with the same length of the latency of the adders.

In Stage B the following operation are computed in parallel:
\begin{equation*}
    \begin{array}{cc}
        B_1 & = A_2 \times \Re(W_N)\\
        B_2 & = A_4 \times \Im(W_N)\\
        B_3 & = A_2 \times \Im(W_N)\\
        B_4 & = A_4 \times \Re(W_N)
    \end{array}
\end{equation*}
while $A_1$ and $A_3$ are delayed to output stage. Finally in Stage C the following operations are performed:
\begin{equation*}
    \begin{array}{cc}
        C_1 & = B_1 - B_4\\
        C_2 & = B_2 + B_3
    \end{array}
\end{equation*}
and the results are delivered to the output.
    
Among these stages data is registered in order to ease resource placement and achieve a higher working frequency.
Adders have a programmable latency among 0 and 14 clock cycles and multipliers among 0 and 12, and they are all implemented in order to allow operands to be applied on every clock cycle (one cycle per operation). The computational complexity of this radix-2 butterfly implementation is 10 Floating Point operations per cycle.
Defining the latency of the three stages $l_A$, $l_B$ and $l_C$, then the total latency of the butterfly operation block $l_{but}$ is:
\begin{equation}
    l_{but}=l_A+l_B+l_C+4
\end{equation}
being 4 number of registration cycles among the stages.
Varying $l_{but}$ has an impact on working clock period $t_{clk}$, thus resulting to be a adaptable parameter.

\section{Scalable 1D FFT implementation}

\begin{figure}[!t]
\centering
\includegraphics[width=\textwidth]{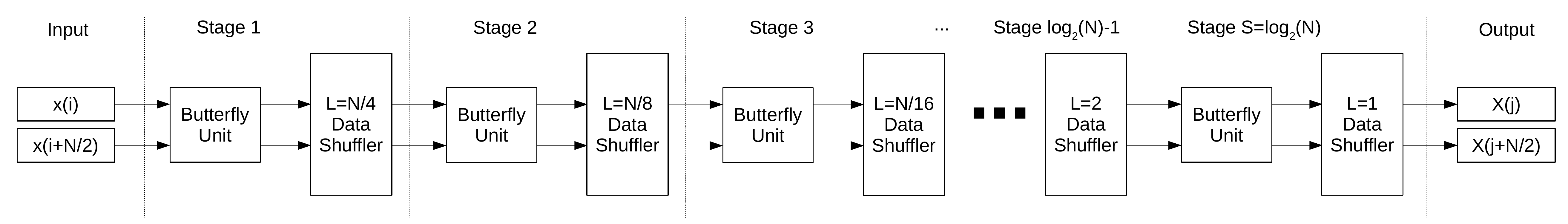}
\caption{Architecture of the pipelined engine for a N-point FFT with R=1.}
\label{pipeline}
\end{figure}

In order to implement a complete row of the parallel-pipelined architecture, we use the butterfly blocks of Fig.\ref{butterfly} chained in $S=log_2(N)$ stages, as shown in Fig. \ref{pipeline}. Inputs at first stage are time multiplexed; twiddle factors $W_N^m$ are fetched from a predefined ROM table, resulting to be a function of the pipeline step and of the row.

\begin{figure}[!t]
\centering
\includegraphics[width=.7\textwidth]{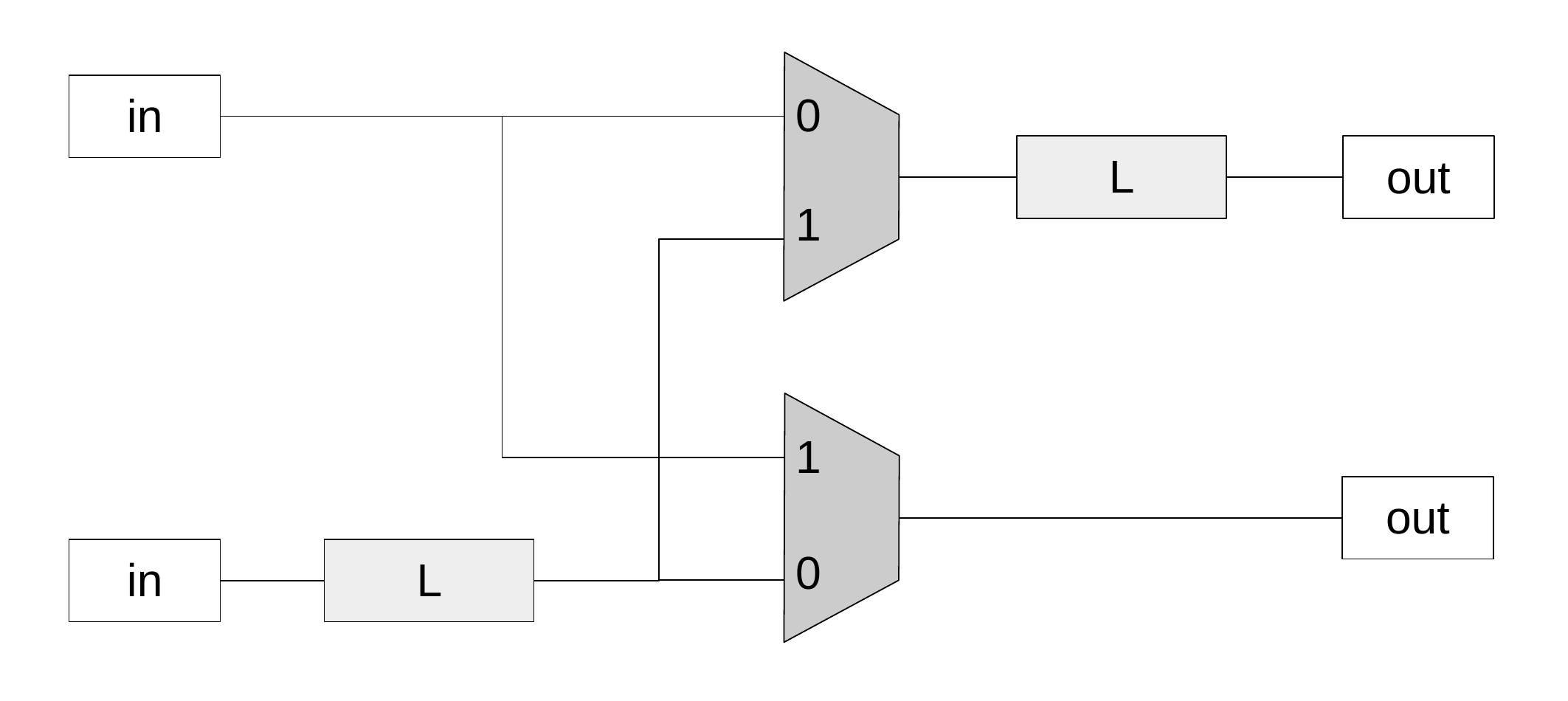}
\caption{Circuit representing the data shuffler block for a generic length L.}
\label{shuffler}
\end{figure}

Between one stage and the other it is necessary to perform a data shuffling phase, in order to reproduce the DIF FFT flow graph on a single row, in this case. Although there could be several ways to implement this logic function, we found the most appropriate one in our context and for FPGAs, in terms of simplicity and minimal delay time and resource usage, to be a data shuffling circuit described in Fig. \ref{shuffler}, already proposed and used in \cite{he1998design}, \cite{gold1973parallelism}, \cite{swartzlander1984radix}, \cite{oppenheim1978applications}, \cite{garrido2009pipelined} and \cite{garrido2013pipelined}. This circuit is composed by two two-port multiplexer and two shift registers of length L; in the upper data stream the shift register delays the data after the multiplexer, in the lower stream it is placed before the multiplexer. The select pin of the multiplexer is driven by the most significant bit of a counter ranging from 0 to $2L-1$.

As said, the data shuffling block is placed after every butterfly unit at each stage, with the length of the shift registers expressed as a function of the stage $L=N/2^{s+1}$ having $s=1, \dots, log_2N-1$. As a result, this block introduces a delay $l_{reord}=L(s)+1$ at each stage $s$, with the last term due to an extra registration cycle in our hardware. We can then model the total latency of the FFT block as:
\begin{equation}
    l_{FFT} = l_{but} log_2N + \sum_{s=1}^{log_2N-1}L(s) + log_2N = (l_{but} + 1) log_2N + \frac{N}{2}-1
    \label{eq:lfft}
\end{equation}

\begin{figure}[!t]
\centering
\includegraphics[width=\textwidth]{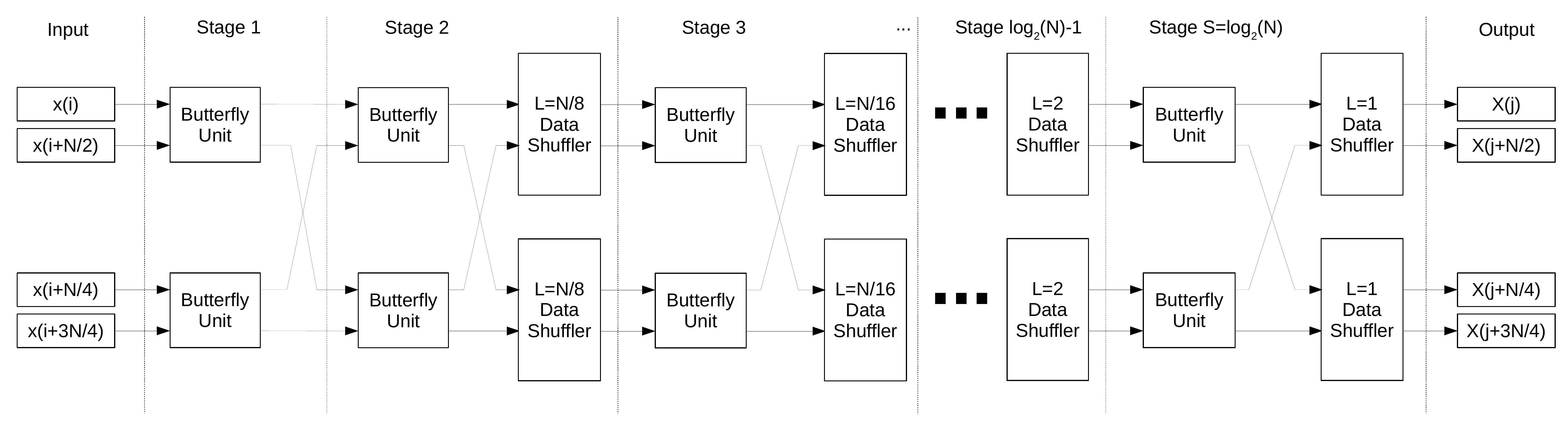}
\caption{Architecture of the pipelined engine for a N-point FFT with R=2.}
\label{4p_pipeline}
\end{figure}

\begin{figure}[!t]
\centering
\includegraphics[width=\textwidth]{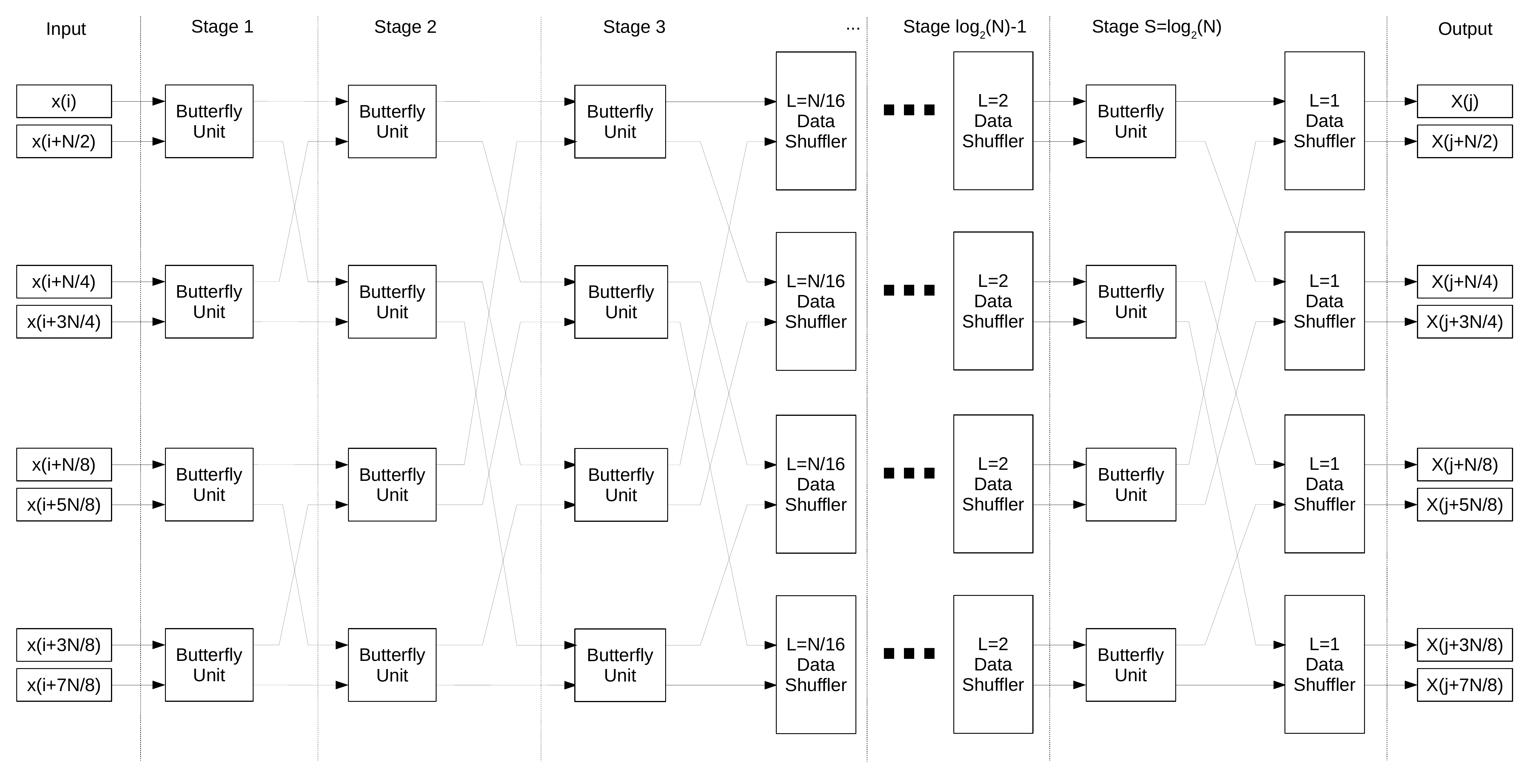}
\caption{Architecture of the pipelined engine for a N-point FFT with R=4.}
\label{8p_pipeline}
\end{figure}

This single row (R=1) architecture can be easily ported to generic R implementations. In our work, we have implemented the case for R=2 and R=4, that are shown respectively in Fig. \ref{4p_pipeline} and Fig. \ref{8p_pipeline}. As it can be noted, the amount of on-chip memory required to implement the shift registers is almost preserved regardless to R and it is easy to demonstrate that it is proportional to $N-2R$ \cite{garrido2013pipelined}.

\section{FFT Implementation results}

The proposed FFT double precision engine in the case of $R=1,2,4$ has been implemented on a Xilinx Virtex Ultrascale+ VU37P device for full synthesis and simulation with Vivado 2018.3 version. The double precision floating point adders, substractors and multipliers \cite{xilinx_pg060} are instanced for high speed architecture optimization and full usage of DSP blocks (3 DSP48E2 for add/substract and 7 for multipliers). Each butterfly implies a usage of a total of 46 DSP blocks.

Several synthesis attempts have been performed changing the size of the FFT transform (in the range of interesting size $N=512, \dots, 8192$) and the latency of the floating point operators $l_{op}=l_A=l_B=l_C$. We characterize the engine by measuring the usage of the main resource types (CLB LUTs, CLB Registers \cite{xilinx_ug574}, BRAM blocks \cite{xilinx_ug573}, DSP blocks \cite{xilinx_ug579}, the maximum operation frequency $f_{max}$ and the power consumption $W$ (estimated using the power using Xilinx Power Analyzer tool). We can then estimate several quantities such as the latency of a single FFT $l_{FFT}$ as defined in Eq.~\ref{eq:lfft}, the time to compute an FFT $T_{FFT}$ as defined in Eq.~\ref{eq:TFFT}, the required data throughput $B_{FFT}$ as defined in Eq.~\ref{eq:BFFT}. Another significant quantity is the amount of (double precision) floating point per second (GFLOPS) the engine is capable to compute, according the following formula:
\begin{equation}
    GFLOPS=\frac{10 R \log_2 N}{t_{clk}}
\end{equation}
Lastly, a measure of power efficiency of the engine can be estimated with the $GFLOPS/W$ ratio.

The synthesis and simulation results are reported in Table~\ref{tab:2p_resource} and \ref{tab:2p_perf} for the $R=1$ implementation; in Table~\ref{tab:4p_resource} and \ref{tab:4p_perf} for the $R=2$ implementation; in Table~\ref{tab:8p_resource} and \ref{tab:8p_perf} for the $R=4$ implementation.

The following considerations can be asserted:
\begin{itemize}
    \item increasing the latency of the floating point operators strongly affects the operating frequency (and thus giving an hint to the upper limit of $t_{clk}$, allowing $f_{max}$ over 380 MHz for the maximum allowable latency;
    \item the side effect of increasing the operators latency is a proportional increase of Registers specialized logic blocks, which have also an impact on power consumption;
    \item increasing the number of rows R is a tangible way to exploit the amount of DSP blocks in the device, with a sustainable increase of LUTs and Registers logic blocks amount of resources;
    \item hardware efficiency has a benefit in increasing R in terms of W and GFLOPS/W
    \item the counterpart of increasing R is the proportional increase of $B_{FFT}$, which easily reaches excessive values for current network link technologies; a moderation of $t_{clk}$ is thus needed, in order to balance data throughput.
    \item the maximum value of instantiable engines can be set to $Q=4$ even for $R=4$, with the exception at larger values of $N=8192$ where LUT Blocks and DSP Blocks amount is exceeded.
\end{itemize}

It is worth noting that exceeding of resources might be prevented by tuning the number of DSP blocks usage for multiplier operator, and reducing the number of registration steps in the butterfly architecture pursuing a more optimized design.

\begin{figure}[!t]
\centering
\includegraphics[width=\textwidth]{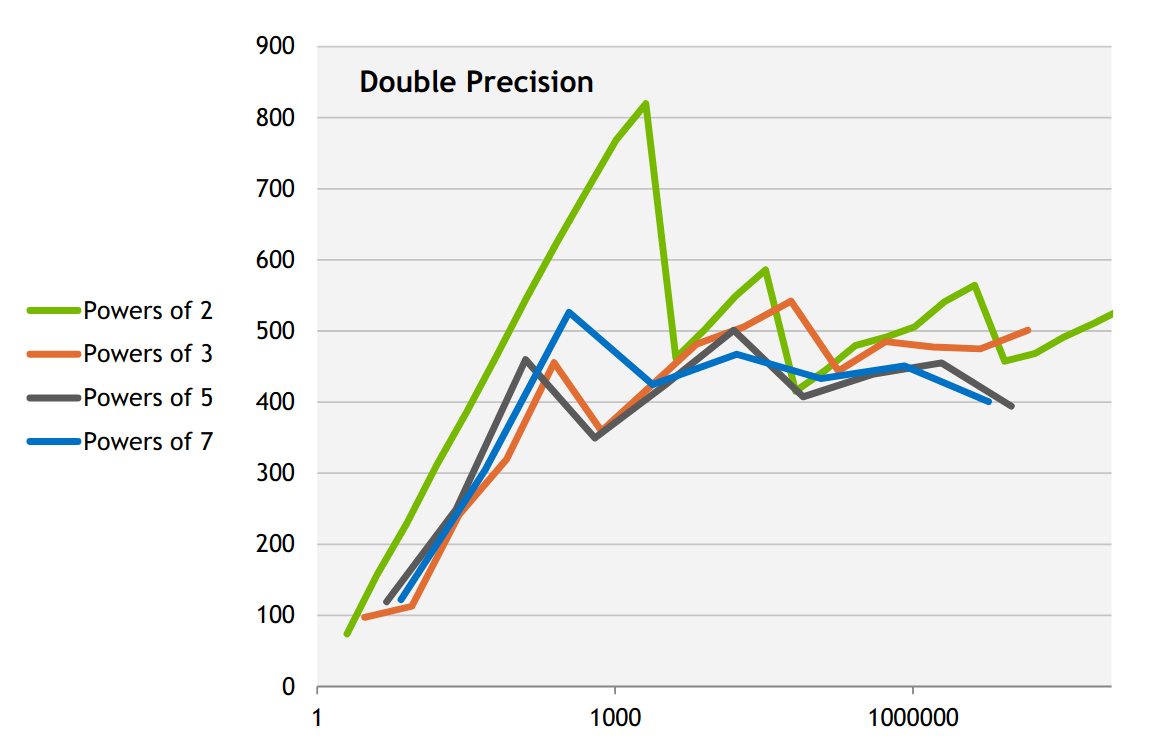}
\caption{Double precision performance (GFLOPS versus N) on a single NVIDIA P100 device with cuFFT library, taken from \cite{nvidia_perf}.}
\label{fig:cuFFT}
\end{figure}

Furthermore, if we compare the resulting GFLOPS performance for $R=4$, in the hypothesis of being capable of instancing $Q=4$ engines, with the NVIDIA CUDA cuFFT \cite{nvidia_cufft} optimized library performance results \cite{nvidia_perf} on a state of the art NVIDIA P100 device (plot shown in Fig.~\ref{fig:cuFFT}), we can see that they are substantially similar in terms of sustained GFLOPS, but at a fraction of the power consumption. However, a quantitative comparison requires a more accurate study.

\begin{table}[]
    \centering
    \begin{tabular}{|c|c|c|c|c|c|c|c|c|}
        \hline
        R & N & $l_{op}$ & \% LUTs & \% REGs & \% BRAM & \% DSP & $f_{max}$ & W \\
        & & & & & & & (MHz) & (Watt)\\
        \hline
          & 512  &   & 3.74 & 1.65 & 0.79 & 4.59 & 250 & 4.142 \\
          & 1024 &   & 4.30 & 1.80 & 0.89 & 5.10 & 247 & 4.612 \\
        1 & 2048 & 3 & 5.01 & 2.01 & 1.19 & 5.61 & 251 & 5.262 \\
          & 4096 &   & 6.04 & 2.22 & 2.58 & 6.12 & 244 & 6.02  \\
          & 8192 &   & 6.45 & 2.46 & 5.90 & 6.63 & 236 & 6.967 \\
        \hline
          & 512  &   & 3.51 & 2.31 & 0.00 & 4.59 & 348 & 4.908 \\
          & 1024 &   & 4.42 & 2.54 & 0.00 & 5.10 & 345 & 5.824 \\
        1 & 2048 & 6 & 4.75 & 2.87 & 1.19 & 5.61 & 346 & 6.894 \\
          & 4096 &   & 5.76 & 3.22 & 2.58 & 6.12 & 323 & 8.016 \\
          & 8192 &   & 6.15 & 3.49 & 6.30 & 6.63 & 344 & 9.119 \\
        \hline
          & 512  &   & 3.49 & 2.63 & 0.00 & 4.59 & 379 & 5.729 \\
          & 1024 &   & 4.41 & 2.91 & 0.00 & 5.10 & 376 & 6.66 \\
        1 & 2048 & 9 & 4.74 & 3.26 & 1.19 & 5.61 & 379 & 7.781 \\
          & 4096 &   & 6.05 & 3.62 & 2.58 & 6.12 & 371 & 8.997 \\
          & 8192 &   & 6.13 & 3.92 & 6.30 & 6.63 & 355 & 10.21 \\
        \hline
          & 512  &         & 3.57 & 3.60 & 0.50 & 4.59 & 380 & 6.941 \\
          & 1024 &         & 4.42 & 4.00 & 0.50 & 5.10 & 380 & 8.131 \\
        1 & 2048 & 14 (12) & 4.75 & 4.42 & 1.19 & 5.61 & 380 & 8.858 \\
          & 4096 &         & 5.76 & 4.91 & 2.58 & 6.12 & 365 & 10.318 \\
          & 8192 &         & 6.15 & 5.32 & 6.30 & 6.63 & 345 & 11.81 \\
        \hline
    \end{tabular}
    \caption{Resource usage and hardware characterization of a single FFT engine on a Xilinx VU37P device for R=1 (hardware architecture depicted in Fig.~\ref{pipeline}).}
    \label{tab:2p_resource}
\end{table}

\begin{table}[]
    \centering
    \begin{tabular}{|c|c|c|c|c|c|c|c|c|}
        \hline
        R & N & $l_{op}$ & latency & $l_{FFT}$ & $T_{FFT}$ & $B_{FFT}$ & GFLOPS & GFLOPS/W \\
          &   &          & cycles  & ($\mu s$) & ($\mu s)$ & ($GB/s$)  &        &          \\
        \hline
          & 512  &   & 382  & 1.53 & 2.55 & 7.45 & 22.5 & 5.43 \\
          & 1024 &   & 652  & 2.64 & 4.71 & 7.36 & 24.7 & 5.36 \\
        1 & 2048 & 3 & 1178 & 4.69 & 8.77 & 7.48 & 27.61 & 5.25 \\
          & 4096 &   & 2216 & 9.08 & 17.48 & 7.27 & 29.28 & 4.86 \\
          & 8192 &   & 4278 & 18.13& 35.48 & 7.03 & 30.68 & 4.40 \\
        \hline
          & 512  &   & 463  & 1.33 & 2.07 & 10.37 & 31.32 & 6.38 \\
          & 1024 &   & 742  & 2.15 & 3.63 & 10.28 & 34.5 & 5.92 \\
        1 & 2048 & 6 & 1277 & 3.69 & 6.65 & 10.31 & 38.06 & 5.52 \\
          & 4096 &   & 2324 & 7.20 & 13.54 & 9.63 & 38.76 & 4.84 \\
          & 8192 &   & 4395 & 12.78& 24.68 & 10.25 & 44.72 & 4.90 \\
        \hline
          & 512  &   & 544  & 1.44 & 2.11 & 11.30 & 34.11 & 5.95 \\
          & 1024 &   & 832  & 2.21 & 3.57 & 11.21 & 37.6 & 5.65 \\
        1 & 2048 & 9 & 1376 & 3.63 & 6.33 & 11.30 & 41.69 & 5.36 \\
          & 4096 &   & 2432 & 6.56 & 12.08 & 11.06 & 44.52 & 4.95 \\
          & 8192 &   & 4512 & 12.71& 24.25 & 10.58 & 46.15 & 4.52 \\
        \hline
          & 512  &         & 661  & 1.74 & 2.41 & 11.32 & 34.2 & 4.93 \\
          & 1024 &         & 962  & 2.53 & 3.88 & 11.32 & 38 & 4.67 \\
        1 & 2048 & 14 (12) & 1519 & 4.00 & 6.69 & 11.32 & 41.8 & 4.72 \\
          & 4096 &         & 2588 & 7.09 & 12.70 & 10.88 & 43.8 & 4.25 \\
          & 8192 &         & 4681 & 13.57& 25.44 & 10.28 & 44.85 & 3.80 \\
        \hline
    \end{tabular}
    \caption{Hardware performances of a single FFT engine on a Xilinx VU37P device for R=1 (hardware architecture depicted in Fig.~\ref{pipeline}) assuming $t_{clk}=1/f_{max}$ with $f_{max}$ taken from the corresponding entry in Table~\ref{tab:2p_resource}.}
    \label{tab:2p_perf}
\end{table}

\begin{table}[]
    \centering
    \begin{tabular}{|c|c|c|c|c|c|c|c|c|}
        \hline
        R & N & $l_{op}$ & \% LUTs & \% REGs & \% BRAM & \% DSP & $f_{max}$ & W \\
        & & & & & & & (MHz) & (Watt)\\
        \hline
          & 512  &   & 7.30  & 3.12 & 1.54  & 9.18  & 238 & 6.791 \\
          & 1024 &   & 8.26  & 3.44 & 1.74  & 10.20 & 232 & 7.543 \\
        2 & 2048 & 3 & 9.38  & 3.80 & 2.28  & 11.21 & 234 & 8.248 \\
          & 4096 &   & 10.81 & 4.19 & 4.76  & 12.23 & 230 & 9.538 \\
          & 8192 &   & 12.87 & 4.65 & 10.32 & 13.25 & 244 & 11.026 \\
        \hline
          & 512  &   & 6.87  & 4.54 & 0.79 & 9.18  & 343 & 10.265 \\
          & 1024 &   & 8.22  & 5.02 & 0.60 & 10.20 & 344 & 11.775 \\
        2 & 2048 & 6 & 9.05  & 5.50 & 0.99 & 11.21 & 345 & 12.508 \\
          & 4096 &   & 10.94 & 6.06 & 4.76 & 12.23 & 341 & 15.244 \\
          & 8192 &   & 15.18 & 6.73 & 8.04 & 13.25 & 330 & 16.82 \\
        \hline
          & 512  &   & 6.84  & 5.16 & 0.79 & 9.18  & 379 & 11.465 \\
          & 1024 &   & 8.34  & 5.77 & 0.60 & 10.20 & 378 & 13.593 \\
        2 & 2048 & 9 & 9.05  & 6.37 & 1.39 & 11.21 & 378 & 14.997 \\
          & 4096 &   & 10.22 & 6.97 & 5.16 & 12.23 & 378 & 16.852 \\
          & 8192 &   & 15.60 & 7.69 & 7.54 & 13.25 & 377 & 21.084 \\
        \hline
          & 512  &         & 6.91  & 7.09  & 1.14 & 9.18  & 380 & 13.14 \\
          & 1024 &         & 8.13  & 7.87  & 1.29 & 10.20 & 392 & 15.61 \\
        2 & 2048 & 14 (12) & 9.09  & 8.67  & 1.39 & 11.21 & 392 & 16.747 \\
          & 4096 &         & 11.92 & 9.51  & 2.98 & 12.23 & 380 & 20.231 \\
          & 8192 &         & 15.71 & 10.48 & 7.94 & 13.25 & 380 & 24.05 \\
        \hline
    \end{tabular}
    \caption{Resource usage and hardware characterization of a single FFT engine on a Xilinx VU37P device for R=2 (hardware architecture depicted in Fig.~\ref{4p_pipeline}).}
    \label{tab:4p_resource}
\end{table}

\begin{table}[]
    \centering
    \begin{tabular}{|c|c|c|c|c|c|c|c|c|}
        \hline
        R & N & $l_{op}$ & latency & $l_{FFT}$ & $T_{FFT}$ & $B_{FFT}$ & GFLOPS & GFLOPS/W \\
          &   &          & cycles  & ($\mu s$) & ($\mu s)$ & ($GB/s$)  &        &          \\
        \hline
          & 512  &   & 254  & 1.07 & 1.61 & 14.19 & 42.84 & 6.31 \\
          & 1024 &   & 396  & 1.71 & 2.81 & 13.83 & 46.4 & 6.15 \\
        2 & 2048 & 3 & 666  & 2.85 & 5.03 & 13.95 & 51.48 & 6.24 \\
          & 4096 &   & 1192 & 5.18 & 9.63 & 13.71 & 55.2 & 5.79 \\
          & 8192 &   & 2230 & 9.14 & 17.53 & 14.54 & 63.44 & 5.75 \\
        \hline
          & 512  &   & 335  & 0.98 & 1.35 & 20.44 & 61.74 & 6.01 \\
          & 1024 &   & 486  & 1.41 & 2.16 & 20.50 & 68.8 & 5.84 \\
        2 & 2048 & 6 & 765  & 2.22 & 3.70 & 20.56 & 75.9& 6.07 \\
          & 4096 &   & 1300 & 3.81 & 6.82 & 20.33 & 81.84 & 5.37 \\
          & 8192 &   & 2347 & 7.11 & 13.32 & 19.67 & 85.8 & 5.10\\
        \hline
          & 512  &   & 416  & 1.10 & 1.44 & 22.59 & 68.22 & 5.95 \\
          & 1024 &   & 576  & 1.52 & 2.20 & 22.53 & 75.6 & 5.56 \\
        2 & 2048 & 9 & 864  & 2.29 & 3.64 & 22.53 & 83.16 & 5.55 \\
          & 4096 &   & 1408 & 3.72 & 6.43 & 22.53 & 90.72 & 5.38 \\
          & 8192 &   & 2464 & 6.54 & 11.97 & 22.47 & 98.8 & 4.65 \\
        \hline
          & 512  &         & 533  & 1.40 & 1.74 & 22.65 & 68.4 & 5.21 \\
          & 1024 &         & 706  & 1.80 & 2.45 & 23.37 & 78.4 & 5.02 \\
        2 & 2048 & 14 (12) & 1007 & 2.57 & 3.88 & 23.37 & 86.24 & 5.15 \\
          & 4096 &         & 1564 & 4.12 & 6.81 & 22.65 & 91.2 & 4.51 \\
          & 8192 &         & 2633 & 6.93 & 12.32 & 22.65 & 98.8 & 4.11 \\
        \hline
    \end{tabular}
    \caption{Hardware performances of a single FFT engine on a Xilinx VU37P device for R=2 (hardware architecture depicted in Fig.~\ref{4p_pipeline}) assuming $t_{clk}=1/f_{max}$ with $f_{max}$ taken from the corresponding entry in Table~\ref{tab:4p_resource}.}
    \label{tab:4p_perf}
\end{table}

\begin{table}[]
    \centering
    \begin{tabular}{|c|c|c|c|c|c|c|c|c|}
        \hline
        R & N & $l_{op}$ & \% LUTs & \% REGs & \% BRAM & \% DSP & $f_{max}$ & W \\
        & & & & & & & (MHz) & (Watt)\\
        \hline
          & 512  &   & 14.41 & 6.09 & 2.38 & 18.35 & 226 & 10.581 \\
          & 1024 &   & 16.16 & 6.73 & 2.48 & 20.39 & 231 & 12.809\\
        4 & 2048 & 3 & 18.08 & 7.40 & 3.32 & 22.43 & 230 & 14.453\\
          & 4096 &   & 20.30 & 8.10 & 7.34 & 24.47 & 222 & 15.646 \\
          & 8192 &   & 23.17 & 8.88 & 15.67 & 26.51 & 231 & 19.39 \\
        \hline
          & 512  &   & 13.65 & 8.77 & 1.59 & 18.35 & 337 & 16.928 \\
          & 1024 &   & 15.30 & 9.81 & 2.08 & 20.39 & 337 & 22.456 \\
        4 & 2048 & 6 & 17.24 & 10.79 & 1.79 & 22.43 & 332 & 20.552\\
          & 4096 &   & 21.69 & 11.82 & 5.56 & 24.47 & 333 & 27.346 \\
          & 8192 &   & 25.52 & 12.93 & 9.92 & 26.51 & 254 & 34.343 \\
        \hline
          & 512  &   & 13.61 & 10.27 & 1.98 & 18.35 & 379 & 22.436 \\
          & 1024 &   & 15.23 & 11.38 & 1.98 & 20.39 & 377 & 25.801 \\
        4 & 2048 & 9 & 17.15 & 12.52 & 1.98 & 22.43 & 376 & 22.198 \\
          & 4096 &   & 20.80 & 13.71 & 4.96 & 24.47 & 378 & 32.156 \\
          & 8192 &   & 27.36 & 15.00 & 11.90 & 26.51 & 291 & 38.973 \\
        \hline
          & 512  &         & 13.56 & 14.05 & 1.79 & 18.35 & 379 & 23.047 \\
          & 1024 &         & 15.59 & 15.60 & 2.23 & 20.39 & 379 & 26.619 \\
        4 & 2048 & 14 (12) & 17.20 & 17.17 & 2.28 & 22.43 & 379 & 21.274 \\
          & 4096 &         & 20.66 & 18.77 & 6.15 & 24.47 & 384 & 28.746 \\
          & 8192 &         & 27.07 & 20.56 & 9.52 & 26.51 & 308 & 40.541 \\
        \hline
    \end{tabular}
    \caption{Resource usage and hardware characterization of a single FFT engine on a Xilinx VU37P device for R=4 (hardware architecture depicted in Fig.~\ref{8p_pipeline}).}
    \label{tab:8p_resource}
\end{table}

\begin{table}[]
    \centering
    \begin{tabular}{|c|c|c|c|c|c|c|c|c|}
        \hline
        R & N & $l_{op}$ & latency & $l_{FFT}$ & $T_{FFT}$ & $B_{FFT}$ & GFLOPS & GFLOPS/W \\
          &   &          & cycles  & ($\mu s$) & ($\mu s)$ & ($GB/s$)  &        &          \\
        \hline
          & 512  &   & 190  & 0.84 & 1.12 & 26.94 & 81.36  & 7.69 \\
          & 1024 &   & 268  & 1.16 & 1.71 & 27.54 & 92.4   & 7.21 \\
        4 & 2048 & 3 & 410  & 1.78 & 2.90 & 27.42 & 101.2  & 7.00 \\
          & 4096 &   & 680  & 3.06 & 5.37 & 26.46 & 106.56 & 6.81 \\
          & 8192 &   & 1206 & 5.22 & 9.65 & 27.54 & 120.12 & 6.19 \\
        \hline
          & 512  &   & 271  & 0.80 & 0.99 & 40.17 & 121.32 & 7.17 \\
          & 1024 &   & 358  & 1.06 & 1.44 & 40.17 & 134.8  & 6.00 \\
        4 & 2048 & 6 & 509  & 1.53 & 2.30 & 39.58 & 146.08 & 7.11 \\
          & 4096 &   & 788  & 2.37 & 3.90 & 39.70 & 159.84 & 5.85 \\
          & 8192 &   & 1323 & 5.21 & 9.24 & 30.28 & 132.08 & 3.85 \\
        \hline
          & 512  &   & 352  & 0.93 & 1.10 & 45.18 & 136.44 & 6.08 \\
          & 1024 &   & 448  & 1.19 & 1.53 & 44.94 & 150.8  & 5.84 \\
        4 & 2048 & 9 & 608  & 1.62 & 2.30 & 44.82 & 165.44 & 7.45 \\
          & 4096 &   & 896  & 2.37 & 3.72 & 45.06 & 181.44 & 5.64 \\
          & 8192 &   & 1440 & 4.95 & 8.47 & 34.69 & 151.32 & 3.88 \\
        \hline
          & 512  &         & 469  & 1.24 & 1.41 & 45.18 & 136.44 & 5.92 \\
          & 1024 &         & 578  & 1.53 & 1.86 & 45.18 & 151.6  & 5.70 \\
        4 & 2048 & 14 (12) & 751  & 1.98 & 2.66 & 45.18 & 166.76 & 7.84 \\
          & 4096 &         & 1052 & 2.74 & 4.07 & 45.78 & 184.32 & 6.41 \\
          & 8192 &         & 1609 & 5.22 & 8.55 & 36.72 & 160.16 & 3.95 \\
        \hline
    \end{tabular}
    \caption{Hardware performances of a single FFT engine on a Xilinx VU37P device for R=4 (hardware architecture depicted in Fig.~\ref{8p_pipeline}).}
    \label{tab:8p_perf}
\end{table}

\section{FPGA external memory integration}

As demonstrated in the FFT architecture modelling discussion, a large size memory block is needed in order the handle the X-Y and Y-Z transposition phase. Due to the FFT engine data throughput $B_{FFT}$ shown in Tables~\ref{tab:2p_perf}, \ref{tab:4p_perf} and \ref{tab:8p_perf}, in the order of tens of GB/s, a viable solution is to use the novel HBM technology, instead of standard DDR4 banks. HBM allows greater bandwidth than DDR4, but it is limited in size (8 or 16 GB), while DDR4 size limit is much higher and multiple banks can be placed.

FPGAs with HBM memory has been recently announced by major vendors and are nowadays available as engineering samples. In late 2019 it is expected to have devices on production silicon. Synthesis and simulation tools are not yet fully supporting this technology, and this has not permitted to include a fully working block in this work.

However, we can sketch an architecture for the DMA Controller block that is responsible to handle the data flowing from the Network controller towards the HBM memory controller, and from the HBM memory controller towards the Y or Z FFT engines.

\begin{figure}[!t]
\centering
\includegraphics[width=\textwidth]{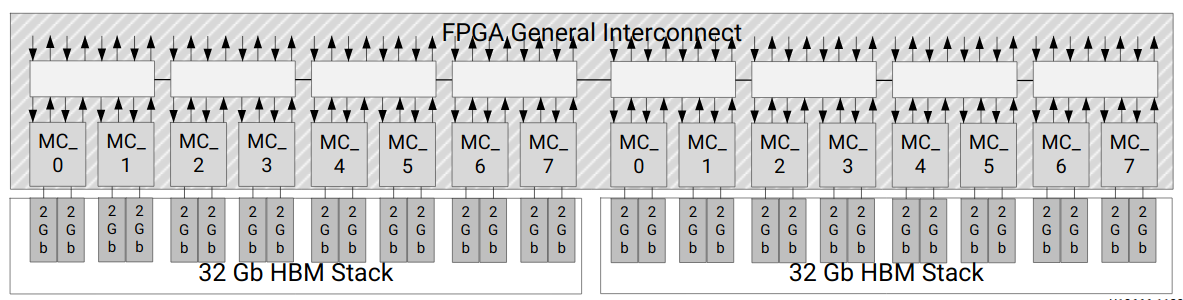}
\caption{Xilinx HBM Controller in a two stack configuration.}
\label{hbm}
\end{figure}

The Xilinx HBM memory controller \cite{xilinx_pg276} provides access to one or two HBM memory stack, for a total of 8 GB of size and 32 independent AXI channels. Each channel is 256-bit wide and runs up to 450 MHz, depending on the speed grade of the device, while memory clock runs at 900 MHz. The internal architecture the IP Core is depicted in Fig.~\ref{hbm}. Each port can optionally address the entire HBM address space (global addressing) with an internal $32 \times 32$ crossbar to allow direct access across both HBM stacks.

When global addressing is enabled, a minimum latency of 128 memory clock cycles is declared. Moreover, burst length has a possible range between 32 and 512 bytes, implying a great fragmentation for large read or write transactions. In order to exploit the full available bandwidth, it is thus very important to make use of all 32 available AXI ports.

\begin{figure}[!t]
\centering
\includegraphics[width=\textwidth]{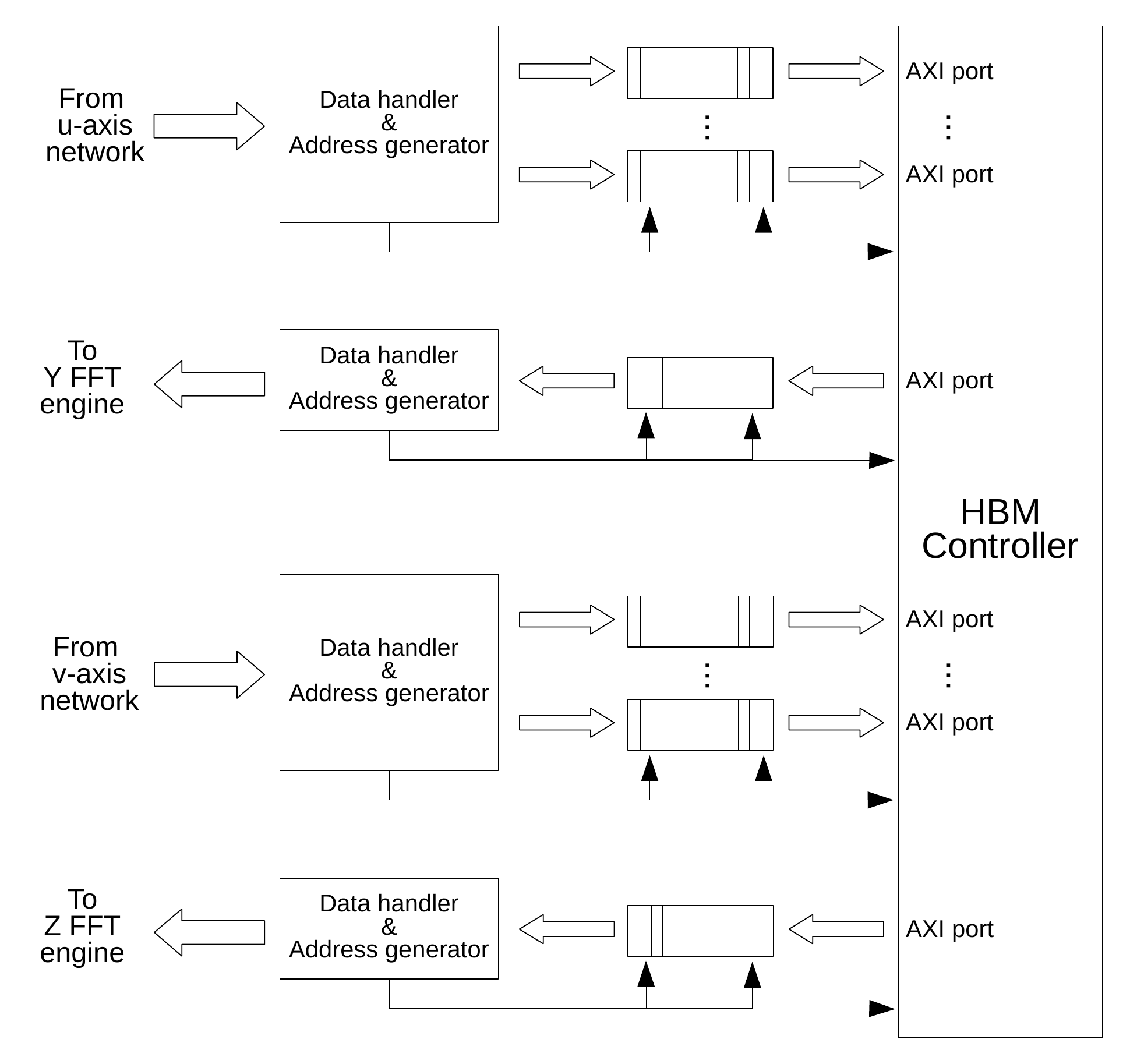}
\caption{Logic architecture of the DMA Controller interfacing the HBM controller.}
\label{dma_ctrl}
\end{figure}

In our context, assuming the pipelined architecture as the hardware computational model, we should divide the 32 ports among the 4 peers:
\begin{itemize}
    \item X-transformed data write with stride N: a data word of 16 byte is written every $16N$ bytes in the address space;
    \item X-transposed data read with stride 1: a data word of 16 byte is read in consecutive addresses;
    \item Y-transformed data write with stride $N^2$: a data word of 16 byte is written every $16N^2$ bytes in the address space;
    \item Y-transposed data read with stride 1: a data word of 16 byte is read in consecutive addresses;
\end{itemize}
and a data write transaction with stride N or $N^2$ matches with the request of data fragmentation and splitting over multiple ports. Of course, due to the latency for setting up a single transaction, some effort must be accomplished in order to aggregate consecutive data by means of special logic blocks with FIFOs and perform maximum burst length transactions. A scheme of these specialized controller is shown on Fig.~\ref{dma_ctrl}, where a series of FIFOs are used primarily for data aggregation, and secondary to decouple clock domain between FFT logic and HBM controller. The exact number of AXI ports for each data flow must be tuned when reliable simulation models will be available.

With this kind of mechanism as DMA Controller, it is expected to guarantee the full data throughput $B_{FFT}$ requirements in the above-mentioned pipelined architecture.

\section{Network integration}

\begin{figure}[!t]
\centering
\includegraphics[width=.8\textwidth]{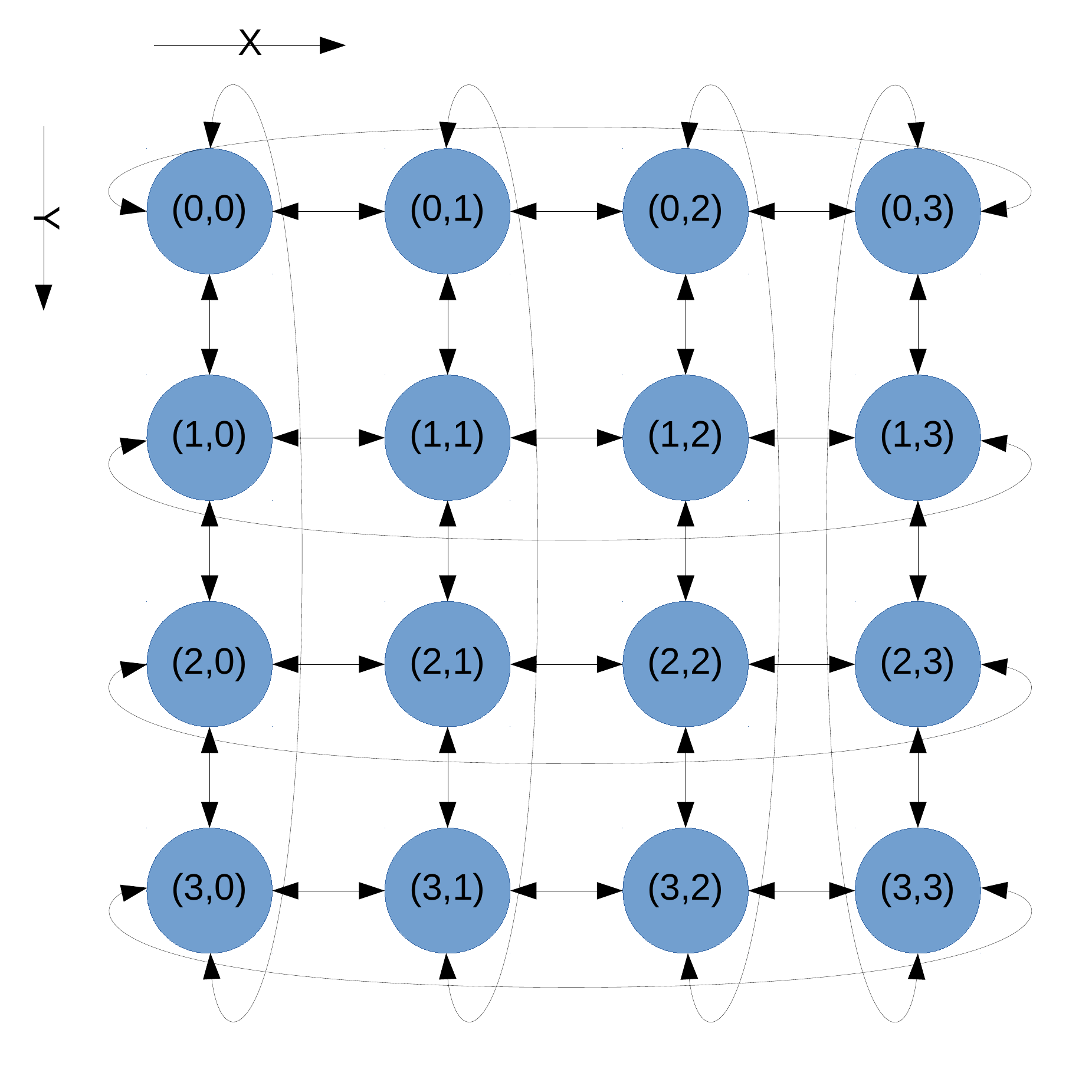}
\caption{Two-dimensional Torus Network.}
\label{2d_mesh}
\end{figure}

\begin{figure}[!t]
\centering
\includegraphics[width=.8\textwidth]{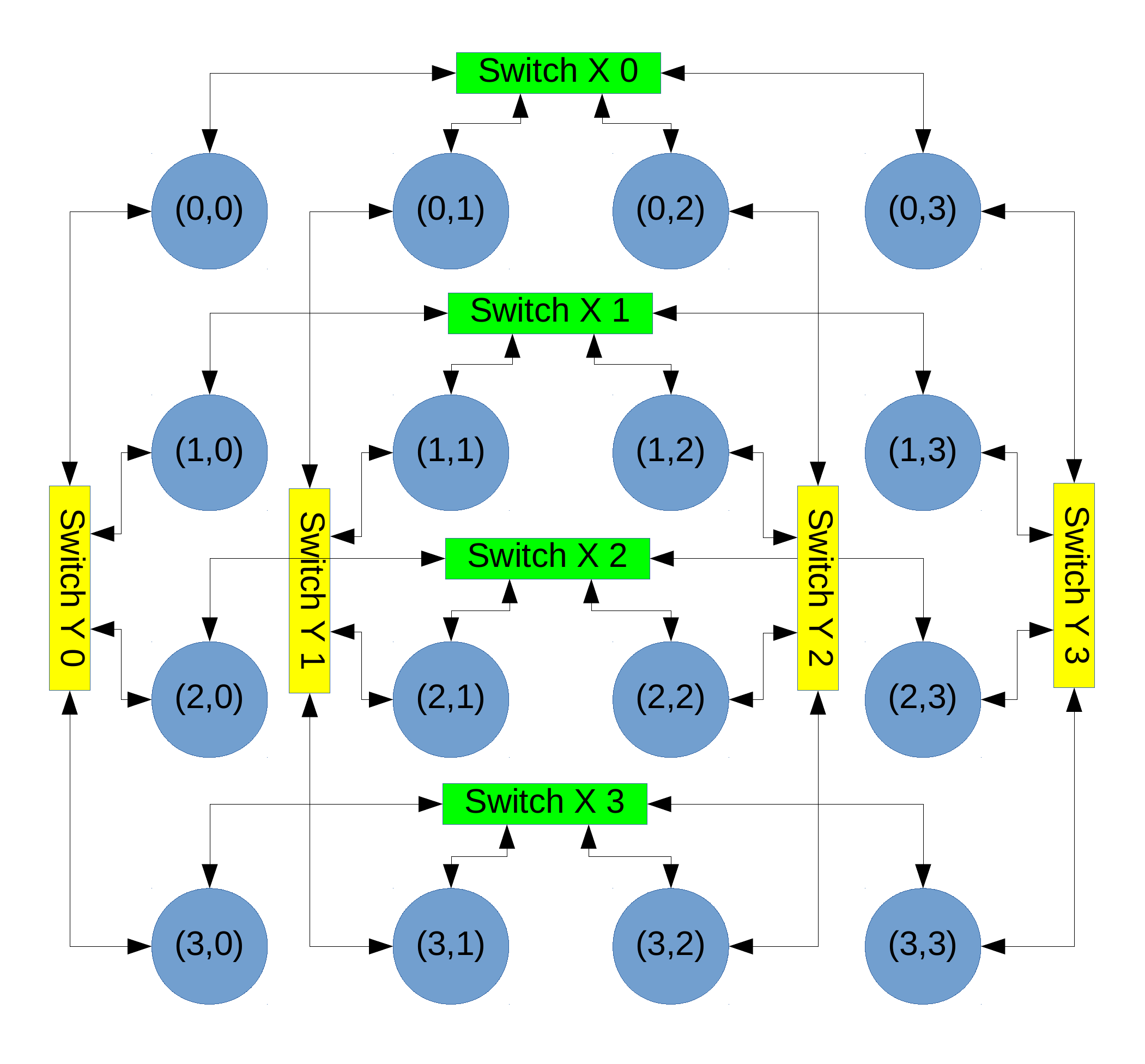}
\caption{Two-dimensional mesh of switches.}
\label{2d_switch}
\end{figure}

In this section we describe the network controller block, calculating the network required bandwidth in two cases:
\begin{itemize}
    \item a two dimensional torus network of bi-directional point-to-point links, such as \apenet described in Sec.~\ref{sec:apenet};
    \item a two dimensional switched network, \textit{i.e.} a two dimensional grid of external switch devices connecting nodes in the same row and in the same column. In this case the switch can be based on standard Ethernet technology (\textit{e.g.} a product like in \cite{fs_switch}), to be used with the UDP/IP core described in Sec.~\ref{sec:updip}.
\end{itemize}
These two topologies are shown respectively in Fig.~\ref{2d_mesh} and Fig.~\ref{2d_switch}. Generally speaking we can describe these topologies of $P$ computing nodes as a grid of size $P_u \times P_v$ along the two Cartesian axis $u$ and $v$. We will assume $P_u=P_v=\sqrt{P}$ in the following for the sake of simplicity.

The main difference between the two topologies is in the number of required Network Interfaces on each computing node, that is on each FPGA: 4 for the torus and 2 for the switched network. Another difference is that a torus network, requiring no external devices, has hypothetically no limits in system size, while in the other case we are limited by existing devices; nevertheless 32-port switches are common off-the-shelf products, and this enables building grids of up to 1024 computing nodes, which is enough for our study.

We can then model the network required bandwidth for the two cases. When using a switched network, assuming that the switch can guarantee full bisection bandwidth \footnote{In computer networking, if the network is bisected into two partitions, the bisection bandwidth of a network topology is the bandwidth available between the two partitions.}, the network bandwidth is simply the data throughput $B_{FFT}$ multiplied by the fraction of data that must be transferred over the network, thus resulting:

\begin{equation}
    B_{Net}^{switch}=B_{FFT} \frac{\sqrt{P}-1}{\sqrt{P}} = \frac{4 s R}{t_{clk}} \frac{\sqrt{P}-1}{\sqrt{P}}
\end{equation}
being $s$ the size in byte of the data word (8 bytes).

When using a torus network, we have to consider the bandwidth degradation due to multi-hop communication of distant nodes. This will produce a further $\sqrt{P}/2$ factor to the previous model, thus resulting:
\begin{equation}
    B_{Net}^{torus}=\frac{2 s R}{t_{clk}} (\sqrt{P}-1)
\end{equation}

\begin{figure}[!t]
\centering
\includegraphics[width=\textwidth]{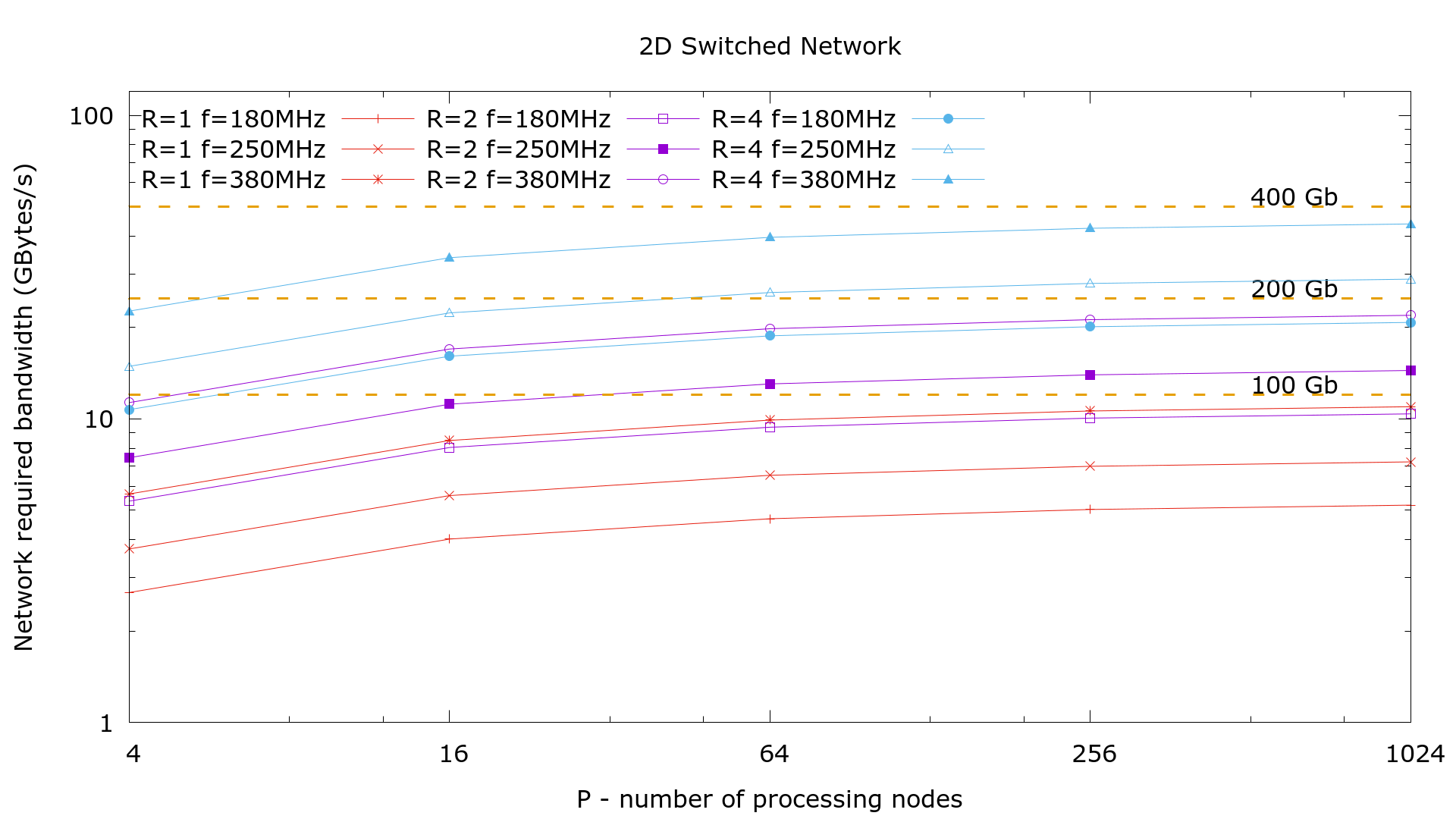}
\caption{Network required bandwidth comparison for a 2D Switched topology model.}
\label{net_bw_switch}
\end{figure}

\begin{figure}[!t]
\centering
\includegraphics[width=\textwidth]{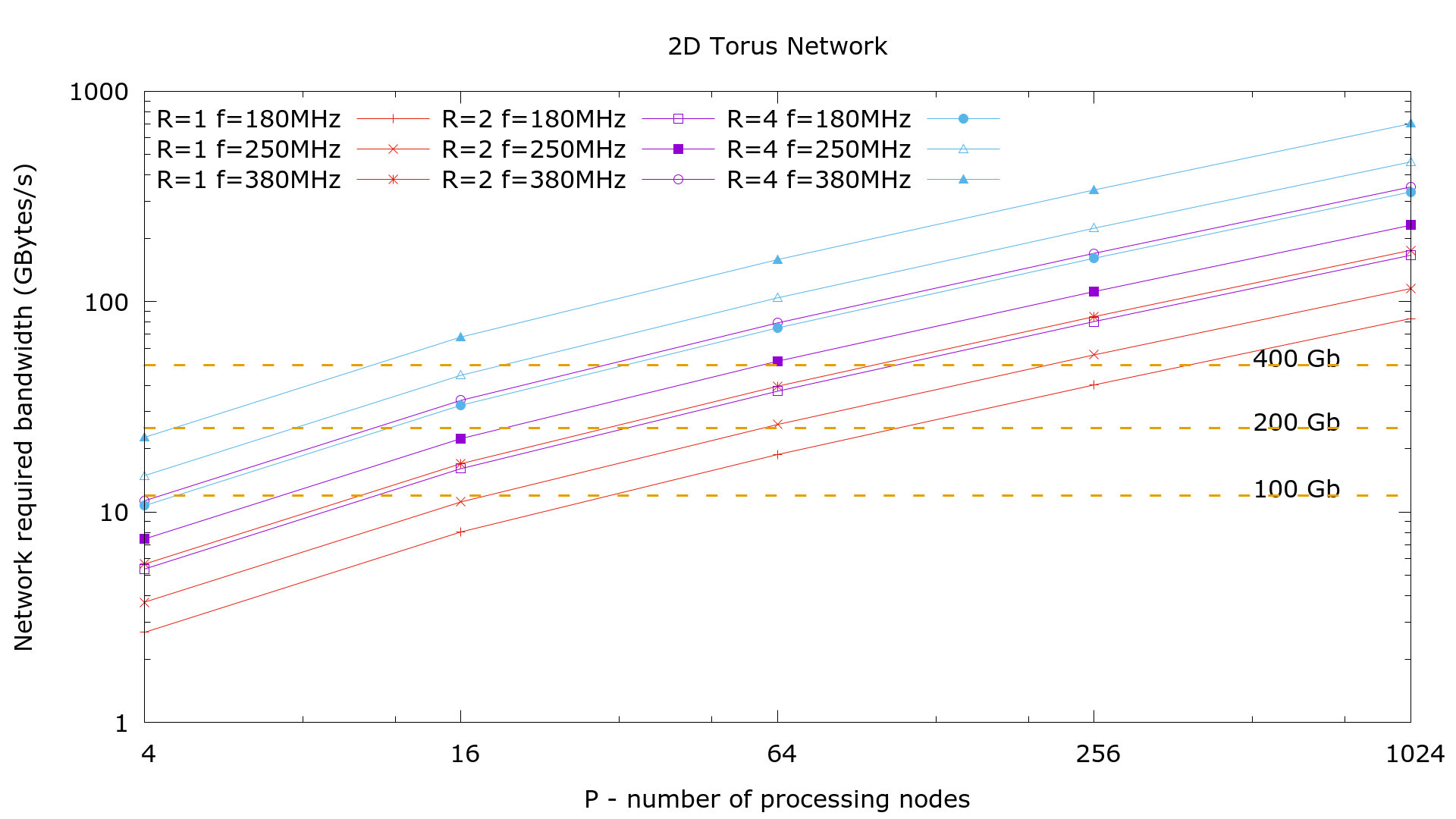}
\caption{Network required bandwidth comparison for a 2D Torus topology model.}
\label{net_bw_torus}
\end{figure}

For a comparison of the two equations above, we have drawn in Fig.~\ref{net_bw_switch} --- for the switched topology --- and Fig.~\ref{net_bw_torus} --- for the torus topology --- the cases for $R=1,2,4$ plotted at three working frequencies $f=1/t_{clk}=180, 250, 380$ MHz. The three working regimes are taken as an example of, respectively, a slow, a standard and a very fast implementation of the FFT core. In the plots are also shown, as a reference, the 100 and 200 Gb/s state of the art peak network channel capacities on target FPGA. It is also shown the 400 Gb/s link limit, as a possible, near future advance, already announced by main FPGA vendors in the next generation devices.

It can be noted from the trend of the plots, the torus network suffers the penalty of multi-hop routing as the size of the network get larger, and the switched network seems to be the only scalable solution. Nevertheless, when exploiting maximum amount of DSP blocks, \textit{i.e.} using $R=4$ implementation, the working frequency must be moderated to a slow value, in order the limit of the network required bandwidth to a sustainable value of the network channel capacity.

We can draw the conclusion that the torus network can be considered a good solution for small size of the network (\textit{i.e.} $\sqrt{P} \leq 4$), while the switched network ensures a sustainable scalability up to $\sqrt{P} \leq 32$, being 32 the maximum available size of a switch guaranteeing full bisection bandwidth.

\section{Global 3D FFT expected performances}

We finally envision the expected performances (in terms of calculation time) of an entire system solving an $N^3$ FFT problem, composed of P processing nodes assembled with a two dimensional 200 Gb/s class $\sqrt{P}\times\sqrt{P}$ switched network. We explore the range of $P=1, \dots, 1024$. The computing engines are organized following the pipelined architecture model with multiplicity $Q=4$ (2 engines for X FFT, 1 for Y FFT, 1 for Z FFT); the FFT engines are instanced with $R=4$ and working at a very conservative $f=180$ MHz frequency. With these parameters we have demonstrated that neither the network nor the memory represent a bottleneck in the pipelined flow, thus the global behaviour can be considered as ideal.

In this framework we can estimate, according to equation \ref{eq:pipe_model}, the overall calculation time for the global 3D FFT solution for $\mu = 1,3$ dimensional observables. Results are shown in Table~\ref{tab:time_perf}.

\begin{table}[]
    \centering
    \begin{tabular}{|c|c|c|c|c|c|c|}
        \hline
        \large{N} & \multicolumn{6}{c|}{\large{P}}              \\ \cline{2-7} 
           & \textbf{1} & \textbf{4}   & \textbf{16}  & \textbf{64}  & \textbf{256}   & \textbf{1024}  \\ 
           &  & $(2\times2)$ & $(4\times4)$ & $(8\times8)$ & $(16\times16)$ & $(32\times32)$ \\
        \hline
        \multicolumn{7}{|c|}{\cellcolor[HTML]{EFEFEF} $\mu = 1$} \\
        \hline
        \textbf{512}  & 0.17 & 0.047 & 0.011 & 0.0029 & 0.00073 & 0.00018 \\
        \textbf{1024} &      & 0.37  & 0.093 & 0.023  & 0.0058  & 0.0014 \\
        \textbf{2048} &      &       & 0.74  & 0.19   & 0.047   & 0.012  \\
        \textbf{4096} &      &       &       &        & 0.37    & 0.093  \\
        \textbf{8192} &      &       &       &        &         & 0.75   \\
        \hline
        \multicolumn{7}{|c|}{\cellcolor[HTML]{EFEFEF} $\mu = 3$} \\
        \hline
        \textbf{512}  & 0.37 & 0.093 & 0.023 & 0.0058 & 0.0015 & 0.00036 \\
        \textbf{1024} &      & 0.75  & 0.19  & 0.047  & 0.012  & 0.0029 \\
        \textbf{2048} &      &       & 1.49  & 0.37   & 0.093  & 0.023  \\
        \textbf{4096} &      &       &       &        & 0.75   & 0.19   \\
        \textbf{8192} &      &       &       &        &        & 1.49   \\
        \hline
    \end{tabular}
    \caption{Expected calculation time (in seconds) for a 3D FFT for a scalar and a three-dimensional observable. Empty cells are cases where local FPGA memory size exceeds 8 GB limit.}
    \label{tab:time_perf}
\end{table}


\begin{table}[]
    \centering
    \begin{tabular}{|c|c|c|c|c|c|c|c|}
        \hline
        \large{N} & \multicolumn{7}{c|}{\large{P} (cores)}              \\ \cline{2-8}
          & \textbf{8}   & \textbf{16}   & \textbf{32}   & \textbf{64}   & \textbf{128}  & \textbf{256}   & \textbf{512} \\
          & (512) & (1024) & (2048) & (4096) & (8192) & (16348) & (32768) \\
        \hline
        \textbf{1024} & 1.20 & 0.67 & 1.61 & 0.29 & 0.18 &      &      \\
        \textbf{2048} &      & 48.2 & 3.75 & 2.26 & 4.90 & 0.74 & 0.41 \\
        \hline

    \end{tabular}
    \caption{Measured calculation time (in seconds) for a 3D FFT solution for a scalar observable on a 64-core Xeon Phy cluster.}
    \label{tab:phy_perf}
\end{table}

For comparison we report in Table~\ref{tab:phy_perf} the 3D FFT performance measured on the Cineca Marconi cluster \cite{marconi}. Here the computing node is consisting of a 64 core Intel Xeon Phy accelerator equipped with 100 Gb/s-class Intel OmniPath. Although the compared systems are very different in architecture and computing node concept and notwithstanding the reduced comparable data set, these results show that our proposed system has higher strong scaling performances. Of course, a fair and complete comparison should use properly measured timing performances on comparable test-beds.

%% file: sections/conclusions.tex
\chapter*{\textsc{Concluding remarks}} 
\label{chap:conclusion}
\addcontentsline{toc}{chapter}{Concluding Remarks}  
\rhead{ }

Nowadays FPGAs are increasingly used as accelerators in computing facilities. The main idea in this work is to couple the networking capabilities (high throughput, low latency, custom topologies, OS bypass) of FPGAs, with the enormous potential they have in accelerate computational tasks. The scientific problem addressed is the solution of the 3D FFT for large sizes, in the range $N=512, \dots, 8192$, driven by numerical simulations of computational fluid dynamics in turbulent regime, where the motion equations are solved with the pseudo-spectral method.

A computing architecture have been proposed, comprised of an collection of $P$ nodes consisting of an FPGA with its own HBM memory bank and a private custom network. The computational problem is divided in specific hardware tasks, and different organizations of these tasks are discussed. A performance and resource model is proposed in order to compare pros and cons of the different approaches.

Two network interface types, derived from the \apenet and \nanet projects, have been described and discussed as building block of the network interconnect of the proposed system. Within this work, it has been developed and used in the context of various experiments, a multi-protocol cross-platform 1/10/40/100 Gb Ethernet core.

The implementation of the parallel-pipelined double precision floating point FFT engine is described, developed with particular regard on hardware resource scalability and timing performance. In particular it has been described a streaming architecture with 2 or 4 or 8 (complex double precision) data word per clock cycle and how this parameter, combined with pipeline latency of each floating point operator, impact on global performances and hardware resource usage.

This kind of architecture allows to make use of most of the floating point ideal capability of the designed target device, and the expected performance, in terms of GFLOPS, seems very promising if compared to current state of the art accelerators. The critical point, when increasing the performances, is the network required bandwidth, which follows this trend. A specific network infrastructure is then proposed in order to exploit the peculiarities of the communication patterns for the 3D FFT problem. We can then build a two-dimensional switched network closely attached to the FFT engines and the memory controllers within the FPGAs, in order to orchestrate FFT calculation and data transposition among the computational phases.

Such a designed system allows very promising performances both on small scale and on large scale, with $P \leq 1024$.

\chapter*{\textsc{Acknowledgment}} 
\label{chap:ack}
\addcontentsline{toc}{chapter}{Acknowledgment}

This work was carried out with the support from the TurboNet project and from the \nanet project. Both are funded
by the INFN Fifth National Scientific Committee for technological research.

The author would like to thank Fabio Bonaccorso, Michele Buzzicotti, Gaetano Salina and Pierpaolo Loreti for valuable discussions that have made this work possible; Luca Biferale for the access at the NewTurb and Marconi clusters.